\documentclass[twocolumn,showkeys,showpacs,preprintnumbers,prd,superscriptaddress,nofootinbib,aps,10pt]{revtex4-1}
\usepackage{graphicx,epsf,bm,amsmath,amsfonts,amssymb,epstopdf,natbib,color,verbatim,multirow,bm,mathtools,mathrsfs,braket,bbold,float}
\usepackage{hyperref}
\hypersetup{colorlinks=true,urlcolor=blue,citecolor=blue,linkcolor=blue,menucolor=blue,anchorcolor=blue,filecolor=blue}
\date{\today}
            
\begin{document}

\title{Imprints of screened dark energy on nonlocal quantum correlations}

\author{Fabiano Feleppa}
\email{ffeleppa@unisa.it}
\affiliation{Dipartimento di Fisica “E.R. Caianiello”, Università di Salerno, Via Giovanni Paolo II 132, I-84084 Fisciano (SA), Italy \looseness=-1}
\affiliation{INFN Sezione di Napoli, Gruppo Collegato di Salerno, Italy \looseness=-1}
\affiliation{Nordic Institute for Theoretical Physics (NORDITA), Hannes Alfv\'{e}ns v\"{a}g 12, SE-114 19 Stockholm, Sweden \looseness=-1}

\author{Gaetano Lambiase}
\email{lambiase@sa.infn.it}
\affiliation{Dipartimento di Fisica “E.R. Caianiello”, Università di Salerno, Via Giovanni Paolo II 132, I-84084 Fisciano (SA), Italy \looseness=-1}
\affiliation{INFN Sezione di Napoli, Gruppo Collegato di Salerno, Italy \looseness=-1}

\author{Sunny Vagnozzi}
\email{sunny.vagnozzi@unitn.it}
\affiliation{Department of Physics, University of Trento, Via Sommarive 14, 38123 Povo (TN), Italy}
\affiliation{Trento Institute for Fundamental Physics and Applications (TIFPA)-INFN, Via Sommarive 14, 38123 Povo (TN), Italy}

\begin{abstract}
\noindent We investigate how screening mechanisms, reconciling light scalar fields driving cosmic acceleration with local fifth force constraints, can be probed via their impact on nonlocal quantum correlations between entangled spin pairs, whose evolution on a curved background is affected by General Relativity (GR) and screened modified gravity effects. We consider a gedankenexperiment featuring a pair of massive, spin-$1/2$ particles orbiting the Earth, evaluating their nonlocal correlations through spin observables associated to the Clauser-Horne-Shimony-Holt (CHSH) inequality. Using a general formalism developed earlier for curved space-time spin evolution, we compute the effects of screening on the CHSH inequality, finding its degree of violation to be suppressed relative to the flat space-time case. Applying this formalism to the chameleon, symmetron, and dilaton mechanisms, we identify currently unconstrained regions of parameter space where the screening contribution is comparable to that of GR. While detecting these effects will be challenging, our work provides a proof-of-principle for testing screened dark energy through quantum nonlocality.
\end{abstract}


\maketitle

\section{Introduction}
\label{sec:introduction}

One of the most pressing open questions in fundamental physics is the origin of the dark energy (DE) component responsible for the observed late-time accelerated expansion of the Universe, initially inferred from Type Ia supernovae~\cite{SupernovaSearchTeam:1998fmf,SupernovaCosmologyProject:1998vns}, and now corroborated by a wide range of probes~\cite{Nadathur:2020kvq,DiValentino:2020evt,Escamilla:2023oce}. The simplest possibility where DE takes the form of a cosmological constant $\Lambda$ is plagued by multiple issues, ranging from the severe fine-tuning required if $\Lambda$ is interpreted as zero-point vacuum energy density of quantum fields~\cite{Carroll:2000fy,Weinberg:1988cp} (see, however, also Refs.~\cite{Moreno-Pulido:2022phq,SolaPeracaula:2022hpd}), to a growing disagreement with several recent cosmological observations~\cite{Yang:2018euj,Guo:2018ans,Vagnozzi:2019ezj,Visinelli:2019qqu,Alestas:2020mvb,DiValentino:2021izs,Dainotti:2021pqg,Vagnozzi:2023nrq,Akarsu:2024qiq,DESI:2024mwx,Cortes:2024lgw,Colgain:2024xqj,Carloni:2024zpl,Giare:2024smz,Gomez-Valent:2024tdb,Yang:2024kdo,DESI:2024kob,Toda:2024ncp,Giare:2024gpk,Jiang:2024xnu,RoyChoudhury:2024wri,Giare:2024oil,Gomez-Valent:2024ejh,Giare:2025pzu,CosmoVerse:2025txj,RoyChoudhury:2025dhe,Scherer:2025esj}. An alternative plausible and intriguing possibility is one where the origin of cosmic acceleration lies in additional, typically (pseudo)scalar degrees of freedom, either due to new fundamental particles and fields, or modifications to General Relativity (GR)~\cite{Ratra:1987rm,Wetterich:1987fm,Caldwell:1997ii,Zlatev:1998tr,Capozziello:2002rd,Li:2004rb,Nojiri:2006ri,Bamba:2012cp,Chamseddine:2013kea,Rinaldi:2014yta,Sebastiani:2016ras,Nojiri:2017ncd,Dutta:2017fjw,Casalino:2018tcd,Saridakis:2018unr,Visinelli:2018utg,Odintsov:2019evb,Saridakis:2020zol,Odintsov:2020zct,Oikonomou:2020qah,Motta:2021hvl,Drepanou:2021jiv,Oikonomou:2022wuk,Trivedi:2024inb,Lin:2025gne}, whose signatures are among the science goals of ongoing and upcoming cosmological surveys~\cite{SimonsObservatory:2018koc,SimonsObservatory:2019qwx}. Such scenarios do not come without challenges, since these degrees of freedom would need to be extremely light (at least on cosmological scales, in order to recover the observed properties of DE), and will typically couple to matter at least with gravitational strength, unless a fundamental protection mechanism or symmetry is at play~\cite{Carroll:1998zi,Amendola:1999er}. Among other interesting effects, this would give rise to yet unobserved, and therefore unwanted, long-range fifth forces.

Unless one is willing to accept a significant fine-tuning in the model parameters, a mechanism which dynamically suppresses these fifth forces on local scales, while allowing for sizeable deviations from GR on cosmological scales, is clearly required.~This is exactly what is achieved by so-called \textit{screening mechanisms}, which dynamically adjust the fifth force's properties (e.g., its range or strength) in order to suppress the force in high-density environments, and evade local gravity tests (see, e.g., Ref.~\cite{Brax:2021wcv} for a recent review). Well-known examples of screening mechanisms include the chameleon~\cite{Khoury:2003aq,Khoury:2003rn}, symmetron~\cite{Hinterbichler:2010es}, dilaton~\cite{Damour:1994zq}, and Vainshtein mechanisms~\cite{Vainshtein:1972sx}. Several experimental strategies have been proposed to test these mechanisms, ranging from cosmological and astrophysical observations, to a variety of laboratory experiments (e.g.\ involving fifth force searches, or precision atomic and/or neutron tests)~\cite{Brax:2021wcv}. While significant portions of parameter space are excluded by current constraints, especially in the case of the chameleon mechanism, a number of open windows still remain (the reader is referred to Ref.~\cite{Burrage:2017qrf} for a review on the subject). Given the importance of screening mechanisms, the task of developing new tests that either complement existing probes or can access currently open windows in parameter space is undoubtedly a crucial and timely endeavor. Since most screening mechanisms can be viewed as modifications of gravity within scalar-tensor theories~\cite{Sakstein:2014jrq,Brax:2021wcv}, several precision tests of gravity can be reinterpreted as probes of screening mechanisms. In this context, a novel possibility towards testing deviations from GR has recently emerged, exploiting the impact of space-time curvature on the evolution of nonlocal quantum correlations~\cite{Bittencourt:2020lgu}. The nonlocal nature of quantum mechanics, once considered problematic~\cite{Einstein:1935rr}, is now recognized as a fundamental feature of the theory~\cite{Aspect:1982fx}. It is captured by violations of Bell's inequalities~\cite{Bell:1964kc,Clauser:1969ny,Brunner:2013est}, a phenomenon that can only arise in entangled states. In the presence of space-time curvature, and considering an Einstein-Podolsky-Rosen (EPR) spin singlet, it was realized in the seminal work by Terashima and Ueda~\cite{Terashima:2003rjs} that gravity-induced spin precession leads to an apparent reduction in the amount of nonlocality between the two spin states, quantified by the degree of violation of Bell-type inequalities. This occurs because the entangled spin states undergo a succession of infinitesimal local Lorentz transformations, leading to a net Wigner rotation that alters the degree of nonlocality~\cite{Czachor:1996cj,Gingrich:2002ota,Terashima:2003ds,Friis:2009va,LeviSaid:2009yra}. Since screening mechanisms can be recast in the language of scalar-tensor theories (specifically, through a relationship between the Einstein-frame and Jordan-frame metrics, with matter fields minimally coupled to the latter~\cite{Brax:2021wcv}) it is natural to ask whether they can leave an imprint on nonlocal quantum correlations. If so, this would open a new avenue for testing screening mechanisms.~\footnote{There has recently been significant interest in novel tests of fundamental physics using nonlocal quantum correlations, as well as in exploring such correlations in contexts other than the usual ones, such as colliders experiments or cosmology (see e.g.\ Refs.~\cite{Abel:1992kz,Ancochea:1998nx,Bramon:1998nz,Hiesmayr:2000rm,Bertlmann:2001sk,Genovese:2001pk,Gallicchio:2013iva,Maldacena:2015bha,Choudhury:2016cso,Chen:2016xqz,Choudhury:2016pfr,Kanno:2017dci,Martin:2017zxs,Choudhury:2017bou,Feng:2018ebt,Ali:2021jch,Fabbrichesi:2021npl,Choudhury:2021mht,Barr:2021zcp,Takubo:2021sdk,Gong:2021bcp,Severi:2021cnj,Espinosa-Portales:2022yok,Aguilar-Saavedra:2022uye,Aguilar-Saavedra:2022wam,Sinha:2022crx,Peruzzo:2022tog,Choudhury:2022mch,Fabbrichesi:2023cev,Dale:2023fnp,Ghosh:2023rpj,Dudal:2023pbc,Petruzziello:2023xhb,Fabbrichesi:2023idl,Dong:2023xiw,Morales:2023gow,Dudal:2023mij,Moradpour:2023gsn,Fabbri:2023ncz,Bernal:2023ruk,Bi:2023uop,DeFabritiis:2023tkh,Ma:2023yvd,Tselentis:2023kwh,Han:2023fci,Cheng:2023qmz,Ehataht:2023zzt,Li:2024luk,Guedes:2024tcq,Barr:2024djo,Morales:2024jhj,Sou:2024tjv,Wu:2024qhd,Fabbrichesi:2024wcd,Bernal:2024xhm,DeFabritiis:2024jfy,Guo:2024jch,Wu:2024asu,Fabbrichesi:2024rec,Cheng:2024btk,Chen:2024drt,Gabrielli:2024kbz,Ruzi:2024cbt,Du:2024sly,Wu:2024ovc,Grabarczyk:2024wnk,Cheng:2025cuv,Han:2025ewp,Varma:2025gkp,Dale:2025nhc,Zhang:2025mmm,Afik:2025grr,Qi:2025onf,Lin:2025eci,Abel:2025skj}).}

In the present pilot study, we address this question explicitly. Using a general formalism previously developed in Ref.~\cite{Terashima:2003rjs}, we compute the effects of screening mechanisms on Bell-type inequalities, focusing on the Clauser-Horne-Shimony-Holt (CHSH) inequality; our approach is rather general and formulated in terms of the post-Newtonian parameters $\beta$ and $\gamma$, as well as the effective gravitational constant ${\cal G}$, all of which determine the Jordan frame metric. We then illustrate step by step how to apply this formalism in practice to screening mechanisms of interest, specializing to the chameleon, symmetry, and dilaton mechanisms for concreteness. In a nutshell, we find that appreciable deviations in the degree of violation of the CHSH inequality, relative to the expectations within GR, can be observed within regions of parameter space currently unconstrained by other tests. While reaching the experimental sensitivity required to observe these effects is extremely challenging, our results provide a proof-of-principle for testing screening mechanisms using tools rooted within quantum theory, motivating future experimental efforts to pursue such tests: to the best of our knowledge, ours represents the first proposal to test screened scalar fields through quantum nonlocality, setting up a new bridge at the emerging and particularly fertile interface between gravitation and (experimental) quantum information~\cite{Rovelli:1990pi,Capozziello:2013wea,Giacomini:2017zju,Luongo:2019ztg,Belenchia:2019gcc,Castro-Ruiz:2019nnl,Galley:2020qsf,Giacomini:2021aof,Belfiglio:2022cnd,Christodoulou:2022knr,Overstreet:2022zgq,Belfiglio:2022yvs,Galley:2023byb,Luongo:2023jyz,Belfiglio:2023moe,Chen:2024xvm,Belfiglio:2024qsa,Capozziello:2024mxh,Belfiglio:2024wel,Cortes:2024jvb,Trivedi:2024qys,Bagchi:2024try,Odintsov:2024ipb,Abreu:2025mxa,Belfiglio:2025cst}.

The rest of this work is then organized as follows. In Sec.~\ref{sec:spincurved}, we review the evolution of spin states in curved space-time, with particular attention to the Wigner rotation which spin-$1/2$ states undergo. Sec.~\ref{sec:nonlocalityscreening} is then devoted to presenting the formalism we employ. In particular, in Sec.~\ref{subsec:scalartensortheories} we begin by expressing the general Jordan frame metric describing the gravitational field around a point-like mass at first post-Newtonian order within screened theories; in Sec.~\ref{subsec:nonlocal}, we then evaluate the degree of nonlocality of a pair of spins evolving on this background, providing explicit general expressions for the degree of reduction in violation of the CHSH inequality relative to both the flat space-time and GR cases; finally, in Sec.~\ref{subsec:application}, after introducing the chameleon, symmetron, and dilaton mechanisms, we specialize the previous general results to these three cases (Sec.~\ref{subsubsec:chameleonapplication}, Sec.~\ref{subsubsec:symmetronapplication}, and Sec.~\ref{subsubsec:dilatonapplication}, respectively), identifying regions of parameter space where these effects could be significant.~In Sec.~\ref{sec:results}, we discuss potential detection prospects by introducing an EPR gedankenexperiment around the Earth and evaluating the degree of violation of the CHSH inequality across the relevant parameter spaces for the three mechanisms (Sec.~\ref{subsec:chameleonresults}, Sec.~\ref{subsec:symmetronresults}, and Sec.~\ref{subsec:dilatonresults}, respectively), followed by a brief qualitative discussion of experimental feasibility in Sec.~\ref{subsec:detectability}. We summarize our main findings and outline future directions in Sec.~\ref{sec:conclusions}. Unless otherwise specified, we adopt units where $c=\hbar=1$.

\section{Spin evolution in curved space-time}
\label{sec:spincurved}

We now briefly review the formalism required for treating the evolution of spin states in curved space-time. The main difficulty is that, unlike in flat space-time, there is no global inertial frame to serve as a universal reference for defining spin. Instead, we need to rely on the concept of local rotational symmetry defined at each point of the manifold, within the corresponding local inertial frame (LIF). As a particle moves along its worldline in curved space-time, its spin undergoes a continuous succession of infinitesimal Wigner rotations (WRs),  induced by local Lorentz transformations (LLTs) between successive LIFs. This precession arises both from the particle's acceleration caused by external forces (if present) and from continuous changes in LIFs along its path.Our goal is to review the formalism necessary to compute the net WR angle obtained by integrating these infinitesimal rotations. We will then apply this formalism to scenarios where screening mechanisms are at play, in order to assess how they modify GR's predictions. For an exhaustive treatment of the formalism, we encourage the reader to consult Refs.~\cite{Terashima:2003rjs,Bittencourt:2020lgu}.

The starting point for treating spin in a general curved space-time, characterized by a metric tensor $g_{\mu\nu}(x)$, is to establish a LIF at each point of our manifold. This is naturally achieved within the tetrad formalism. Specifically, at each point of the manifold we introduce a LIF by using the vierbein/tetrad field $e_a^{\mu}(x)$, as well as its inverse $e_{\mu}^a$. These are defined by the following relations:
\begin{gather}
e_a^{\mu}(x) e_b^{\nu}(x) g_{\mu\nu}(x) = \eta_{ab}\,,
\label{eq:flatcurved}\\
e_a^{\mu}(x) e_{\nu}^a = \delta_{\nu}^{\mu}, \quad e_{\mu}^a(x) e_b^{\mu} = \delta_b^a\,,
\label{eq:vierbeininverses}
\end{gather}
where $\eta_{ab} = {\text{diag}}(-1,1,1,1)$ is the Minkowski metric. In what follows, we use Latin indices ($a,b,\ldots=0,1,2,3$) as local inertial coordinate labels, whereas the Greek alphabet is used for general coordinate indices. In the general coordinate system, indices are lowered and raised using the metric $g_{\mu\nu}(x)$ and its inverse $g^{\mu\nu}(x)$, respectively. Similarly, in the LIF indices are lowered and raised using the Minkowski metric $\eta_{ab}$ and its inverse $\eta^{ab}$ respectively, which share the same matrix components. We note that Eq.~(\ref{eq:flatcurved}) represents a map between general curved space-time and flat LIFs at each point of the manifold; specifically, this map allows us to define quantities locally, by converting a generic tensorial object to its LIF counterpart using vierbeins. For example, a rank $(1,1)$ tensor $V_{\nu}^{\mu}$ can be converted to its LIF counterpart $V^a_b$ via $V^{\mu}_{\nu} \to V^a_b=e^a_{\mu}e^{\nu}_bV_{\nu}^{\mu}$, and analogously for tensors of arbitrary rank. It is important to note that the LIF is not uniquely determined, as it remains inertial under a Lorentz transformation. In other words, the vierbeins possess a residual gauge freedom associated to LLTs, which in what follows we assume to be described by $\Lambda^a_b(x)$.

We now consider spin states in curved space-time. For definiteness, in what follows we specialize to spin-$1/2$ particles, but the tools we discuss can be adapted to particles of different spin. In flat space-time, the spin state of a particle is defined by its transformation under a particular representation of the Lorentz group. In curved space-time, it is natural to define the spin state of a spin-$1/2$ particle within the corresponding representation of the LLT group. Let us now consider a massive spin-$1/2$ particle whose rest mass is $m$, and whose worldline is parametrized by its proper time $\tau$. The particle's four-velocity is then $u^{\mu}(x)=dx^{\mu}/d\tau$, while its four-momentum is $p^{\mu}(x)=mu^{\mu}(x)$. In a given LIF, its four-momentum is given by $p^a(x)=e_{\mu}^a(x)p^{\mu}(x)$. We then denote the quantum state of this particle as $\ket{p^a(x),\sigma;x}$, where $\sigma=\{\uparrow, \downarrow\}$ is the third component of the spin; in other words, this is a local spin eigenstate labeled by momentum and spin projection along the third direction in the LIF. We now recall that (local) Lorentz transformations act as unitary operators on quantum states. Under a LLT described by $\Lambda$,~\footnote{For simplicity, we sometimes omit LIF indices when their presence is clear from the context, i.e.\ we refer to $\Lambda^a_b(x)$ as $\Lambda$.} the state is acted upon by the operator $U[\Lambda(x)]$ and transforms as follows:
\begin{equation}
U[\Lambda(x)]\ket{p^a(x),\sigma;x} = \sum_{\sigma^\prime} D^{(1/2)}_{\sigma^\prime \sigma}[W(x)] \ket{\Lambda p^a(x),\sigma^\prime;x}\,.
\label{eq:statetransformation}
\end{equation}
In the above, $D_{\sigma^\prime \sigma}^{(1/2)}[W(x)]$ is the so-called Wigner $D$-matrix, a $2 \times 2$ unitary matrix acting on the space of spin states, corresponding to the spin-$1/2$ irreducible representation of the WR matrix $W(x) \equiv W(\Lambda(x), p(x))$. This rotation captures how the spin transforms under a Lorentz transformation $\Lambda$, and the explicit form of $D_{\sigma^\prime \sigma}^{(1/2)}[W(x)]$ will be specified later on. The WR matrix $W(x)$ appearing in Eq.~(\ref{eq:wignerrotationmatrix}) can then be derived by considering a transformation from a generic reference frame to the particle's rest frame, and is found to be given by the following expression (see, e.g., Refs.~\cite{Terashima:2003rjs,Bittencourt:2020lgu}):
\begin{equation}
W(x)=L^{-1}[\Lambda(x) p(x)]\Lambda(x)L[p(x)]\,,
\label{eq:wignerrotationmatrix}
\end{equation}
where $L[p(x)]$ denotes the Lorentz boost that takes the particle from its rest frame, where its four-momentum reads $p_{\text{rest}}^a=(m,0,0,0)$, to a generic frame with four-momentum $p(x)=p^a(x)$.

Our discussion thus far has been completely generic. To make progress, we now need to derive a specific form for Eq.~(\ref{eq:statetransformation}), expressed in terms of the metric coefficients (encoding the curvature of space-time) and, where relevant, the particle's acceleration due to external forces. To do so, we assume that over the course of an infinitesimal proper time interval $d\tau$, the particle moves from $x^{\mu}$ to $x^{\prime\mu}=x^{\mu}+u^{\mu}d\tau$, whereas its four-momentum changes from $p^a(x)$ to $p^a(x^{\prime})$. The variation in four-momentum $\delta p^a(x)$ is then given by the following:
\begin{equation}
\delta p^a(x) = \delta p^{\mu}(x)e_{\mu}^a(x) + p^{\mu}(x)\delta e_{\mu}^a(x)\,.
\label{eq:deltapx}
\end{equation}
In particular, the first term in Eq.~(\ref{eq:deltapx}) accounts for the change in momentum due to external forces (if present), where $\delta p^{\mu}(x)$ reads as follows:
\begin{equation}
\delta p^{\mu}(x) = -\frac{p^{\nu}(x)}{m} \left [ a^{\mu}(x)p_{\nu}(x) - p^{\mu}(x)a_{\nu}(x) \right ] d\tau\,,
\label{eq:deltapmux}
\end{equation}
with $a^{\mu}(x)$ being the particle’s four-acceleration. The second term in Eq.~(\ref{eq:deltapx}) instead reflects the change in LIF and reads as follows:
\begin{equation}
\delta e_{\mu}^a(x) = -u^{\nu}(x)\omega_{\nu b}^a(x)e_{\mu}^b(x) d\tau \equiv \chi_b^ae_{\mu}^b(x) d\tau,
\label{eq:deltaemuax}
\end{equation}
where $\omega_{\mu b}^a = e_{\nu}^a(x) \nabla_{\mu}e_b^{\nu}(x)$ is the spin connection, reflecting the effects of space-time curvature. Additionally, we have defined $\chi^a_b \equiv -u^{\nu}(x) \omega_{\nu b}^a(x)$. Combining Eqs.~(\ref{eq:deltapx},\ref{eq:deltapmux},\ref{eq:deltaemuax}), the effect of changing momentum and local frame can be expressed in the following compact form:
\begin{equation}
\delta p^a(x) = \lambda_b^a(x) p^b(x)\,,
\end{equation}
where $\lambda_b^a(x)$ is given by the following:
\begin{equation}
\lambda_b^a (x)= \frac{1}{m} \left [ p^a(x)a_b(x)-a^a(x)p_b(x) \right ] + \chi_b^a(x)\,.
\end{equation}
Gathering the previous results we find that, as the particle moves through curved space-time over the course of an infinitesimal proper time interval $d\tau$, its four-momentum transforms as follows:
\begin{equation}
p^a(x^\prime) = \Lambda_b^a(x)p^b(x)\,,
\end{equation}
where $\Lambda_b^a(x)$, which encodes the effects of an infinitesimal LLT, reads as follows:
\begin{equation}
\Lambda_b^a(x) = \delta_b^a + \lambda_b^a(x)d\tau\,,
\label{eq:lambdax}
\end{equation}
with $\delta_b^a$ being the Kronecker delta. Importantly, spin states then transform via representations of the LLT described by $\Lambda^a_b$.

The infinitesimal LLT described by Eq.~(\ref{eq:lambdax}) is associated to an infinitesimal WR, as in Eq.~(\ref{eq:wignerrotationmatrix}). The matrix elements of $W^a_b$ can be written compactly as follows:
\begin{equation}
W_b^a(x) = \delta_b^a + \theta_b^a(x)d\tau\,,
\end{equation}
where the only non-zero $\theta_j^i$ elements are the following ones:
\begin{equation}
\theta_{j}^{i}(x)=\lambda_{j}^{i}(x)+\frac{\lambda_{0}^{i}(x)p_{j}(x)-\lambda_{i0}(x)p^{j}(x)}{p^{0}(x)+m}\,.
\label{eq:theta}
\end{equation}
All the other components $\theta_j^i$ are zero, in other words:
\begin{equation}
\theta_{0}^{0}(x)=\theta_{i}^{0}(x)=\theta_{0}^{i}(x)=0\,.
\end{equation}
In our case, the spin-$1/2$ irreducible representation of this infinitesimal WR is given by the following $D$-matrix:
\begin{multline}
D_{\sigma^\prime \sigma}^{(1/2)}[W(x)] = \\ \mathbb{1}
+\frac{i}{2} \left [ \theta_{23}(x)\sigma_{x}+\theta_{31}(x)\sigma_{y}+\theta_{12}(x)\sigma_{z} \right ] d\tau\,,
\label{eq:spin12repr}
\end{multline}
where $\mathbb{1}$ denotes the $2 \times 2$ identity matrix, whereas $\sigma_x$, $\sigma_y$, and $\sigma_z$ are the Pauli matrices. Inserting Eq.~(\ref{eq:spin12repr}) into Eq.~(\ref{eq:statetransformation}) allows us to study the evolution of a particle's spin state due to an infinitesimal LLT, taking place over the course an infinitesimal proper time interval $d\tau$.

We have so far focused on an infinitesimal LLT. For finite proper time intervals, one needs to integrate a series of such infinitesimal LLTs throughout the particle's path in curved space-time. Let us assume that, in a finite proper time interval $\tau_f-\tau_i$, the particle moves from $x^{\mu}(\tau_{i})$ to $x^{\mu}(\tau_{f})$. Then, it can be shown that this integration becomes a Dyson series which converges to the following expression for the $D$-matrix:
\begin{multline}
D_{\sigma^{\prime} \sigma}^{(1/2)}[W(x)] = T\exp\Bigg\{ \frac{i}{2} \int_{\tau_i}^{\tau_f} d\tau\,\Big[ \theta_{23}(x(\tau))\, \sigma_x \\
+\theta_{31}(x(\tau))\sigma_y+\theta_{12}(x(\tau))\sigma_z \Big] \Bigg\}\,,
\label{eq:spin12reprfinite}
\end{multline}
where $T$ denotes the time-ordering operator. This expression captures the total precession (or, more precisely, evolution) accumulated by the spin state due to continuous changes in both acceleration and space-time curvature throughout the particle's journey (see Refs.~\cite{Terashima:2003rjs,Bittencourt:2020lgu} for further details and discussions on the topic).

\section{nonlocal correlations in screened modified gravity}
\label{sec:nonlocalityscreening}

We now discuss how the tools introduced in the previous section can be applied to the case study of nonlocal correlations in screened modified gravity, specializing to the chameleon, symmetron, and dilaton mechanisms. Since all three mechanisms fall within the class of scalar-tensor theories of gravity, and we later consider a gedankenexperiment around the Earth, we begin by reviewing scalar-tensor theories with a particular focus on the general Jordan-frame metric describing the gravitational field of a point-like mass at first post-Newtonian order, closely following Ref.~\cite{Zhang:2016njn}. We then consider a pair of spin states evolving on this background, evaluate their degree of nonlocal correlations, and specialize the general expression we obtain to the three screening mechanisms we are studying.

\subsection{Scalar-tensor theories}
\label{subsec:scalartensortheories}

Scalar-tensor theories of gravity extend GR by introducing one or more scalar fields alongside the metric tensor~\cite{Fujii:2003pa}. These fields interact with the metric and potentially with matter as well, modifying the dynamics of both the gravitational and matter fields. Such theories naturally emerge in a variety of fundamental contexts, ranging from the low-energy limit of string theory, to attempts to explain cosmic acceleration and/or dark matter. In what follows, we will limit our discussion to the case where only one additional scalar field is present (although bi-scalar extensions have been widely considered in the literature~\cite{Rinaldi:2014gha,Ohashi:2015fma,Saridakis:2016mjd,Petronikolou:2023cwu,Cecchini:2024xoq}).

A key feature of scalar-tensor theories is the freedom to choose a frame, which plays a crucial role in determining how the field couples to gravity and matter. Two of the most widely used frames are the Jordan frame (JF) and the Einstein frame (EF). Whether these two frames are physically equivalent remains an open question: while they are generally considered equivalent at the classical level, it is unclear whether this equivalence persists once quantum corrections are taken into account. This may have important consequences for observables, as predictions for measurable quantities may depend on the chosen frame, a topic which is the subject of ongoing research (see, e.g., Refs.~\cite{Magnano:1993bd,Capozziello:1996xg,Faraoni:1999hp,Faraoni:2006fx,Capozziello:2010sc,Steinwachs:2011zs,Ren:2014sya,Postma:2014vaa,Kamenshchik:2014waa,Domenech:2015qoa,Myrzakulov:2015qaa,Quiros:2015bfa,Sakstein:2015jca,vandeBruck:2015gjd,Banerjee:2016lco,Kamenshchik:2016gcy,Bahamonde:2016wmz,Pandey:2016unk,Jarv:2016sow,Mathew:2017lvh,Karam:2017zno,Azri:2018gsz,Rinaldi:2018qpu,Falls:2018olk,Nashed:2019yto,Giacomini:2020grc,Elizalde:2020icc,Oikonomou:2020oex,Bamonti:2021jmg,Copeland:2021qby,Racioppi:2021jai,Galaverni:2021jcy,Shtanov:2022wpr,Paliathanasis:2022tmt,Nojiri:2022ski,Paliathanasis:2022akr,Karciauskas:2022jzd,DeAngelis:2022qhm,Bamber:2022eoy,Odintsov:2022bpg,Schiavone:2022wvq,Oikonomou:2023dgu,Velasquez:2023jld,Diaz:2023tma,Luongo:2024opv,GiontiSJ:2023tgx,Seleim:2023enf,Belfiglio:2024swy,Galaverni:2025acu,Wang:2025ger}). In the JF, the scalar field $\widetilde{\phi}$, or a function of it, multiplies the Ricci scalar constructed out of the JF metric $\widetilde{g}_{\mu\nu}$, to which matter is minimally coupled. Specifically, in the JF the action takes the following form~\cite{Faraoni:1999hp}:
\begin{multline}
S = \int d^4 x \, \sqrt{-\widetilde{g}} \, \frac{M_{\text{Pl}}^2}{2} 
\Bigg[ \widetilde{\phi}\,\widetilde{R} 
- \frac{\omega(\widetilde{\phi})}{\widetilde{\phi}} 
\widetilde{g}^{\mu\nu}\widetilde{\nabla}_{\mu}\widetilde{\phi}
\widetilde{\nabla}_{\nu}\widetilde{\phi} 
\\
- U(\widetilde{\phi}) \Bigg]
+ \int d^4x \,\sqrt{-\widetilde{g}}\, 
\mathcal{L}_m\!\left(\psi_m,\widetilde{g}_{\mu\nu}\right) \,,
\label{eq:jf}
\end{multline}
where $\widetilde{g}$ is the metric determinant, $\widetilde{R}$ the Ricci scalar constructed from $\widetilde{g}_{\mu\nu}$, $M_{\text{Pl}}=\sqrt{1/8\pi G}$ the reduced Planck mass, $\omega(\widetilde{\phi})$ the coupling function, $U(\widetilde{\phi})$ the JF scalar field potential, and $\mathcal{L}_m$ the matter Lagrangian, which depends on the matter fields $\psi_m$. From Eq.~(\ref{eq:jf}), it is clear that in the JF the scalar field is non-minimally coupled to curvature, whereas matter is minimally coupled to the metric. Consequently, the field equations in the JF differ from the Einstein equations of GR, although matter moves along geodesics of the JF metric.

To express the theory in the EF, we perform a conformal transformation $\widetilde{g}_{\mu\nu} = F^2(\phi)g_{\mu\nu}$ and appropriately redefine the scalar field, so that the action takes the following form~\cite{Faraoni:1999hp}:
\begin{multline}
S = \int d^4x\,\sqrt{-g} \left [ \frac{M_{\text{Pl}}^2}{2}R-\frac{1}{2}g^{\mu\nu}\nabla_{\mu}\phi\nabla_{\nu}\phi -V(\phi) \right ]
\\
+ \int d^4x \sqrt{-g}\,\mathcal{L}_m \left ( \psi_{m},F^2(\phi)g_{\mu\nu} \right ) \,,
\label{eq:ef}
\end{multline}
where $F(\phi)$ and $V(\phi)$ are the (conformal) coupling function and EF scalar field potential, respectively. In the EF, the scalar field couples minimally to gravity, and the gravitational sector of the theory is given by the standard Einstein-Hilbert term of GR. The trade-off for this apparent simplicity is that matter fields now no longer couple directly to the EF metric, but to the JF metric $F^2(\phi)g_{\mu\nu}$. Therefore, in the EF the field equations are those of GR, but matter does not move along geodesics of the EF metric, as it is non-minimally coupled to the scalar field.

In most studies of modified theories of gravity, it is assumed (either explicitly or implicitly) that observations are made in the JF. More precisely, since one is ultimately concerned with measurable quantities, the assumption is that the matter composing the detector of the experiment is minimally coupled to the JF metric. While it is in principle possible to work within the EF, the non-minimal coupling of the scalar field to matter significantly complicates the interpretation of observables such as distances and times, which must be carefully redefined to account for the variation of physical units with the scalar field -- a task that is rarely, if ever, addressed in such studies~\cite{Faraoni:2006fx}. For these reasons, in accordance with standard practice, and mostly as a matter of practical convenience, in what follows we find it significantly more convenient to carry out our analysis in the JF.

As already anticipated, the gedankenexperiment we will consider later is imagined to be located around the Earth, where the gravitational field is weak enough to justify the use of the parametrized post-Newtonian (PPN) formalism. In scalar-tensor theories, at first post-Newtonian (1PN) order, the metric describing the gravitational field of a point-like mass can be written in terms of the PPN parameters $\gamma$ and $\beta$, as well as the effective gravitational ``constant'' $\mathcal{G}$. These three PPN parameters have been calculated in various works: see, e.g., Ref.~\cite{Hohmann:2013rba} for a detailed calculation in the JF, and Ref.~\cite{Scharer:2014kya} where these quantities have been expressed in terms of the conformal coupling function. However, the authors of Ref.~\cite{Zhang:2016njn} pointed out that applying the results of Refs.~\cite{Scharer:2014kya,Hohmann:2013rba} to screening mechanisms may be problematic, due to the inherent approximation of point-like sources. While this approximation is acceptable when solving the equations for the metric field, it is not suitable for the scalar field, whose properties depend on the surrounding matter density.

To address this issue, the authors of Ref.~\cite{Zhang:2016njn} start by working in the EF and deriving an approximate solution for the scalar field in the case of an extended source surrounded by a homogeneous background. Varying the action of Eq.~(\ref{eq:ef}) with respect to the scalar field, and assuming non-relativistic matter, the equation of motion for the scalar field is given by
\begin{equation}
g^{\mu\nu}\nabla_{\mu}\nabla_{\nu}\phi = \frac{dV_{\text{eff}}}{d\phi}\,,
\label{eq:scalareom}
\end{equation}
with the effective potential $V_{\text{eff}}(\phi)$ defined as follows:
\begin{equation}
V_{\text{eff}} = V(\phi)+\rho F(\phi)\,,
\label{eq:veff}
\end{equation}
where $\rho$ is the (conserved) scalar field energy density in the EF~\cite{Zhang:2016njn}.~\footnote{It is worth noting that there exist different conventions for defining the effective potential in the literature, see footnote~3 of Ref.~\cite{Burrage:2017qrf} for further discussions on the matter.} As we can observe, the dynamics of the scalar field are governed by the effective potential $V_{\text{eff}}(\phi)$, which explicitly depends on $\rho$. With appropriately chosen functions $V(\phi)$ and $F(\phi)$, the effective potential may have a minimum (an assumption we will adopt from this point forward) allowing the scalar field to acquire an effective mass that varies with the ambient density. At this stage, our discussion is still deliberately generic; later, we will specialize to different choices of $V(\phi)$ and $F(\phi)$, and thereby to different screening mechanisms. The value of the scalar field at the minimum of the potential, which we denote by $\phi_{\min} = \phi_{\min}(\rho)$, can be obtained by simply solving the following equation:
\begin{equation}
\frac{dV_{\text{eff}}}{d\phi}\bigg\rvert_{\phi_{\min}} = 0\,,
\label{eq:minimumpotential}
\end{equation}
while the effective mass squared of the scalar field, denoted as $m_{\text{eff}}^2$, is defined as follows:
\begin{equation}
m_{\text{eff}}^2 \equiv \frac{d^2V_{\text{eff}}}{d\phi^2}\bigg\rvert_{\phi_{\text{min}}}\,.
\label{eq:effectivemass}
\end{equation}
In Ref.~\cite{Zhang:2016njn}, the analysis is restricted to the case of a static and spherically symmetric source of radius $R$ and total mass $M_E$ (with the subscript $E$ indicating that this is the EF mass). We work in isotropic coordinates, and denote the EF radial coordinate by $\xi$. The matter density profile $\rho(\xi)$ is then set to $\rho_c$ for $\xi<R$, and to $\rho_{\infty}$ for $\xi>R$, where $\rho_{\infty}$ represents the homogeneous matter density surrounding the source. We denote by $\phi_c$ and $\phi_{\infty}$ the values of $\phi$ that minimize the effective potential for $\rho = \rho_c$ and $\rho = \rho_{\infty}$ respectively, with the corresponding scalar field masses being $m_c$ and $m_{\infty}$. By requiring that the solution to Eq.~(\ref{eq:scalareom}) is non-singular at the origin and that the scalar field asymptotically converges to its background value, while neglecting the field backreaction on the metric, it is found that the exterior scalar field profile is given by the following expression~\cite{Zhang:2016njn}:~\footnote{Here we are interested only in the exterior solution, as the gedankenexperiment we will consider later involves spin states orbiting around the Earth.}
\begin{equation}
\phi(\xi) \simeq \phi_{\infty} - \epsilon M_{\text{Pl}}\frac{GM_{E}}{\xi}e^{-m_{\infty}\xi}\,,
\label{eq:sfsol}
\end{equation}
where the parameter $\epsilon \ll 1$, whose physical interpretation is that of scalar charge, takes the form~\cite{Brax:2013ida}
\begin{equation} 
\epsilon \equiv \frac{\phi_{\infty} - \phi_{c}}{M_{\text{Pl}}\Phi_{\text{N}}}\,,
\label{eq:scalarcharge}
\end{equation}
with $\Phi_{\text{N}} = GM_E/R$ being the Newtonian potential at the surface of the body. 

Having addressed the issue of the scalar field profile, we turn our attention to the metric field in the PPN approximation. As previously mentioned, unlike in the evaluation of the scalar field, here we work under the assumption of a point-like source.~\footnote{While reasonable, this approach is inherently an approximation. A calculation of the PPN parameters for a homogeneous, spherical mass source was carried out in Ref.~\cite{Hohmann:2017qje} which, in turn, also constitutes an approximation. Further refinements would require relaxing this assumption and accounting for the density profile of the gravitating source. Such an analysis, however, would be significantly more involved and would require numerical methods, and is beyond the scope of the present pilot study.} Using the solution for the scalar field given by Eq.~(\ref{eq:sfsol}), the authors of Ref.~\cite{Zhang:2016njn} performed a first-order PPN analysis in the EF. However, the PPN parameters $\gamma$ and $\beta$, as well as $\mathcal{G}$, are instead defined in the JF, which is related to the EF by a conformal transformation. The JF metric around a point-like mass at 1PN is found to be given as~\cite{Zhang:2016njn}:~\footnote{As already mentioned, in what follows we work with the JF metric. However, to express the latter in terms of the coupling function's expansion coefficients [see Eq.~\eqref{eq:expansionF}], it is necessary to perform a PPN expansion in the EF.}
\begin{equation}
\begin{aligned}
\widetilde{ds}^2 &= F(\phi)^2 ds^2 = \\
&\phantom{=}\, - \left ( 1 - \frac{2\mathcal{G}(\epsilon)M}{r}+\beta(\epsilon)\frac{4\mathcal{G}^2(\epsilon)M^2}{2r^2} \right ) dt^2 \\
&\phantom{=}\, + \left ( 1 + \gamma(\epsilon) \frac{2\mathcal{G}(\epsilon)M}{r} \right ) \left( dr^2 + r^2 d\Omega^2 \right ) \,,
\end{aligned}
\label{eq:ds2jf}
\end{equation}
where $d\Omega^2 \equiv d\vartheta^2+\sin^2\vartheta d\varphi^2$ is the metric on the unit two-sphere, and the functions $\mathcal{G}(\epsilon)$, $\gamma(\epsilon)$, and $\beta(\epsilon)$ are given by the following expressions:
\begin{align}
&\mathcal{G}(\epsilon) = G F_{\infty}^2 \left [ 1 + \left ( \frac{F_1M_{\text{Pl}}}{F_{\infty}} - \frac{V_1}{M_{\text{Pl}}m_{\infty}^2}\right)\epsilon \right ] \,,
\label{eq:Geps} \\
&\gamma(\epsilon) = 1 - \left ( \frac{2F_1M_{\text{Pl}}}{F_{\infty}} - \frac{3V_1}{2M_{\text{Pl}}m_{\infty}^2} \right ) \epsilon \nonumber \\
& \quad \quad \quad + \left ( \frac{2F_1^2M_{\text{Pl}}^2}{F_{\infty}^2} - \frac{7V_1F_1}{2m_{\infty}^2F_{\infty}} + \frac{3V_1^2}{2M_{\text{Pl}}^2m_{\infty}^4} \right ) \epsilon^2\,,
\label{eq:gammaeps} \\
&\beta(\epsilon) = 1 + \frac{3V_1}{4M_{\text{Pl}}m_{\infty}^2}\epsilon \nonumber \\
&+ \left [ \left ( \frac{F_2}{F_{\infty}} - \frac{F_1^2}{2F_{\infty}^2} \right ) M_{\text{Pl}}^2 - \frac{3V_1F_1}{2m_{\infty}^2F_{\infty}} + \frac{3V_1^2}{4M_{\text{Pl}}^2m_{\infty}^4} \right ] \epsilon^2\,.
\label{eq:betaeps}
\end{align}
In the JF, we use $r$ to denote the radial coordinate and $M$ to denote the mass of the gravitating source. The relation between these quantities in the JF and their EF counterparts is given by~\cite{Zhang:2016njn}
\begin{equation}
M = \frac{M_E}{F_{\infty}}\,, \quad r = F_{\infty}\xi\,.
\label{eq:massjfef}
\end{equation}
In Eqs.~(\ref{eq:Geps}--\ref{eq:betaeps}), which are valid in the context of local tests of gravity (see Ref.~\cite{Zhang:2016njn} for details about the approximations made), $F_{\infty} \equiv F(\phi_{\infty})$, $F_1$ and $F_2$ represent the zeroth-, first- and second-order coefficients, respectively, in the 1PN expansion of the conformal coupling function $F(\phi)$ around the cosmological background value, which we previously denoted by $\phi_{\infty}$. This expansion is performed as follows:
\begin{gather} 
\label{eq:expansionF}
F(\phi) \simeq F_{\infty} + F_1 \left ( \phi - \phi_{\infty} \right ) + F_2 \left ( \phi - \phi_{\infty} \right ) ^2\,, \\
F_1 \equiv \frac{dF(\phi)}{d\phi}\bigg\rvert_{\phi_{\infty}}\,, \quad F_2 \equiv \frac{1}{2}\frac{d^2F(\phi)}{d\phi^2}\bigg\rvert_{\phi_{\infty}}\,. \nonumber
\end{gather}
Similarly, $V_1$ and $V_2$ are the first- and second-order coefficients in the 1PN expansion of the scalar potential $V(\phi)$ around $\phi_{\infty}$:
\begin{gather}
\label{eq:expansionV}
V(\phi) \simeq V_{\infty} + V_1 \left ( \phi - \phi_{\infty} \right ) + V_2 \left ( \phi - \phi_{\infty} \right ) ^2\,, \\
V_1 \equiv \frac{dV(\phi)}{d\phi}\bigg\rvert_{\phi_{\infty}}\,, \quad V_2 \equiv \frac{1}{2}\frac{d^2V(\phi)}{d\phi^2}\bigg\rvert_{\phi_{\infty}}\,. \nonumber
\end{gather}
The zeroth-order coefficient in the expansion of the potential, $V_{\infty} \equiv V(\phi_{\infty})$, serves as an effective cosmological constant which one ultimately expects to be linked to the phenomenon of cosmic acceleration. However, this term does not appear in Eqs.~(\ref{eq:Geps}--\ref{eq:betaeps}) because, assuming it is offset by a constant down to the observed DE density, its effects within the Solar System are negligible. For further details on the derivation of the above formulas, including other approximations used, see Ref.~\cite{Zhang:2016njn}.

For our purposes, we find it convenient to rewrite the line element in Eq.~(\ref{eq:ds2jf}) as follows:
\begin{eqnarray}
\widetilde{ds}^2 &=& F(\phi)^2ds^2 \label{eq:ds2pot} \\
&=& - \left ( 1 + 2\Phi \right ) dt^2 + \left ( 1 - 2\Psi \right ) \left ( dr^2 + r^2d\Omega^2 \right ) \,, \nonumber 
\end{eqnarray}
where the (screened) gravitational potentials $\Phi$ and $\Psi$ take the following forms:
\begin{align}
&\Phi = \Phi(r,\epsilon) \equiv -\frac{\mathcal{G}(\epsilon)M}{r} + \beta(\epsilon)\frac{\mathcal{G}^2(\epsilon)M^2}{r^2}\,,
\label{eq:phiscreen}\\
&\Psi = \Psi(r,\epsilon) \equiv -\gamma(\epsilon)\frac{\mathcal{G}(\epsilon)M}{r}\,.
\label{eq:psiscreen}
\end{align}
We now note that Eq.~(\ref{eq:ds2pot}) is formally identical to Eq.~(25) of Ref.~\cite{Bittencourt:2020lgu}. Therefore, we can rely on the results of Secs.~IV and V therein, since, in the context of the nonlocal correlations we consider, these results are independent of the specific forms of the metric potentials, and hold at first order in $\Phi$ and $\Psi$.

A clarification is in order before moving forward. Screening mechanisms can be interpreted either as modifications to gravity, where the scalar field alters the effective metric coupling to matter (if seen from the JF perspective), or as an additional scalar degree of freedom non-minimally coupled to matter within GR (if seen from the EF perspective). From the JF viewpoint, it is natural to refer to these mechanisms as ``modified gravity'', whereas from the EF viewpoint it is natural to view the system as standard GR plus an appropriately screened fifth force~\cite{Sakstein:2014jrq}. The choice of terminology thus depends on the theoretical perspective, but physically both describe the same phenomenology; in what follows, since as discussed earlier we work within the JF, we will oftentimes refer to screening mechanisms as ``screened modified gravity'', a terminology which is not uncommon in the literature.

\subsection{nonlocal correlations}
\label{subsec:nonlocal}

The gedankenexperiment we propose to test screened modified gravity follows the same setup as described in Ref.~\cite{Terashima:2003rjs,Bittencourt:2020lgu}. Specifically, we consider an EPR source emitting a pair of entangled spin-$1/2$ particles in opposite directions, within the weak gravitational field of the Earth.~In spherical coordinates, the source is located at the azimuthal coordinate $\varphi=0$, while two observers with detectors are situated at $\varphi=\pm\bar{\varphi}$. The particles undergo non-geodesic motion, traveling along a circular orbit confined to the equatorial plane, $\vartheta = \pi/2$ (see, e.g., Fig.~1 of Ref.~\cite{Bittencourt:2020lgu}). Upon reaching the observers, their spin components are measured along specific directions, and the corresponding measurement operators will be introduced later.

Initially, we assume that the two-particle system is in the spin-singlet state, described by the following ket:
\begin{equation}
\ket{\chi} = \frac{1}{\sqrt{2}} \left ( \ket{p_+^a, \uparrow; 0} \ket{p_-^a, \downarrow; 0} - \ket{p_+^a, \downarrow; 0} \ket{p_-^a, \uparrow; 0} \right ) \,,
\label{eq:stateinitial}
\end{equation}
where we recall that $p^a=e_{\mu}^ap^{\mu}$ is the four-momentum in the LIF, in this case given by the following:
\begin{equation}
p_{\pm}^a = \left ( m\cosh\zeta,0,0,\pm m\sinh\zeta \right )\,,
\end{equation}
with $m$ being the particle's rest mass and $\zeta$ the rapidity in the LIF, related to the particle's velocity $v$ in the same LIF by $v=\tanh({\zeta})$. In Eq.~(\ref{eq:stateinitial}), the subscript $+$ refers to the particle moving in the $+\bar{\varphi}$ direction (henceforth particle $A$), and conversely the subscript $-$ refers to the particle moving in the $-\bar{\varphi}$ direction (henceforth particle $B$). The above state is a superposition of the two possible configurations where one particle has spin up and the other spin down, and encodes a perfect anti-correlation in the spins: if the spin of particle $A$ is measured to be up along a certain direction, the spin of particle $B$ will be down along the same direction, and vice versa.

To quantify the degree of quantum nonlocality in the system, we use the CHSH inequality~\cite{Clauser:1969ny}, a Bell-type inequality based on dichotomic measurement operators, which sets an upper limit on correlations which can be explained by any local hidden variable theory. Violations of the inequality indicate that the correlations between the particles are inherently nonlocal. In \textit{flat space-time}, we would perform spin measurements for particle $A$ at $\varphi = \bar{\varphi}$ along the directions $\mathbf{a}=(1,0,0)$ and $\mathbf{a^{\prime}}=(0,1,0)$, associated with operators $\mathcal{\hat{Q}}$ and $\mathcal{\hat{R}}$, respectively. Similarly, for particle $B$ we perform spin measurements at $\varphi = -\bar{\varphi}$ along the directions $\mathbf{b}=(-1,-1,0)/\sqrt{2}$ and $\mathbf{b^{\prime}}=(1,-1,0)/\sqrt{2}$, associated with operators $\mathcal{\hat{S}}$ and $\mathcal{\hat{T}}$, respectively. These measurement directions are chosen to maximize the violation of the CHSH inequality for the entangled state $\ket{\chi}$, leading to the following value of the CHSH parameter $\mathscr{S}$:
\begin{eqnarray}
\mathscr{S}[\ket{\chi}] &\equiv& \left| \langle \chi | \hat{\mathcal{Q}} \hat{\mathcal{S}} | \chi \rangle + \langle \chi | \hat{\mathcal{R}} \hat{\mathcal{S}} | \chi \rangle \right. \nonumber \\
&& \left. + \, \, \langle \chi | \hat{\mathcal{R}} \hat{\mathcal{T}} | \chi \rangle - \langle \chi | \hat{\mathcal{Q}} \hat{\mathcal{T}} | \chi \rangle \right| \nonumber \\
&=& 2\sqrt{2}\,.
\end{eqnarray}
which saturates the Tsirelson bound~\cite{Tsirelson:1980ry}. On the other hand, as discussed in Refs.~\cite{Terashima:2003rjs,Bittencourt:2020lgu}, in the context of the gedankenexperiment at hand, the individual spins of the particles undergo WRs resulting from the combined effects of LLTs and acceleration along their trajectories. This evolution transforms the initial entangled state $\ket{\chi}$ into a new state $\ket{\chi^{\prime}}$, which reduces the degree of violation of the CHSH inequality. For the sake of clarity, we now review the results of Ref.~\cite{Bittencourt:2020lgu}, and specialize them to the context of screened modified gravity.

The specific expression for $\ket{\chi^{\prime}}$ can be easily derived from Eq.~(\ref{eq:statetransformation}), with the spin-$1/2$ representation of the Wigner $D$-matrix given by Eq.~(\ref{eq:spin12reprfinite}). A straightforward calculation shows that, in the particular physical setting we are considering, the only non-vanishing components of $\theta_{j}^{i}$ as given in Eq.~(\ref{eq:theta}) are the following:
\begin{multline}
\theta_{3}^{1} = -\theta_{1}^{3}=
\\
\cosh(\zeta)\sinh(\zeta) \left [ \partial_{r}\Phi - \frac{1 + \Psi(r)-r \partial_{r}\Psi(r)}{r} \right ] \,,
\end{multline}
which are constant in time (see Refs.~\cite{Terashima:2003rjs,Bittencourt:2020lgu} for details). As a consequence, the spin-$1/2$ representation of the Wigner $D$-matrix given in Eq.~(\ref{eq:spin12reprfinite}), and representing the finite WR accumulated by integrating the infinitesimal LLTs along the particles' circular trajectories, reduces to the following expression:
\begin{eqnarray} 
D_{\sigma^{\prime} \sigma}^{(1/2)}[W(x)] &=& \exp \left [ -\frac{i}{2}\theta_{31}\sigma_{y}\left(\tau_{f} - \tau_{i}\right) \right ] \nonumber \\
&\equiv& \exp \left ( -\frac{i}{2}\theta_{31}\sigma_{y}\Delta\tau \right ) \,.
\label{eq:spin12chsh}
\end{eqnarray}
In the specific setting under consideration, the particle four-velocity is given as~\cite{Bittencourt:2020lgu}
\begin{align}
u^{\mu}(x) &= \left((1-\Phi)\cosh(\zeta),0,0,\frac{1+\Psi}{r}\sinh(\zeta)\right)\,,
\end{align}
from which we find that the elapsed proper time $\Delta\tau$ corresponding to an azimuthal displacement of $\pm\bar{\varphi}$ along the circular trajectories takes the following form:
\begin{eqnarray}
\Delta\tau &=& \int_{\varphi_{i}}^{\varphi_{f}}\frac{d\varphi}{u^{\phi}} \simeq \frac{r \left [ 1 - \Psi(r) \right ] }{\sinh(\zeta)} \left ( \varphi_{f} - \varphi_{i} \right ) \nonumber \\
&=& \begin{cases}
\dfrac{r\bar{\varphi} \left ( 1 - \Psi \right ) }{\sinh(\zeta)}\,, & \text{Particle A} \\
-\dfrac{r\bar{\varphi} \left ( 1 - \Psi \right ) }{\sinh(\zeta)}\,. & \text{Particle B}
\end{cases} 
\label{eq:deltatau}
\end{eqnarray}
Inserting Eq.~(\ref{eq:deltatau}) into Eq.~(\ref{eq:spin12chsh}), we obtain the following expression for the $D$-matrix:
\begin{equation} 
D_{\sigma^{\prime}\sigma}^{(1/2)} \left [ W \left ( \pm \bar{\varphi},0 \right ) \right ] = \exp \left ( \mp \frac{i}{2}\sigma_{y}\Theta \right ) \,,
\label{eq:spin12chshcircularparticlesab}
\end{equation}
with the upper (lower) sign in the exponent of Eq.~(\ref{eq:spin12chshcircularparticlesab}) referring to particle $A$ ($B$), and where the angle $\Theta$ is defined as
\begin{equation}
\Theta \equiv \dfrac{r\bar{\varphi} \left ( 1 - \Psi \right ) }{\sinh\zeta}\theta_{31} = \bar{\varphi}\cosh(\zeta) \left [ 1 - r\partial_{r} \left ( \Phi + \Psi \right ) \right ] \,.
\label{eq:thetaangle}
\end{equation}
By inserting Eq.~(\ref{eq:spin12chshcircularparticlesab}) into Eq.~(\ref{eq:statetransformation}) we find that once the particles have reached an azimuthal angle $\varphi = \pm\bar{\varphi}$, the initial state $\ket{\chi}$ has transformed, due to the sequence of successive infinitesimal LLTs, into the following state:
\begin{align}
\ket{\chi^{\prime}} = \frac{1}{\sqrt{2}} \biggl[ &
\cos\Theta \, \ket{p_+^a, \uparrow; \bar{\varphi}} \ket{p_-^a, \downarrow; -\bar{\varphi}} \nonumber \\
& - \, \, \cos\Theta \, \ket{p_+^a, \downarrow; \bar{\varphi}} \ket{p_-^a, \uparrow; -\bar{\varphi}} \nonumber \\
& + \, \, \sin\Theta \, \ket{p_+^a, \uparrow; \bar{\varphi}} \ket{p_-^a, \uparrow; -\bar{\varphi}} \nonumber \\
& + \, \, \sin\Theta \, \ket{p_+^a, \downarrow; \bar{\varphi}} \ket{p_-^a, \downarrow; -\bar{\varphi}} 
\biggr]\,.
\label{eq:chiprime}
\end{align}
The state given in Eq.~(\ref{eq:chiprime}) includes the effect of the trivial (spatial) rotation of the LIFs at $\pm \bar{\varphi}$. Nevertheless, it is useful to factor this out, in order to isolate the WR effect. To do so, we rotate the spin basis at each observer's position around the $2$-axis by angles $\mp \bar{\varphi}$ for particles $A$ and $B$ respectively.~\footnote{The trivial rotation is simply due to the geometry of the setup: the spin directions at $\pm \bar{\varphi}$ are not perfectly aligned in the LIFs of the two observers. This causes the measurements of the spin states at these two points to be misaligned even before considering WR effects [see the discussion below Eq.~(53) of Ref.~\cite{Terashima:2003rjs} for further discussions].} This transformation leads to the following rotated states:
\begin{align}
\ket{p_{\pm}^a, \uparrow; \pm \bar{\varphi}}^{\prime} &= \cos\frac{\bar{\varphi}}{2} \ket{p_{\pm}^a, \uparrow; \pm \varphi} \pm \sin\frac{\bar{\varphi}}{2} \ket{p_{\pm}^a, \downarrow; \pm \bar{\varphi}}\,, \nonumber \\
\ket{p_{\pm}^a, \downarrow; \pm \bar{\varphi}}^{\prime} &= \mp \sin\frac{\bar{\varphi}}{2} \ket{p_{\pm}^a, \uparrow; \pm \bar{\varphi}} + \cos\frac{\bar{\varphi}}{2} \ket{p_{\pm}^a, \downarrow; \pm \bar{\varphi}}\,.
\label{eq:rotatedstate}
\end{align}
Performing this rotation ensures that only WR effects are taken into account, and leads to the following rotated state (to avoid cluttering the notation, we refer to such state as $\ket{\chi^{\prime}}$, as no ambiguity arises):
\begin{align}
\ket{\chi^{\prime}} = \frac{1}{\sqrt{2}} \biggl[\, &
\cos\Delta \, \ket{p_+^a, \uparrow; \bar{\varphi}}^{\prime} \ket{p_-^a, \downarrow; -\bar{\varphi}}^{\prime} \nonumber \\
& - \, \, \cos\Delta \, \ket{p_+^a, \downarrow; \bar{\varphi}}^{\prime} \ket{p_-^a, \uparrow; -\bar{\varphi}}^{\prime} \nonumber \\
& + \, \, \sin\Delta \, \ket{p_+^a, \uparrow; \bar{\varphi}}^{\prime} \ket{p_-^a, \uparrow; -\bar{\varphi}}^{\prime} \nonumber \\
& \textcolor{red}{+} \, \, \sin\Delta \, \ket{p_+^a, \downarrow; \bar{\varphi}}^{\prime} \ket{p_-^a, \downarrow; -\bar{\varphi}}^{\prime} \biggr] \,,
\end{align}
with the angle $\Delta$ defined as
\begin{equation}
\Delta \equiv \Theta - \bar{\varphi} = \bar{\varphi} \left \{ \cosh(\zeta) \left [ 1 - r\partial_{r} \left ( \Phi + \Psi \right ) \right ] - 1 \right \} \,.
\label{eq:delta}
\end{equation}
Considering spin measurements for the two particles along the same directions as in the flat space-time case, we find a reduction in the degree of violation of the CHSH inequality, with the CHSH parameter now given by the following expression:
\begin{eqnarray}
\mathscr{S}[\ket{\chi^{\prime}}] &=& \left| \langle \chi^{\prime} | \hat{\mathcal{Q}} \hat{\mathcal{S}} | \chi^{\prime} \rangle 
+ \langle \chi^{\prime} | \hat{\mathcal{R}} \hat{\mathcal{S}} | \chi^{\prime} \rangle \right. \nonumber \\
&& \left. + \, \, \langle \chi^{\prime} | \hat{\mathcal{R}} \hat{\mathcal{T}} | \chi^{\prime} \rangle 
- \langle \chi^{\prime} | \hat{\mathcal{Q}} \hat{\mathcal{T}} | \chi^{\prime} \rangle \right| \nonumber \\
&=& 2\sqrt{2}\cos^2\Theta\,.
\end{eqnarray}
In the above expression, $\Theta$ is defined in Eq.~(\ref{eq:thetaangle}). This result, however, still includes the effects of trivial (geometrical) rotations. In order to isolate the effects arising from the curvature of space-time, we must not only rotate the states as in Eq.~(\ref{eq:rotatedstate}), but also adjust the measurement directions, rotating them around the 2-axis by angles $-\bar{\varphi}$ and $\bar{\varphi}$ for particles $A$ and $B$, respectively. Accordingly, we perform spin measurements for particle $A$ along the directions $\mathbf{a^{\prime\prime}}=\left ( \cos\bar{\varphi},0,-\sin\bar{\varphi} \right )$ and $\mathbf{a^{\prime\prime\prime}}=\left ( 0,1,0 \right )$, associated with operators $\mathcal{\hat{Q^{\prime}}}$ and $\mathcal{\hat{R^{\prime}}}$, respectively. Similarly, we perform spin measurements for particle $B$ along the directions $\mathbf{b^{\prime\prime}}=\left ( -\cos\bar{\varphi}, -1, -\sin\bar{\varphi} \right )/\sqrt{2}$ and $\mathbf{b^{\prime\prime\prime}}=\left ( \cos\bar{\varphi}, -1, \sin\bar{\varphi} \right )/\sqrt{2}$, associated with operators $\mathcal{\hat{S^{\prime}}}$ and $\mathcal{\hat{T^{\prime}}}$, respectively. As a result, the CHSH parameter results is then given as
\begin{eqnarray}
\mathcal{S^{\prime}}[\ket{\chi^{\prime}}] &=& \left| \langle \chi^{\prime} | \hat{\mathcal{Q^{\prime}}} \hat{\mathcal{S^{\prime}}} | \chi^{\prime} \rangle 
+ \langle \chi^{\prime} | \hat{\mathcal{R^{\prime}}} \hat{\mathcal{S^{\prime}}} | \chi^{\prime} \rangle \right. \nonumber \\
&& \left. + \, \, \langle \chi^{\prime} | \hat{\mathcal{R^{\prime}}} \hat{\mathcal{T^{\prime}}} | \chi^{\prime} \rangle 
- \langle \chi^{\prime} | \hat{\mathcal{Q^{\prime}}} \hat{\mathcal{T^{\prime}}} | \chi^{\prime} \rangle \right| \nonumber \\
&=& 2\sqrt{2}\cos^2\Delta\,,
\label{eq:chsh}
\end{eqnarray}
with $\Delta$ defined in Eq.~(\ref{eq:delta}). The reduction factor $\cos^2\Delta$ reflects how curved space-time impacts observable quantum correlations between two particles, incorporating both GR effects and those of screened modified gravity through the conformal coupling function ($\Delta$ depends on $\Phi$ and $\Psi$ which in turn depend on the expansion coefficients of the coupling functions). We find that the CHSH inequality is violated to a lesser extent compared to flat space-time, indicating that the presence of curvature, as well as screening effects, modifies how nonlocal correlations manifest in the measurements. It is important to emphasize that this reduction does not imply a weakening of the quantum correlations themselves. Rather, it means that the optimal measurement directions required to maximize the violation of the CHSH inequality must be adjusted to account for these effects.

We now explicitly establish the connection to screened modified gravity. Recalling the definitions of the gravitational potentials in Eqs.~(\ref{eq:phiscreen},\ref{eq:psiscreen}), we find that the quantity $\Delta$, defined in Eq.~(\ref{eq:delta}) and determining the degree of violation of the CHSH inequality as in Eq.~(\ref{eq:chsh}), can be written, at first order in $\mathcal{G}M/r$, as
\begin{equation}
\Delta(\epsilon, r) = \bar{\varphi}\left\{ \cosh(\zeta) \left [ 1 - \frac{\mathcal{G}(\epsilon)M(1 + \gamma(\epsilon))}{r} \right ] - 1 \right\}\,.
\label{eq:Deltaeps}
\end{equation}
In the limit $\epsilon \to 0$, the PPN parameters reduce to unity, and $\mathcal{G}$ to the gravitational constant $G$, so that $\Delta$ reduces to the following expression:
\begin{align}
\Delta_{\text{GR}} &\equiv \lim_{\epsilon \to 0}\Delta(\epsilon,r) \nonumber \\
&=\bar{\varphi} \left [ \cosh(\zeta) \left ( 1 - \dfrac{2GM}{r} \right ) - 1\right ] \,,
\label{eq:DeltaGR}
\end{align}
thus recovering the GR result of Ref.~\cite{Bittencourt:2020lgu}. Throughout, we compare $\Delta$ and $\Delta_{\text{GR}}$ for the same value of the Earth’s gravitational parameter measured by satellite orbits, i.e., we identify $GM$ in Eq.~(\ref{eq:DeltaGR}) with $\mathcal{G}M$. Hence, the screening contribution is $\Delta-\Delta_{\text{GR}}=\bar{\varphi}\cosh(\zeta)(1-\gamma)GM/r$.

\subsection{Application to screening mechanisms}
\label{subsec:application}

Our goal is now to apply the formalism laid out previously to the three screening mechanisms of interest, which we first introduce: chameleon, symmetron, and dilaton. More specifically, we will now specialize Eq.~(\ref{eq:Deltaeps}) to these three mechanisms by explicitly computing the PPN parameters $\gamma(\epsilon)$, $\beta(\epsilon)$, and $\mathcal{G}(\epsilon)$ from the specific choices of coupling function $F(\phi)$ and potential $V(\phi)$. We recall that these three mechanisms allow a light scalar field to evade stringent constraints imposed by tests of gravity in high-density regions (such as the Solar System), while still playing a significant cosmological role and potentially driving cosmic acceleration~\cite{Brax:2021wcv}. In the simplest scenario, the fifth force mediated by such a light particle can be characterized by its range, coupling to matter, and scaling with $r$ (i.e., whether the force follows the inverse-square law or drops off more slowly or quickly). In a nutshell, screening operates by dynamically tuning these quantities so as to hide the fifth force in high-density regions. In the subsequent sections, we will closely follow Ref.~\cite{Zhang:2016njn}.

\subsubsection{Chameleon screening}
\label{subsubsec:chameleonapplication}

The chameleon mechanism was introduced by Khoury and Weltman~\cite{Khoury:2003aq,Khoury:2003rn} and has since been extensively studied (see, e.g., Refs.~\cite{Brax:2004ym,Brax:2004qh,Capozziello:2007eu,Brax:2008hh,Brax:2010xq,Brax:2010kv,Gannouji:2010fc,Brax:2011wp,Wang:2012kj,Khoury:2013yya,Erickcek:2013dea,Elder:2016yxm,Brax:2016did,Burrage:2016bwy,Burrage:2017shh,Burrage:2018pyg,Katsuragawa:2019uto,Sakstein:2019qgn,Desmond:2019ygn,Hartley:2019wzu,Vagnozzi:2019kvw,Cai:2021wgv,Vagnozzi:2021quy,Karwal:2021vpk,Katsuragawa:2021wmw,Dima:2021pwx,Benisty:2021cmq,Tamosiunas:2021kth,Briddon:2021etm,Brax:2021owd,Ferlito:2022mok,Yuan:2022cpw,Chakrabarti:2022zvv,Tamosiunas:2022tic,Brax:2022olf,Benisty:2022lox,Boumechta:2023qhd,Elder:2023oar,Benisty:2023dkn,Paliathanasis:2023ttu,Paliathanasis:2023dfz,Benisty:2023vbz,Briddon:2023ayq,Hogas:2023pjz,Zaregonbadi:2023vcv,Benisty:2023clf,Kumar:2024ylj,Baez-Camargo:2024jia,OShea:2024jjw,Paliathanasis:2024sle,Pizzuti:2024hym,Nojiri:2025low,Zaregonbadi:2025ils}). This mechanism operates by allowing the scalar to acquire a density-dependent effective mass $m_{\text{eff}}$, which increases with the ambient density. Since the range of the scalar-mediated fifth force is inversely proportional to its (effective) mass,  this implies that in high-density environments (where such forces are tightly constrained) the interaction becomes sufficiently short-ranged to evade detection. Conversely, on cosmological scales, the scalar can, in principle, play the role of a quintessence field driving the accelerated expansion of the Universe. Chameleon particles can be probed through a wide range of experimental approaches, including but not limited to astrophysical and cosmological observations, precision atomic spectroscopy, torsion balance experiments, and terrestrial laboratory tests~\cite{Burrage:2017qrf}.

We recall that different screening mechanisms are characterized by different choices for the scalar potential $V(\phi)$ and the conformal coupling function $F(\phi)$ relating EF and JF. In the chameleon case, these functions are given by the following expressions:
\begin{align}
V(\phi) &= \frac{\Lambda^{4+n}}{\phi^n}\,,
\label{eq:potcham} \\
F(\phi) &= \exp \left ( \frac{\beta_m \phi}{M_{\text{Pl}}} \right ) \,,
\label{eq:confcham}
\end{align}
where $\Lambda$ is a constant with units of mass, $\beta_m$ is a positive, dimensionless coupling constant, and $n$ is an index that is taken to be positive in this work.~\footnote{The case $n<0$ is also potentially interesting, but works only if $n$ is an even integer. We refer the reader to Ref.~\cite{Burrage:2017qrf} for further discussions on this case.} If the chameleon is to play a role in cosmic acceleration, one would generally expect $\Lambda$ to be close to the so-called DE scale $\Lambda_{\text{DE}} \approx 2.4\,{\text{meV}}$, although other values of $\Lambda$ are possible.~\footnote{We note that a well-known no-go theorem excludes the possibility of chameleon self-acceleration~\cite{Wang:2012kj}, defined as acceleration in the JF but not in the EF, in the complete absence of any cosmological constant. However, this no-go theorem does not exclude the possibility of a chameleon field driving cosmic acceleration \textit{à la} quintessence, i.e., by rolling down the potential rather than through strictly modified gravity effects.} On the other hand, the coupling constant $\beta_m$ sets the energy scale of the chameleon coupling to matter, which is $M_{\text{Pl}}/\beta_m$. The matter coupling $\beta_m$ could in principle be different for various matter species: for simplicity, however, we will assume a single value for all. With these assumptions, the effective potential in Eq.~(\ref{eq:veff}) then reads as follows:
\begin{equation}
V_{\text{eff}}(\phi) = \frac{\Lambda^{4+n}}{\phi^n}+\rho \exp \left ( \frac{\beta_m \phi}{M_{\text{Pl}}} \right ) \,,
\end{equation}
where $\rho$ is the ambient \textit{matter} density, which we recall can be expressed in covariant form as the trace of the energy-momentum tensor associated to matter fields. The above effective potential clearly possesses a minimum, except for the trivial case where $n=0$. Using Eqs.~(\ref{eq:minimumpotential},\ref{eq:effectivemass}), the value of the field at the minimum of the effective potential, as well as the field's effective mass, are obtained as follows:
\begin{align}
\phi_{\text{min}}(\rho) &\simeq \left ( \frac{n M_{\text{Pl}}\Lambda^{4 + n}}{\beta_m \rho} \right ) ^{\frac{1}{n + 1}}\,,
\label{eq:phimincham} \\
m_{\text{eff}}^2(\rho) &\simeq (n + 1)\frac{\beta_m \rho}{M_{\text{Pl}}\phi_{\text{min}}(\rho)}\,,
\label{eq:effmasscham}
\end{align}
where we have assumed $\beta_m \phi/M_{\text{Pl}} \ll 1$, in agreement with terrestrial and astrophysical tests of chameleon screening~\cite{Burrage:2017qrf,Brax:2010kv}. As anticipated earlier, from Eq.~(\ref{eq:effmasscham}) we see that $m_{\text{eff}}$ grows with $\rho$, which allows the fifth force to become very short-ranged in high-density environments, thereby escaping detection.

In order to determine the PPN parameters, we now expand the bare potential in Eq.~(\ref{eq:potcham}) around the background field value $\phi_{\infty} = \phi_{\text{min}}(\rho_{\infty})$. Comparing this expansion with the one in Eq.~(\ref{eq:expansionV}), one can then read off the coefficients $V_{\infty}$, $V_1$, and $V_2$, as follows:
\begin{align}
V_{\infty} &= \frac{\beta_m \rho_{\infty}\phi_{\infty}}{n M_{\text{Pl}}}, 
\\
V_1 &= -\frac{\beta_m \rho_{\infty}}{M_{\text{Pl}}}, 
\\
V_2 &= \frac{(n + 1)\beta_m \rho_{\infty}}{2M_{\text{Pl}}\phi_{\infty}}\,.
\end{align}
A similar expansion for $F(\phi)$, when compared against Eq.~(\ref{eq:expansionF}), allows us to read off $F_{\infty}$, $F_1$, and $F_2$, which are given by the following expressions:
\begin{align}
F_{\infty} &= \exp \left ( \frac{\beta_m \phi_{\infty}}{M_{\text{Pl}}} \right ) \,,
\\
F_1 &= \frac{\beta_m}{M_{\text{Pl}}}\exp \left ( \frac{\beta_m \phi_{\infty}}{M_{\text{Pl}}} \right ) \,, 
\\
F_2 &= \frac{\beta_m^2}{2M_{\text{Pl}}^2}\exp \left ( \frac{\beta_m \phi_{\infty}}{M_{\text{Pl}}} \right ) \,.
\end{align}
By plugging the above expansion coefficients of $V$ and those of $F$ into Eqs.~(\ref{eq:Geps}--\ref{eq:betaeps}), and using Eqs.~(\ref{eq:scalarcharge},\ref{eq:phimincham},\ref{eq:effmasscham}), we arrive at the following expressions for $\mathcal{G}$, $\gamma$, and $\beta$, respectively:
\begin{align}
\mathcal{G}_{\text{cham}} &\simeq G \left ( 1 + \frac{\beta_m \phi_{\infty}}{M_{\text{Pl}}\Phi_{\text{N}}} \right ) \,, \label{eq:Gepscham} \\
\gamma_{\text{cham}} &\simeq 1 - \frac{2\beta_m \phi_{\infty}}{M_{\text{Pl}}\Phi_{\text{N}}}\,,
\label{eq:gammaepscham} \\
\beta_{\text{cham}} &\simeq 1 - \frac{3}{4\Phi_{\text{N}}(n + 1)} \left ( \frac{\phi_{\infty}}{M_{\text{Pl}}} \right ) ^2\,,
\label{eq:betaepscham}
\end{align}
where we have again used the fact that $\beta_m \phi/M_{\text{Pl}} \ll 1$, while also making the well-motivated assumption that $\rho_c \gg \rho_{\infty}$. In Eqs.~(\ref{eq:Gepscham}--\ref{eq:betaepscham}), $\Phi_N$ represents the Newtonian potential in the EF. However, given the relation between masses and radial coordinates in the EF and JF, and the fact that $F_\infty$ is very close to unity, the difference between the Newtonian potentials in these two frames is negligible, and we can simply refer to $\Phi_N$ as being the Newtonian potential.

We can now directly plug Eqs.~(\ref{eq:Gepscham}--\ref{eq:betaepscham}) into Eq.~(\ref{eq:Deltaeps}), obtaining the following specific expression for the parameter $\Delta$: 
\begin{equation}
\Delta \simeq \Delta_{\text{GR}} + \Delta_{\text{cham}} = \Delta_{\text{GR}} \left ( 1 + \frac{\Delta_{\text{cham}}}{\Delta_{\text{GR}}} \right ) \,,
\label{eq:Deltacham1}
\end{equation}
where the GR contribution $\Delta_{\text{GR}}$ is given by Eq.~(\ref{eq:DeltaGR}), whereas we refer to the additional contribution from screened modified gravity as $\Delta_{\text{cham}}$, given by the following expression:
\begin{equation}
\Delta_{\text{cham}} \equiv \frac{2GM\Bar{\varphi}\cosh(\zeta)\beta_m\phi_{\infty}}{rM_{\text{Pl}}\Phi_{\text{N}}} \,,
\label{eq:deltacham}
\end{equation}
Above, $r$ is the distance of the entangled particles from the gravitational source. In closing, we note that Eq.~(\ref{eq:deltacham}) is the leading-order result in $\epsilon$, with corrections suppressed by a further factor $\beta_m\phi_{\infty}/(M_{\text{Pl}}\Phi_{\text{N}})$.

\subsubsection{Symmetron screening}
\label{subsubsec:symmetronapplication}

The symmetron mechanism was introduced by Hinterbichler and Khoury~\cite{Hinterbichler:2010es,Hinterbichler:2011ca} and, like the chameleon mechanism, has been extensively investigated (see, e.g., Refs.~\cite{Davis:2011pj,Brax:2011pk,Winther:2011qb,Upadhye:2012rc,Bamba:2012yf,Silva:2013sla,Burrage:2016rkv,Burrage:2016yjm,Pinho:2017jpk,OHare:2018ayv,Burrage:2018zuj,Contigiani:2018hbn,Cronenberg:2018qxf,Elder:2019yyp,Jenke:2020obe,Pitschmann:2020ejb,Perivolaropoulos:2022txg,Nosrati:2022uzu,Christiansen:2023tfy,Hogas:2023vim,Kading:2023hdb,Christiansen:2024vqv,Xiong:2024vsd,Li:2024ynr,Morales-Navarrete:2025lzp,Udemba:2025csd}). In contrast to the chameleon mechanism, which adjusts the fifth force's range to evade local gravity tests, the symmetron mechanism instead alters the force's strength.

The symmetron mechanism is characterized by the following ($\mathbb{Z}_2$ symmetry breaking) scalar field potential and (quadratic) conformal coupling function~\cite{Hinterbichler:2010es,Hinterbichler:2011ca}:
\begin{align}
V(\phi) &= -\frac{1}{2}\mu^2\phi^2 + \frac{\lambda}{4}\phi^4\,,
\label{eq:vsymmetron} \\
\quad F(\phi) &= 1 + \frac{\phi^2}{2M_{\text{sym}}^2}\,,
\label{eq:fsymmetron}
\end{align}
where $\mu$ and $M_{\text{sym}}$ are two constants with units of mass, and $\lambda$ is a positive, dimensionless coupling constant. The symmetron effective potential, obtained by combining Eqs.~(\ref{eq:veff},\ref{eq:vsymmetron},\ref{eq:fsymmetron}), is then given by
\begin{equation}
V_{\text{eff}}(\phi) = \frac{1}{2} \left ( \frac{\rho}{M_{\text{sym}}^2} - \mu^2 \right ) \phi^2 + \frac{\lambda}{4}\phi^4\,.
\end{equation}
From the above expression we deduce that, depending on the local matter density, the quadratic term can become negative, leading to spontaneous breaking of the $\mathbb{Z}_2$ symmetry. More specifically, if the coefficient of $\phi^2$ is positive, i.e., if $\rho > M_{\text{sym}}^2\mu^2$ (in high-density regions), the effective potential has a single minimum at $\phi = \phi_{\min} = 0$, corresponding to a symmetric phase where the $\mathbb{Z}_2$ symmetry remains unbroken: in other words, the scalar field is trapped near, or exactly at, $\phi=0$. On the other hand, if the coefficient of $\phi^2$ is negative, i.e., if $\rho < M_{\text{sym}}^2\mu^2$ (in low-density regions), the potential takes on a Mexican hat shape, with minima at the following field values:
\begin{equation}
\phi_{\min}(\rho) = \pm \frac{\mu}{\sqrt{\lambda}} \sqrt{1 - \frac{\rho}{M_{\text{sym}}^2\mu^2}} \,,
\label{eq:fieldminld}
\end{equation}
as can be easily obtained from Eq.~(\ref{eq:minimumpotential}). A key feature of the symmetron mechanism is that the coupling between field fluctuations and matter fields is proportional to the local background field value, $\phi_{\infty}$. In high-density environments, where $\phi_{\infty}$ is zero or very close to zero, the coupling to matter is suppressed, thus dynamically reducing the strength of the fifth force. In contrast, in low-density environments, $\phi_{\infty}$ can become large. From Eq.~(\ref{eq:effectivemass}) and working within the regime $\rho<M_{\text{sym}}^2\mu^2$, it is found that the effective symmetron mass takes the following form:
\begin{equation}
m_{\text{eff}}^2(\rho)=2\mu^2 \left ( 1 - \frac{\rho}{M_{\text{sym}}^2\mu^2} \right ) \,.
\label{eq:effmsym}
\end{equation}
From Eq.~(\ref{eq:fieldminld}), we note that the value of the field at the minimum of $V_{\text{eff}}$ can take two different values. Without loss of generality, we assume the scalar field background to be positive, thus obtaining the following value:~\footnote{As we shall see, our final results turn out to depend on $\phi_{\infty}^2$.}
\begin{equation}
\phi_{\infty} = \phi_{\text{min}}(\rho_{\infty}) = \frac{\mu}{\sqrt{\lambda}} \sqrt{1 - \frac{\rho_{\infty}}{M_{\text{sym}}^2\mu^2}} \,,
\label{eq:phiinftysymmetron}
\end{equation}
which we recall is valid if $\rho_{\infty} < M_{\text{sym}}^2\mu^2$.

As we did in Sec.~\ref{subsubsec:chameleonapplication} for chameleons, we now expand the bare potential and coupling function in Eqs.~(\ref{eq:vsymmetron},\ref{eq:fsymmetron}) around $\phi_{\infty}$, comparing them with the expansions in Eq.~(\ref{eq:expansionF}) and Eq.~(\ref{eq:expansionV}), respectively. This allows us to read off the expansion terms, given by the following expressions:
\begin{align}
F_{\infty} &= 1 + \frac{\phi_{\infty}^2}{2M_{\text{sym}}^2}\,, 
\\
F_1 &= \frac{\phi_{\infty}}{M_{\text{sym}}^2}\,,
\\
F_2 &= \frac{1}{2M_{\text{sym}}^2}\,,
\\
V_{\infty} &= -\frac{M_{\text{sym}}^4\mu^4 - \rho_{\infty}^2}{4\lambda M_{\text{sym}}^4}\,, \\
V_1 &= -\frac{\rho_{\infty}\phi_{\infty}}{M_{\text{sym}}^2}\,,
\\
V_2 &= \mu^2 - \frac{3\rho_{\infty}}{2M_{\text{sym}}^2}\,.
\end{align}
To obtain the PPN parameters, we plug the above results into Eqs.~(\ref{eq:Geps}--\ref{eq:betaeps}), making use of Eqs.~(\ref{eq:scalarcharge},\ref{eq:fieldminld},\ref{eq:effmsym}), which leads to the following expressions for $\mathcal{G}$, $\gamma$, and $\beta$:
\begin{align}
\mathcal{G}_{\text{sym}} &\simeq G \left ( 1 + \frac{\phi_{\infty}^2}{M_{\text{sym}}^2\Phi_{\text{N}}} \right ) \,,
\label{eq:Gepssym} \\
\gamma_{\text{sym}} &\simeq 1 - \frac{2\phi_{\infty}^2}{M_{\text{sym}}^2\Phi_{\text{N}}}\,,
\label{eq:gammaepssym} \\
\beta_{\text{sym}} &\simeq 1 + \frac{1}{2} \left ( \frac{\phi_{\infty}}{M_{\text{sym}}\Phi_{\text{N}}} \right ) ^2\,,
\label{eq:betaepssym}
\end{align}
where we have assumed $\phi \ll M_{\text{sym}}$~\cite{Hinterbichler:2011ca}.

Finally, by plugging Eqs.~(\ref{eq:Gepssym}--\ref{eq:betaepssym}) into Eq.~(\ref{eq:Deltaeps}), we obtain $\Delta$ as follows:
\begin{equation}
\Delta \simeq \Delta_{\text{GR}} + \Delta_{\text{sym}} = \Delta_{\text{GR}} \left ( 1 + \frac{\Delta_{\text{sym}}}{\Delta_{\text{GR}}} \right ) \,,
\end{equation}
where, again, the standard GR contribution $\Delta_{\text{GR}}$ is given by Eq.~(\ref{eq:DeltaGR}), while the symmetron contribution $\Delta_{\text{sym}}$ takes the following form:
\begin{equation}
\Delta_{\text{sym}} \equiv \frac{2GM\Bar{\varphi}\cosh(\zeta)\phi_{\infty}^2}{rM_{\text{sym}}^2\Phi_{\text{N}}}\,,
\label{eq:deltasym}
\end{equation}
Of course, as in the chameleon case, $r$ denotes the distance between the entangled particles and the gravitational source, i.e.\ the Earth in our gedankenexperiment.

\subsubsection{Dilaton screening}
\label{subsubsec:dilatonapplication}

The last type of screening mechanism we consider is the environmental dilaton mechanism~\cite{Brax:2010gi,Brax:2011ja}, also extensively studied in recent years~\cite{Brax:2012nk,Brax:2017wcj,Hartley:2018lvm,Burgess:2021qti,Brax:2022vlf,Fischer:2023eww,Fischer:2024coj,Reyes:2024lzu,Fischer:2024gni,Smith:2024ayu,Smith:2024ibv,Kading:2024jqe,Smith:2025grk,Bachs-Esteban:2025mxl,Brax:2025ahm}. The theoretical basis for this mechanism is firmly rooted in string theory arguments, and essentially provides an environmentally-dependent analogue of the Damour-Polyakov mechanism~\cite{Damour:1994zq}. Its potential origin in string theory makes the environmental dilaton particularly compelling compared to its chameleon and symmetron counterparts, which instead are more phenomenological in nature. Like the chameleon and symmetron mechanisms, the environmental dilaton is amenable to a wide range of experimental tests. Note that we will omit the ``environmental'' part in what follows.

In the dilaton screening mechanism, the bare potential is usually taken to be exponentially decreasing as conjectured for the string dilaton in the strong coupling regime ($\phi \to \infty$). Specifically, the potential and conformal coupling function are given by
\begin{align}
V(\phi) &= \mathcal{V}_{0}\exp \left ( -\frac{\phi}{M_{\text{Pl}}} \right ) \,,
\label{eq:vdilaton} \\
F(\phi) &= 1 + \frac{ \left ( \phi - \phi_{\star} \right ) ^2}{2M_{\text{dil}}^2}\,,
\label{eq:fdilaton}
\end{align}
where $\mathcal{V}_0$ is a constant with units of energy density, $M_{\text{dil}}$ is a coupling constant with units of mass, and $\phi_{\star}$ is a constant field value at which the coupling of the dilaton to matter weakens. Combining Eqs.~(\ref{eq:veff},\ref{eq:vdilaton}), one finds that the effective potential of the dilaton takes the following form:
\begin{equation}
V_{\text{eff}}(\phi) = \mathcal{V}_{0}\exp \left ( -\frac{\phi}{M_{\text{Pl}}} \right )  + \rho \left [ 1 + \frac{ \left ( \phi - \phi_{\star} \right ) ^2}{2M_{\text{dil}}^2} \right ] \,.
\end{equation}
The model is conceptually similar to the symmetron mechanism introduced in Sec.~\ref{subsubsec:symmetronapplication}. In the symmetron case, the coupling between field fluctuations and matter is proportional to the local background value, which undergoes a phase transition and is driven to zero in high-density environments.~\footnote{For a detailed explanation of how the dilaton mechanism works, along with various subtleties (such as how, in certain regions of parameter space, the dilaton can employ a chameleon-like screening mechanism and thus behave as a hybrid of the chameleon and symmetron mechanisms), we refer the reader to Refs.~\cite{Brax:2010gi,Sakstein:2014jrq,Fischer:2023koa}.}~In the dilaton case, the field-matter coupling also approaches zero in dense environments, but this is achieved by minimizing the function $F(\phi)$ at a value $\phi_{\star}$, as can be deduced from Eq.~(\ref{eq:fdilaton}). By using Eqs.~(\ref{eq:minimumpotential},\ref{eq:effectivemass}), and taking into account that $\phi/M_{\text{Pl}} \ll 1$ (i.e., we are in the strong coupling regime), we obtain both the value of the dilaton field at the minimum of $V_{\text{eff}}$, as well as its effective mass:
\begin{align}
\phi_{\text{min}}(\rho) &\simeq \phi_{\star} + \frac{M_{\text{dil}}^2\mathcal{V}_{0}}{M_{\text{Pl}}\rho}\exp \left ( -\frac{\phi_{\star}}{M_{\text{Pl}}} \right ) \,,
\label{eq:phimindil} \\
m_{\text{eff}}^2(\rho) &\simeq \frac{\rho}{M_{\text{dil}}^2} + \frac{\mathcal{V}_{0}}{M_{\text{Pl}}^2}\exp \left ( -\frac{\phi_{\star}}{M_{\text{Pl}}} \right ) \,.
\label{eq:effmassdil}
\end{align}
From Eq.\ \eqref{eq:phimindil}, we deduce that as the density increases, $\phi_{\text{min}} \to \phi_{\star}$, and consequently $F(\phi)\to 1$, indicating a decoupling of the dilaton from matter.

We now expand both $F(\phi)$ and $V(\phi)$ around $\phi_{\infty}$, and compare these expansions with Eqs.~(\ref{eq:expansionF},\ref{eq:expansionV}) to read off $F_\infty,F_1,F_2,V_\infty,V_1$ and $V_2$, as follows:
\begin{align}
F_{\infty} &= 1 + \frac{ \left ( \phi_{\infty} - \phi_{\star} \right ) ^2}{2M_{\text{dil}}^2}\,,
\\
F_1 &= \frac{\phi_{\infty} - \phi_{\star}}{M_{\text{dil}}^2}\,,
\\
F_2 &= \frac{1}{2M_{\text{dil}}^2}\,,
\\
V_{\infty} &= \mathcal{V}_{0}\exp \left ( -\frac{\phi_{\infty}}{M_{\text{Pl}}} \right ) \,,
\\
V_1 &= -\frac{\mathcal{V}_{0}}{M_{\text{Pl}}}\exp \left ( -\frac{\phi_{\infty}}{M_{\text{Pl}}} \right ) \,, 
\\
V_2 &= \frac{\mathcal{V}_{0}}{2M_{\text{Pl}}^2}\exp \left ( -\frac{\phi_{\infty}}{M_{\text{Pl}}} \right ) \,.
\end{align}
Proceeding as in the last two sections, we plug the above expressions into Eqs.~(\ref{eq:Geps}--\ref{eq:betaeps}), and make use of Eqs.~(\ref{eq:scalarcharge},\ref{eq:phimindil},\ref{eq:effmassdil}), leading to the following expressions for $\mathcal{G}$, $\gamma$, and $\beta$, respectively:
\begin{align}
\mathcal{G}_{\text{dil}} &\simeq G \left [ 1 + \frac{ \left ( \phi_{\infty} - \phi_{\star} \right ) ^2}{M_{\text{dil}}^2\Phi_{\text{N}}} \right ] \,,
\label{eq:Gepsdil} \\
\gamma_{\text{dil}} &\simeq 1 - \frac{2 \left ( \phi_{\infty} - \phi_{\star} \right ) ^2}{M_{\text{dil}}^2\Phi_{\text{N}}}\,,
\label{gammaepsdil} \\
\beta_{\text{dil}} &\simeq 1 + \frac{1}{2} \left ( \frac{\phi_{\infty} - \phi_{\star}}{M_{\text{dil}}\Phi_{\text{N}}} \right ) ^2\,,
\label{eq:betaepsdil}
\end{align}
Finally, plugging Eqs.~(\ref{eq:Gepsdil}--\ref{eq:betaepsdil}) into Eq.~(\ref{eq:Deltaeps}) gives the following expression:
\begin{equation}
\Delta \simeq \Delta_{\text{GR}} + \Delta_{\text{dil}} = \Delta_{\text{GR}} \left ( 1 + \frac{\Delta_{\text{dil}}}{\Delta_{\text{GR}}} \right ) \,.
\end{equation}
Once again, the standard GR contribution $\Delta_{\text{GR}}$ is given by Eq.~(\ref{eq:DeltaGR}), whereas the dilaton contribution $\Delta_{\text{dil}}$ is given as
\begin{equation}
\Delta_{\text{dil}} \equiv \frac{2GM\Bar{\varphi}\cosh(\zeta)\left(\phi_{\infty}-\phi_{\star}\right)^2}{rM_{\text{dil}}^2\Phi_{\text{N}}}\,.
\label{eq:deltadil}
\end{equation}

\section{Results and discussion}
\label{sec:results}

So far, we have evaluated the expected suppression of the CHSH inequality violation due to screened modified gravity, focusing on the three screening mechanisms examined. To be concrete, we now aim to connect these theoretical results to the type of EPR gedankenexperiment considered in Refs.~\cite{Terashima:2003rjs,Bittencourt:2020lgu}. Our goal in this pilot study is to provide a preliminary, order-of-magnitude estimate of the screened modified gravity contribution to this effect (relative to the GR contribution), in order to assess whether there exist regions of parameter space, especially those still open to other probes, that may represent promising targets for future experiments, albeit likely only in a somewhat futuristic context. Specifically, as a concrete realization of the setup proposed in Refs.~\cite{Terashima:2003rjs,Bittencourt:2020lgu}, we consider a hypothetical experiment conducted in orbit around the Earth, approximately $500 \, \text{km}$ above its surface, where the matter density can be approximated as $\rho_{\infty} \sim 10^{-6} \,\text{eV}^4$~\cite{Khoury:2003aq,Khoury:2003rn,Zhang:2016njn}. An EPR source located at $\varphi = 0$ then emits a pair of entangled spin-$1/2$ particles in opposite directions. The two particles travel along circular orbits with a rapidity set to $\zeta \sim 5 \times 10^{-5}$ and eventually reach $\varphi = \pm \bar{\varphi}$, where two detectors are situated. Their spin components along the four directions associated with the operators $\mathcal{\hat{Q^{\prime}}}$, $\mathcal{\hat{R^{\prime}}}$, $\mathcal{\hat{S^{\prime}}}$, and $\mathcal{\hat{T^{\prime}}}$ (defined in Sec.~\ref{subsec:nonlocal}) are then measured to calculate the CHSH parameter $\mathscr{S}$ from Eq.~(\ref{eq:chsh}), where potential effects of screened modified gravity are captured, as per Eqs.~(\ref{eq:DeltaGR}, \ref{eq:deltacham}, \ref{eq:deltasym}, \ref{eq:deltadil}). As for the experimental specifications we have chosen, we stress that these are purely representative, and a detailed study of their impact is deferred to future work. Nevertheless, a height of $500 \, \text{km}$ is comparable to that of the satellite-based entanglement experiment of Ref.~\cite{Yin:2017ghw}, while the rapidity corresponds to a velocity of $v \sim 15 \,\text{km}/\text{s}$.

\subsection{Chameleon screening}
\label{subsec:chameleonresults}

\begin{figure}[!t]
\includegraphics[width=1.0\linewidth]{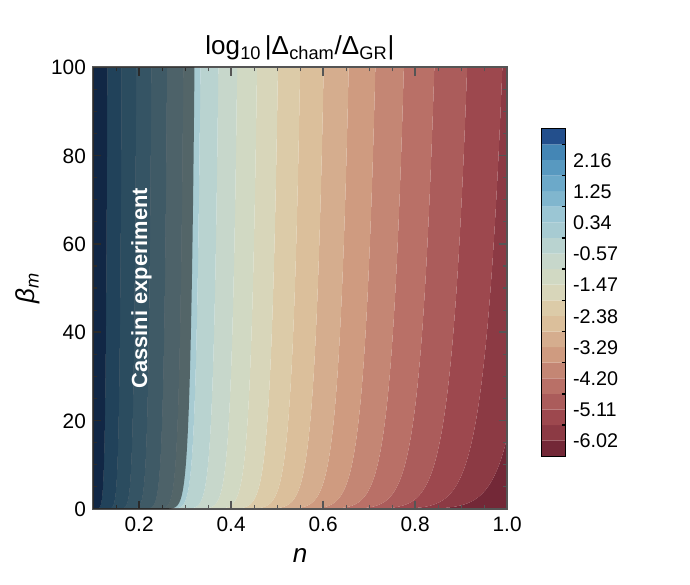}
\centering
\caption{Contour plot of $\log_{10}|\Delta_{\text{cham}}/\Delta_{\text{GR}}|$ as a function of $n$ and $\beta_m$, where $n$ is the index determining the power-law behavior of the chameleon potential, and $\beta_m$ is the strength of the chameleon coupling to matter. $\Delta_{\text{cham}}$ and $\Delta_{\text{GR}}$ are the chameleon and pure General Relativity contributions to the parameter $\Delta$, respectively, with $\cos^2\Delta$ controlling the degree of violation of the CHSH inequality in our gedankenexperiment. For what concerns the parameters of our gedankenexperiment, as discussed in the main text, we set $\rho_{\infty}\sim 10^{-6}\,\text{eV}^4 $, $\Lambda=2.4\,\text{meV}$, and $\zeta\simeq 5 \times 10^{-5}$. The dark blue shaded region of the plot represents the area excluded by the \textit{Cassini} probe \cite{Zhang:2016njn}. We see that there is a narrow region of parameter space where the chameleon contribution is significant compared to that of GR.}
\label{fig:deltaratiochameleon}
\end{figure}

We begin by considering the chameleon screening mechanism described in Sec.~\ref{subsubsec:chameleonapplication}.~In its most basic realization, the model contains three free parameters: the potential power-law index $n$, the potential scale $\Lambda$, and the matter coupling $\beta_m$, with $\Lambda$ entering into Eqs.~(\ref{eq:phimincham},\ref{eq:effmasscham}).~\footnote{In principle, other parameters of the chameleon model could include the coupling $\beta_{\gamma}$, controlling the strength of the inevitable chameleon coupling to photons~\cite{Brax:2010uq}, as well as possible disformal couplings to matter fields.} In what follows, we set $\Lambda$ to the DE scale, $\Lambda=\Lambda_{\text{DE}} \approx 2.4\,{\text{meV}}$. This case is not only the most extensively studied in the literature,  as it allows the chameleon to drive cosmic acceleration \textit{\`{a} la quintessence} on cosmological scales, but it also allows the parameter space to be reduced to just two dimensions, then characterized by $n$ and $\beta_m$.

Current limits on $n$ and $\beta_m$ when $\Lambda$ is fixed to $\Lambda_{\text{DE}}$ are shown for instance in Fig.~6 of Ref.~\cite{Burrage:2017qrf}, where the $y$-axis corresponds to $-\log_{10}\beta_m$. As we can see, for $n \geq 1$, a significant portion of parameter space is already excluded by torsion balance experiments, atom interferometry tests, and astrophysical observations, leaving open the region of parameter space $1 \lesssim \beta_m \lesssim 100$. In the specific case of $n=1$, which is the most commonly studied one, most of this region has actually been recently excluded by levitated force sensors~\cite{Yin:2022geb}, leaving open the region $1 \lesssim \beta_m \lesssim 5$ (we stress that this limit only applies to the case $n=1$). With these considerations in mind, and given that the ratio $\Delta_{\text{cham}}/\Delta_{\text{GR}}$ is strongly suppressed for $n > 1$, we focus on the range $0.1 < n < 1$, which is partially unconstrained by current probes, as can again be seen from Fig.~6 of Ref.~\cite{Burrage:2017qrf}. We note that it is not meaningful to consider $n = 0$, in which case the potential reduces to a cosmological constant and the effective potential has no minimum for $\phi \neq 0$; this is why we set the lower cut at $n = 0.1$. In line with atom interferometry constraints, for the matter coupling we instead consider the range $0 \leq \beta_m \leq 100$. We note in passing that, for $n < 1$, values of $\beta_m \lesssim 1$ are completely unconstrained. This range of $\beta_m$ has been argued to be theoretically questionable based on quantum gravity considerations, as it would correspond to a trans-Planckian energy scale~\cite{Brax:2019rwf}. Such a scenario would lead to a force weaker than gravity, which is at odds with the weak gravity conjecture (WGC), and, more generally, with a series of related string-inspired conjectures broadly encompassed by the Swampland program~\cite{Palti:2019pca}. From a purely phenomenological perspective, however, we find no \textit{a priori} reason to exclude values of $\beta_m \lesssim 1$ from our analysis. In fact, one could argue that probing this region is particularly interesting, as a possible detection of a non-zero $\beta_m \lesssim 1$ could provide the first direct counterexample to the WGC, with significant implications for our understanding of quantum gravity.

In Fig.~\ref{fig:deltaratiochameleon}, we plot $\log_{10}|\Delta_{\text{cham}}/\Delta_{\text{GR}}|$ as a function of the chameleon parameters $n$, i.e., the index determining the power-law behavior of the chameleon potential, and $\beta_m$, namely the strength of the chameleon coupling to matter. We recall that $\Delta_{\text{GR}}$ and $\Delta_{\text{cham}}$ encode the GR and screened modified gravity contributions to the parameter $\Delta = \Delta_{\text{GR}} + \Delta_{\text{cham}}$, with $\cos^2 \Delta$ modulating the degree of violation of the CHSH inequality (relative to the flat space-time case) within the specific gedankenexperiment we consider. The region shaded in dark blue is excluded by stringent constraints on the PPN parameter $\gamma$ obtained from the \textit{Cassini} probe, with $\vert \gamma - 1 \vert < 2.3 \times 10^{-5}$. This constraint, based on Shapiro time-delay measurements made during \textit{Cassini}'s solar conjunction in 2002, tightly limits several modified gravity theories~\cite{Bertotti:2003rm}. From Fig.~\ref{fig:deltaratiochameleon}, we see that as $n$ increases, the chameleon contribution is suppressed relative to that of GR, as can be directly inferred from Eq.~(\ref{eq:deltacham}). However, for sufficiently small values of $n$, the chameleon contribution becomes comparable to that of GR and even exceeds it for values of $n$ in the interval $0.1 \lesssim n \lesssim 0.3$. Nevertheless, in the region allowed by \textit{Cassini} the chameleon contribution is at most comparable to that of GR.

\subsection{Symmetron screening}
\label{subsec:symmetronresults}

\begin{figure*}[!t]
\centering
\includegraphics[width=0.477\textwidth]{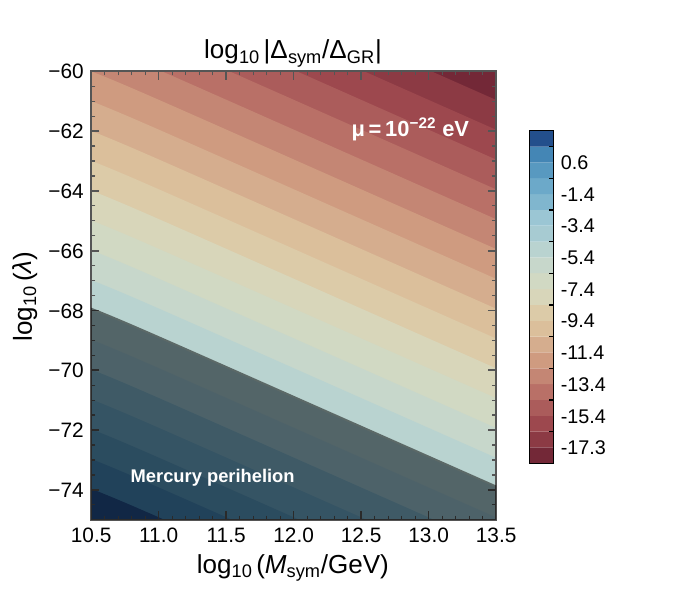}\hfill\includegraphics[width=0.477\textwidth]{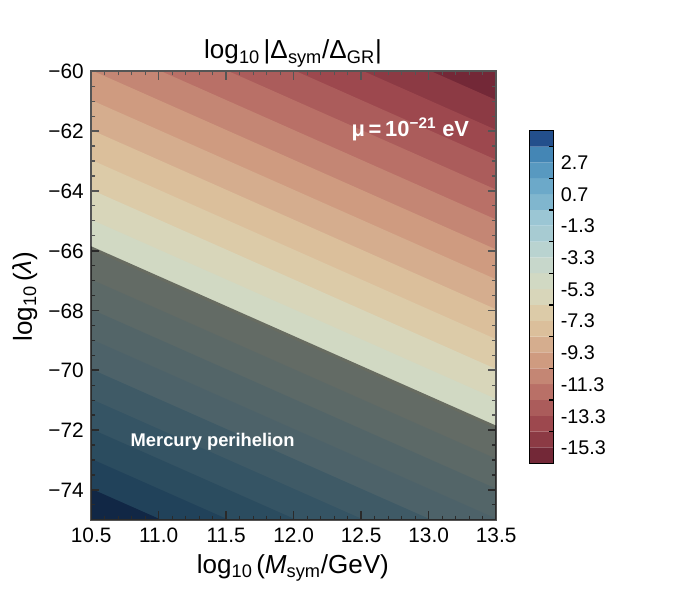} \vskip 0.5cm
\includegraphics[width=0.477\textwidth]{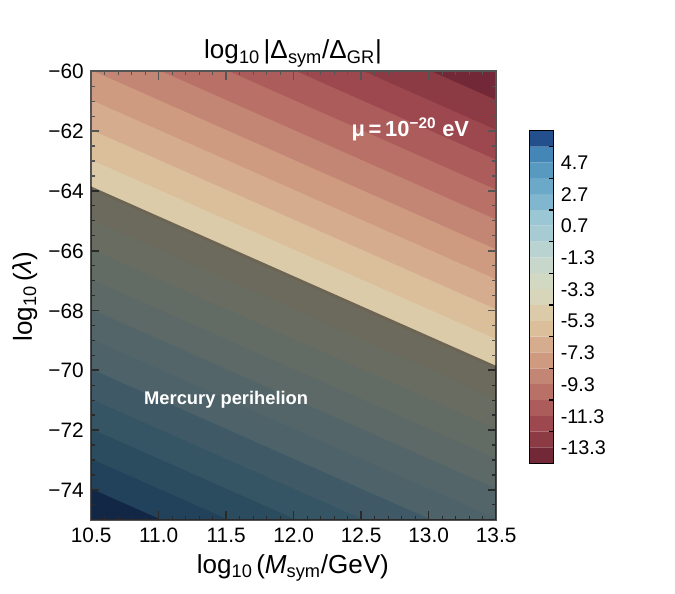}\hfill\includegraphics[width=0.477\textwidth]{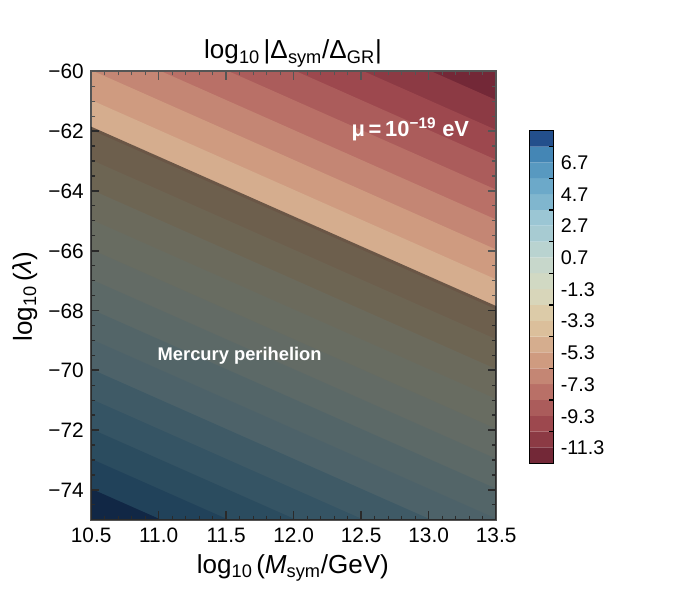}
\caption{As in Fig.~\ref{fig:deltaratiochameleon}, but for $\log_{10}|\Delta_{\text{sym}}/\Delta_{\text{GR}}|$ as a function of $\log_{10}(M_{\text{sym}}/{\text{GeV}})$ and $\log_{10}\lambda$, where $M_{\text{sym}}$ is the scale of the symmetron coupling to matter and $\lambda$ is the dimensionless symmetron self-coupling. The four panels above correspond to different representative values of the tachyonic mass:~$\mu=10^{-22}\,{\text{eV}}$ (upper left panel), $10^{-21}\,{\text{eV}}$ (upper right panel), $10^{-20}\,{\text{eV}}$ (lower left panel), and $10^{-19}\,{\text{eV}}$ (lower right panel). For $\mu=10^{-22}\,{\text{eV}}$, there is a narrow region of parameter space where the symmetron contribution is up to $\sim 3\times 10^{-5}$ of that of GR. The dark blue shaded region represents the area excluded by \textit{Mercury perihelion} \cite{Zhang:2016njn}.}
\label{fig:deltaratiosymmetron}
\end{figure*}

We now move on to the symmetron-screened modified gravity case, described in Sec.~\ref{subsubsec:symmetronapplication}. This model is characterized by three parameters: the dimensionless symmetron self-coupling $\lambda$, the tachyonic mass $\mu$, and the scale of the symmetron coupling to matter $M_{\text{sym}}$. Following Ref.~\cite{Fischer:2024eic}, we consider the parameter space spanned by $M_{\text{sym}}$ and $\lambda$, while fixing the tachyonic mass to four representative values:~$\mu=10^{-22}\,{\text{eV}}$, $10^{-21}\,{\text{eV}}$, $10^{-20}\,{\text{eV}}$, and finally $10^{-19}\,{\text{eV}}$. In this work, we focus on the region where $10.5 \leq \log_{10}(M_{\text{sym}}/\text{GeV}) \leq 13.5$ and $-75 \leq \log_{10}(\lambda) \leq -60$; here the symmetron is in its broken phase at $\rho_{\infty}$, its range $m_{\text{eff}}^{-1}(\rho_{\infty}) \geq (\sqrt{2}\mu)^{-1}$ exceeds Solar-System scales, and the Earth and the Sun are screened, so that Eqs.~(\ref{eq:Geps}--\ref{eq:betaeps}) and the \textit{Mercury-perihelion} bound of Ref.~\cite{Zhang:2016njn}, both derived in the approximation $m_{\infty}r \ll 1$, apply. By contrast, for the laboratory-scale values of $\mu$ used in Ref.~\cite{Fischer:2024eic}, the symmetron effects at the orbit are exponentially suppressed.

As in the chameleon case, we plot $\log_{10}|\Delta_{\text{sym}}/\Delta_{\text{GR}}|$ as a function of $\log_{10}(M_{\text{sym}}/\text{GeV})$ and $\log_{10}\lambda$ in Fig.~\ref{fig:deltaratiosymmetron}, with each of the four panels corresponding to a different value of the tachyonic mass $\mu$, as indicated on the plots. The \textit{Mercury-perihelion} constraints are shown as shaded regions. We see a clear trend as $\mu$ increases: as is evident from Eq.~(\ref{eq:phiinftysymmetron}), $\phi_{\infty}$ grows, which in turn enhances the effective coupling and therefore increases the ratio $\Delta_{\text{sym}}/\Delta_{\text{GR}}$, signaling weaker screening. As a result, since the required suppression of deviations from GR scales with $\mu^2$~\cite{Zhang:2016njn}, the \textit{Mercury-perihelion} constraint becomes more stringent. Conversely, for smaller $\mu$, screening becomes more efficient and the constraint is relaxed, thereby leading to a wider region of viable parameter space.

From Fig.~\ref{fig:deltaratiosymmetron}, we deduce that there exists a narrow region of the parameter space where the symmetron contribution can reach $\sim 3\times 10^{-5}$ of the GR one. More specifically, considering the case $\mu=10^{-22}\,{\text{eV}}$, we see that the \textit{Mercury-perihelion} constraint corresponds approximately to the region defined by
\begin{equation}
\log_{10}\beta_{\text{sym}} \gtrsim -47\,, \quad \beta_{\text{sym}} \equiv \lambda \left ( \frac{M_{\text{sym}}}{{\text{GeV}}} \right ) ^2\,,
\label{eq:cassinisym}
\end{equation}
with smaller values of $\beta_{\text{sym}}$ being excluded.~Values of $M_{\text{sym}}$ and $\lambda$ that saturate the constraint given by Eq.~(\ref{eq:cassinisym}) are also those that maximize the symmetron contribution relative to the GR one in the $\mu = 10^{-22} \, \text{eV}$ case. Finally, we note that for this choice of $\mu$ the symmetron range, $\sim 10^{4}\,$AU, largely exceeds Solar-System scales.

\subsection{Dilaton screening}
\label{subsec:dilatonresults}

\begin{figure*}[!t]
\centering
\includegraphics[width=0.47\textwidth]{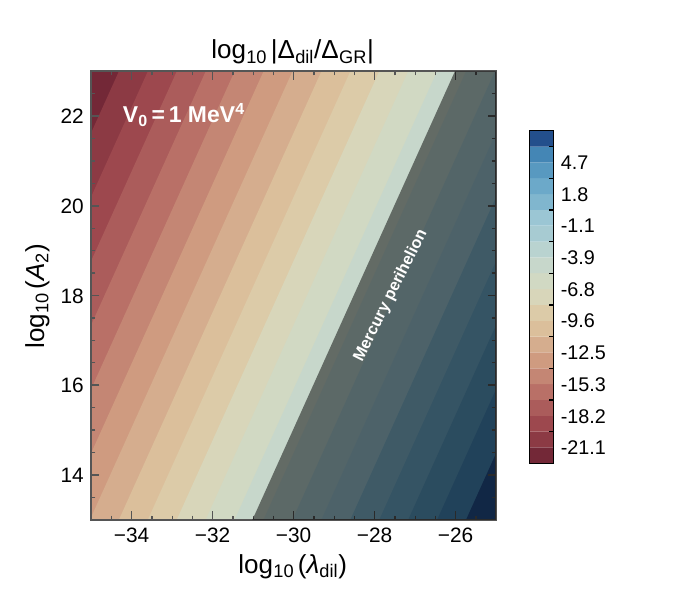}\hfill\includegraphics[width=0.47\textwidth]{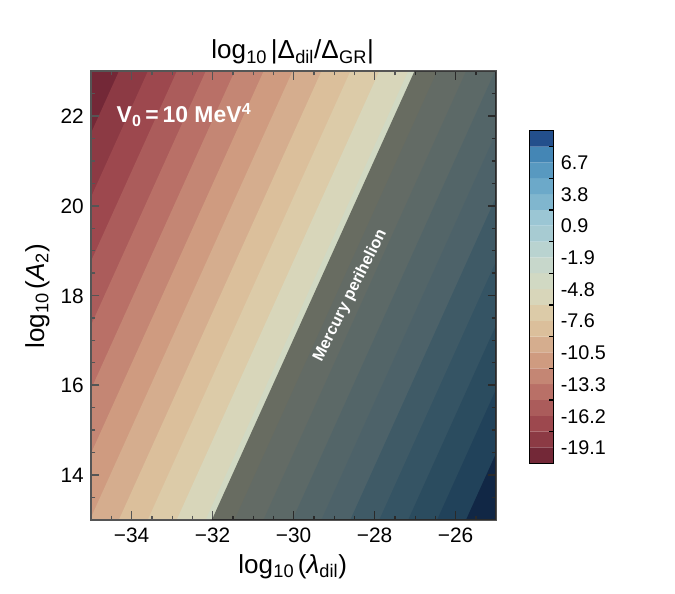}
\caption{As in Fig.~\ref{fig:deltaratiochameleon}, but for $\log_{10}|\Delta_{\text{dil}}/\Delta_{\text{GR}}|$ as a function of $\log_{10}\lambda_{\text{dil}}$ and $\log_{10}A_2$, the two dimensionless parameters determining the effects of dilaton screening. The two panels correspond to different representative values of the potential scale: $V_0=1\,{\text{MeV}}^4$ (left panel) and $10\,{\text{MeV}}^4$ (right panel). For $V_0=1\,{\text{MeV}}^4$, there is a narrow region of parameter space where the dilaton contribution is up to $\sim 3\times 10^{-5}$ of that of GR. The dark blue shaded region represents the area excluded by \textit{Mercury perihelion} \cite{Zhang:2016njn}.}
\label{fig:deltaratiodilaton}
\end{figure*}

We finally move on to the dilaton-screened modified gravity case, described in Sec.~\ref{subsubsec:dilatonapplication}. Also in this case there are three free parameters: the potential scale $\mathcal{V}_0$, whose dimensions are of energy density, the field value $\phi_{\star}$ at which the dilaton coupling to matter weakens, and the dilaton mass scale $M_{\text{dil}}$. For ease of comparison to other recent works in the literature, following Ref.~\cite{Fischer:2024eic}, we consider an alternative reparametrization of the effects of dilaton screening in terms of two dimensionless constants: a dimensionless self-coupling $\lambda_{\text{dil}}$, and a dimensionless constant $A_2 \equiv \lambda_{\text{dil}}^2 M_{\text{Pl}}^2/M_{\text{dil}}^2$ controlling the strength of the coupling to matter. This can be easily achieved by considering the substitution $\lambda_{\text{dil}} \, \tilde{\phi}=\phi-\phi_{\star}$ in Eq.~(\ref{eq:vdilaton}) and defining $V_{0} \equiv \mathcal{V}_0e^{-\phi_{\star}/M_{\text{Pl}}}$, see Ref.~\cite{Kading:2023mdk} for further details. With these redefinitions, the effective potential can then be rewritten as
\begin{equation}
V_{\text{eff}}(\phi) = V_{0}\exp \left ( -\frac{\lambda_{\text{dil}} \, \phi}{M_{\text{Pl}}} \right )  + \rho \left ( 1+{A_2}\frac{\phi^2}{2M_{\text{Pl}}^2} \right ) \,,
\end{equation}
where, for simplicity, we have renamed $\tilde{\phi}=\phi$. As in Refs.~\cite{Brax:2022uyh,Fischer:2023koa}, we take this field to be canonically normalized, so that the above effective potential defines the model used in our analysis, rather than a rescaling of the field in Eq.~(\ref{eq:ef}); its background mass, $m_{\infty}^2 \simeq A_2\rho_{\infty}/M_{\text{Pl}}^2$, then satisfies $m_{\infty}r \ll 1$ on Solar-System scales for the parameter values considered below, so that, as for the symmetron, Eqs.~(\ref{eq:Geps}--\ref{eq:betaeps}) and the \textit{Mercury-perihelion} bound apply. Starting from the above expression, we obtain the reparametrized version of Eq.~(\ref{eq:deltadil}), which now reads
\begin{equation}
\Delta_{\text{dil}} \equiv \frac{2GM\Bar{\varphi}\cosh(\zeta)A_2\phi_{\infty}^2}{rM_{\text{Pl}}^2\Phi_{\text{N}}}\,,
\end{equation}
with $\phi_\infty = \lambda_{\text{dil}} M_{\text{Pl}}V_0/(A_2 \rho_\infty)$; following Ref.~\cite{Zhang:2016njn}, we focus on the region of parameter space where $-35 \leq \log_{10}(\lambda_{\text{dil}}) \leq -25$ and $13 \leq \log_{10}(A_2) \leq 23$. By contrast, for the larger values of $A_2$ considered in Ref.~\cite{Fischer:2023koa}, the dilaton range is much shorter than the orbit of Mercury. Following in particular Ref.~\cite{Fischer:2023koa}, we consider the representative value $V_0=1\,{\text{MeV}}^4$ and, to further illustrate the effects of changing the scale of the dilaton potential, we also consider the case where $V_0=10\,{\text{MeV}}^4$. For the cosmological motivation of this model, see Ref.~\cite{Brax:2022uyh}, where $V_0$ is tied to the DE density.

In Fig.~\ref{fig:deltaratiodilaton}, we plot $\log_{10}|\Delta_{\text{dil}}/\Delta_{\text{GR}}|$ as a function of $\log_{10}\lambda_{\text{dil}}$ and $\log_{10}A_2$, with the two panels corresponding to the two different choices of $V_0$, and the \textit{Mercury-perihelion} constraints once again corresponding to the shaded regions. The trend as $V_0$ increases can be understood in analogy to the symmetron case. From Eq.~(\ref{eq:phimindil}), we see that as $V_0$ increases, $\phi_{\infty}$ increases and therefore screening weakens, leading to a tighter \textit{Mercury-perihelion} constraint. For $V_0 = 1\,\text{MeV}^{4}$, there is a region of the parameter space where the dilaton effects are up to $\sim 3\times 10^{-5}$ of those of GR.

\subsection{Detectability of proposed effects}
\label{subsec:detectability}

So far, we have derived theoretical predictions for the expected value of the parameter $\Delta$ within the three screened modified gravity models considered. We recall once again that $\cos^2\Delta$ modulates the degree of violation of the CHSH inequality, relative to the flat space-time case, within the specific gedankenexperiment we consider. We have focused on the relative contribution of screened modified gravity with respect to standard GR, finding that there are viable regions of parameter space where this contribution can be significant (see {Fig.~\ref{fig:deltaratiochameleon}). The remaining question to address is whether these effects can be detected at all, assuming that a certain sensitivity on the CHSH parameter $\mathscr{S}$ can be achieved. Here, we provide a brief and deliberately qualitative discussion of this issue.

To address this question, we assume that, within a given setting, $\mathscr{S}$ can be experimentally determined with a relative precision of $\delta\mathscr{S}/\mathscr{S} < 1$. In the $\Delta \ll 1$ limit (which we will shortly argue is appropriate), basic error propagation shows that this translates into a relative precision on $\Delta$ given as
\begin{equation}
\frac{\delta\Delta}{\Delta} \approx \frac{1}{2\Delta^2}\frac{\delta\mathscr{S}}{\mathscr{S}}\,.
\label{eq:relativeprecision}
\end{equation}
The fact that the relative precision on $\Delta$ depends on $\Delta$ itself, i.e., that $\Delta$ appears on the right-hand side of Eq.~(\ref{eq:relativeprecision}), is a common feature in indirect measurements, particularly in the regime where the derivative of the observable ($\mathscr{S}$) with respect to the parameter of interest ($\Delta$) becomes small.~ We use Eq.~(\ref{eq:relativeprecision}) to address the following question: given a certain currently achievable $\delta\mathscr{S}/\mathscr{S}$, what is the minimum value of $\Delta$ which can be resolved? The answer can be found by requiring that the relative uncertainty on $\Delta_{\min}$ is of order unity, i.e., setting $\delta\Delta_{\min}/\Delta_{\min} \sim 1$ in Eq.~(\ref{eq:relativeprecision}), leading to the following:
\begin{equation}
\Delta_{\min} \approx \sqrt{\frac{1}{2}\frac{\delta\mathscr{S}}{\mathscr{S}}}\,,
\label{eq:deltamin}
\end{equation}
which we again emphasize is obtained in the small-$\Delta$ limit. This limit is indeed appropriate for the experimental setting under consideration. In fact, as explicitly shown in Ref.~\cite{Terashima:2003rjs}, in the non-relativistic limit ($v \to 0$ and $r_s/r \to 0$, with $r_s$ the Schwarzschild radius of the source), the GR contribution is $\Delta \sim {\cal O}(10^{-9})$.\footnote{See Fig.~3 of Ref.~\cite{Terashima:2003rjs}, as well as the discussion around the end of Sec.~3.1 of the same work.} As we have seen, within viable regions of parameter space, the screened modified contribution to $\Delta$ is at best comparable to the GR one. Therefore, the small-$\Delta$ approximation adopted in deriving Eqs.~(\ref{eq:relativeprecision}, \ref{eq:deltamin}) is fully justified, since within the \textit{Cassini/Mercury-perihelion} constraints we expect $\Delta \lesssim 10^{-8}$.

Setting aside the question of whether our gedankenexperiment is feasible from an experimental standpoint, the next relevant issue concerns the precision $\delta\mathscr{S}/\mathscr{S}$ that can realistically be achieved. To the best of our knowledge, the most precise measurement to date is reported in Ref.~\cite{Storz:2023jjx}, which finds $\mathscr{S} = 2.0747 \pm 0.0033$, corresponding to a relative precision of the order of one part in $1000$, i.e., $\delta\mathscr{S}/\mathscr{S} \sim {\cal O}(10^{-3})$. While the measurement of Ref.~\cite{Storz:2023jjx} was performed using microwave photons in superconducting circuits, whereas our gedankenexperiment involves massive particles such as electrons, it is nonetheless reasonable to take $\delta\mathscr{S}/\mathscr{S} \sim 10^{-3}$ as a benchmark for the relative precision in the CHSH parameter $\mathscr{S}$ that is realistically achievable with current technology. From Eq.~(\ref{eq:deltamin}), this translates into $\Delta_{\min} \sim 10^{-2}$, which is far from the level required to disentangle screened modified gravity effects from those of GR. Conversely, detecting values of $\Delta$ as small as $\sim 10^{-9}$ would require a relative precision of $\delta\mathscr{S}/\mathscr{S} \sim 10^{-18}$, a sensitivity that is well beyond what can realistically be achieved in the near future.

At first glance, it would therefore appear that the prospects for testing screened DE in the proposed setting are elusive at best. Nevertheless, it is worth reminding the reader of an important caveat regarding this negative conclusion. Most entanglement-based tests of Bell-type inequalities make use of photons. In fact, the authors of Ref.~\cite{Yin:2017ghw} recently reported $\%$-level measurements of violations of Bell-type inequalities around the Earth, achieved through the distribution of entangled photon pairs via a satellite-based entanglement system with a baseline of up to $\approx 2400 \,\text{km}$. This suggests that an experimental setup similar to that of our gedankenexperiment is, in principle, feasible. In this pilot study, however, we have, for simplicity, considered massive particles (for instance, electrons), following closely the spirit of Ref.~\cite{Bittencourt:2020lgu}. Extending our analysis to massless particles such as photons is far from trivial: a significant part of the results of Sec.~\ref{sec:spincurved} would require revision and, more importantly, the inevitable coupling of photons to the scalar field would need to be taken into account~\cite{Brax:2010uq}. A dedicated study of the extension to massless particles is deferred to future work.

We also stress that the benchmark $\delta\mathscr{S}/\mathscr{S} \sim 10^{-3}$ reflects the precision achieved in a specific current experiment. However, significant improvements in the precision of future Bell-type inequality measurements are anticipated in the coming years. These improvements may result from increased detection efficiency (driven, for instance, by advances in superconducting nanowire single-photon detectors) and from improvements in entangled photon sources, particularly in terms of both brightness and fidelity (Sagnac-based sources could play an important role), among other technological advances. Our earlier deliberately pessimistic estimate regarding detection prospects should therefore be viewed as overly conservative, and is likely to improve considerably in the future as alternative experimental settings and technologies continue to advance.

\section{Conclusions}
\label{sec:conclusions}

The origin of cosmic acceleration remains one of the most pressing open problems in fundamental physics. In the absence of clear observational guidance, a broad interdisciplinary approach, exploring a wide range of detection strategies, becomes essential, and it is within this spirit that our work is framed. Building on recent studies of the interplay between gravitation and quantum nonlocality~\cite{Bittencourt:2020lgu}, we take a novel step forward by investigating how screening mechanisms affect the behavior of entangled quantum systems, in particular the degree of violation of Bell-type inequalities in curved space-time backgrounds. Screening mechanisms (or screened modified gravity theories, depending on the adopted frame) are well-motivated theoretical constructions that explain cosmic acceleration through new (ultra)light scalar degrees of freedom, while remaining consistent with local tests of gravity. Our pilot study is the first to propose testing screening mechanisms through quantum nonlocality, thereby establishing a new bridge between gravitation and quantum information.

Using a general formalism previously developed for curved space-time spin evolution, we compute the effects of screened modified gravity on Bell-type inequalities. The degree of violation of these inequalities is suppressed due to the net Wigner rotation accumulated by entangled states as they propagate through curved space-time, resulting from a succession of infinitesimal local Lorentz transformations along their worldlines. We apply this formalism to the CHSH inequality, considering three well-motivated screening mechanisms (the chameleon, symmetron, and environmental dilaton models) and explicitly calculate the suppression of the inequality’s violation in each case. As a possible experimental test of these effects, we adopt the gedankenexperiment of Refs.~\cite{Terashima:2003rjs,Bittencourt:2020lgu}, featuring an EPR pair of spin-$1/2$ particles propagating in a circular trajectory around the Earth. We find that, for the chameleon mechanism, effects associated with screening mechanisms can be significant compared to those arising from pure General Relativity, within regions of parameter space that remain unconstrained by current observations. This serves as a proof-of-principle for testing screening mechanisms through their subtle imprints on nonlocal quantum correlations. Prospects for detecting these effects with current CHSH inequality measurements, however, remain extremely challenging.

We emphasize that this work should be regarded as a pilot study and a first step toward understanding the potential impact of screened modified gravity on nonlocal quantum correlations. Despite the formidable challenges outlined above, our findings provide grounds for cautious optimism, which future work could strengthen. In particular, a crucial next step is extending the analysis to massless particles, not only because most entanglement-based Bell tests use photons, but also because we expect screening effects to be significantly larger in this case. Furthermore, for simplicity, we have considered specific experimental parameters (particle velocities, orbital distances, and initial states); exploring whether ``sweet spots'' in parameter space could maximize screening effects (similar to the analysis of Ref.~\cite{Bittencourt:2020lgu}) would be of great interest. Finally, a more ambitious extension of this pilot study could aim to distinguish between Einstein and Jordan frames through nonlocal quantum correlations (see Ref.~\cite{Chakraborty:2023kel} for a recent proposal in this direction), thereby shedding light on one of the most fundamental open questions in gravitational physics. These are all highly promising research directions for which our pilot study provides a foundation, making even our cautious conclusions a meaningful step forward at the intersection of gravitation and quantum information. We plan to explore these issues in future follow-up works.

\begin{acknowledgments}
\noindent We thank Philipp Hauke, Leonardo Mastrototaro, Anupam Mazumdar, Valerio Faraoni, and Jeremy Sakstein for useful discussions. FF and GL acknowledge support from the Istituto Nazionale di Fisica Nucleare (INFN) through the Commissione Scientifica Nazionale 4 (CSN4) Iniziativa Specifica ``Quantum Universe'' (QGSKY). SV acknowledges support from the University of Trento and the Provincia Autonoma di Trento (PAT, Autonomous Province of Trento) through the UniTrento Internal Call for Research 2023 grant ``Searching for Dark Energy off the beaten track'' (DARKTRACK, grant agreement no.\ E63C22000500003), and from the INFN CSN4 Iniziativa Specifica ``Quantum Fields in Gravity, Cosmology and Black Holes'' (FLAG). This publication is based upon work from the COST Action CA21106 ``COSMIC WISPers in the Dark Universe: Theory, astrophysics and experiments'' (Cosmic WISPers) and CA21136 ``Addressing observational tensions in cosmology with systematics and fundamental physics'' (CosmoVerse), both supported by COST (European Cooperation in Science and Technology).
\end{acknowledgments}

\bibliography{nonlocality}

@article{Terashima:2003rjs,
    author = "Terashima, Hiroaki and Ueda, Masahito",
    title = "{Einstein-Podolsky-Rosen correlation in a gravitational field}",
    eprint = "quant-ph/0307114",
    archivePrefix = "arXiv",
    doi = "10.1103/PhysRevA.69.032113",
    journal = "Phys. Rev. A",
    volume = "69",
    pages = "032113",
    year = "2004"
}

@article{Scharer:2014kya,
    author = {Sch\"arer, Andreas and Ang\'elil, Raymond and Bondarescu, Ruxandra and Jetzer, Philippe and Lundgren, Andrew},
    title = "{Testing scalar-tensor theories and parametrized post-Newtonian parameters in Earth orbit}",
    eprint = "1410.7914",
    archivePrefix = "arXiv",
    primaryClass = "gr-qc",
    doi = "10.1103/PhysRevD.90.123005",
    journal = "Phys. Rev. D",
    volume = "90",
    number = "12",
    pages = "123005",
    year = "2014"
}

@article{Hohmann:2013rba,
    author = "Hohmann, Manuel and Jarv, Laur and Kuusk, Piret and Randla, Erik",
    title = "{Post-Newtonian parameters $\gamma$ and $\beta$ of scalar-tensor gravity with a general potential}",
    eprint = "1309.0031",
    archivePrefix = "arXiv",
    primaryClass = "gr-qc",
    doi = "10.1103/PhysRevD.88.084054",
    journal = "Phys. Rev. D",
    volume = "88",
    number = "8",
    pages = "084054",
    year = "2013",
    note = "[Erratum: Phys.Rev.D 89, 069901 (2014)]"
}

@article{Zhang:2016njn,
    author = "Zhang, Xing and Zhao, Wen and Huang, He and Cai, Yifu",
    title = "{Post-Newtonian parameters and cosmological constant of screened modified gravity}",
    eprint = "1603.09450",
    archivePrefix = "arXiv",
    primaryClass = "gr-qc",
    doi = "10.1103/PhysRevD.93.124003",
    journal = "Phys. Rev. D",
    volume = "93",
    number = "12",
    pages = "124003",
    year = "2016"
}

@article{Burrage:2017qrf,
    author = "Burrage, Clare and Sakstein, Jeremy",
    title = "{Tests of Chameleon Gravity}",
    eprint = "1709.09071",
    archivePrefix = "arXiv",
    primaryClass = "astro-ph.CO",
    doi = "10.1007/s41114-018-0011-x",
    journal = "Living Rev. Rel.",
    volume = "21",
    number = "1",
    pages = "1",
    year = "2018"
}

@article{Brax:2013ida,
    author = "Brax, Philippe",
    title = "{Screening mechanisms in modified gravity}",
    doi = "10.1088/0264-9381/30/21/214005",
    journal = "Class. Quant. Grav.",
    volume = "30",
    pages = "214005",
    year = "2013"
}

@article{Hohmann:2017qje,
    author = {Hohmann, Manuel and Sch\"arer, Andreas},
    title = "{Post-Newtonian parameters \ensuremath{\gamma} and \ensuremath{\beta} of scalar-tensor gravity for a homogeneous gravitating sphere}",
    eprint = "1708.07851",
    archivePrefix = "arXiv",
    primaryClass = "gr-qc",
    doi = "10.1103/PhysRevD.96.104026",
    journal = "Phys. Rev. D",
    volume = "96",
    number = "10",
    pages = "104026",
    year = "2017"
}

@article{Bittencourt:2020lgu,
    author = "Bittencourt, Victor A. S. V. and Blasone, Massimo and Illuminati, Fabrizio and Lambiase, Gaetano and Luciano, Giuseppe Gaetano and Petruzziello, Luciano",
    title = "{Quantum nonlocality in extended theories of gravity}",
    eprint = "2012.15331",
    archivePrefix = "arXiv",
    primaryClass = "gr-qc",
    doi = "10.1103/PhysRevD.103.044051",
    journal = "Phys. Rev. D",
    volume = "103",
    number = "4",
    pages = "044051",
    year = "2021"
}

@article{Clauser:1969ny,
    author = "Clauser, John F. and Horne, Michael A. and Shimony, Abner and Holt, Richard A.",
    title = "{Proposed experiment to test local hidden variable theories}",
    doi = "10.1103/PhysRevLett.23.880",
    journal = "Phys. Rev. Lett.",
    volume = "23",
    pages = "880--884",
    year = "1969"
}

@article{Tsirelson:1980ry,
    author = "Tsirelson, B. S.",
    title = "{QUANTUM GENERALIZATIONS OF BELL'S INEQUALITY}",
    doi = "10.1007/BF00417500",
    journal = "Lett. Math. Phys.",
    volume = "4",
    pages = "93--100",
    year = "1980"
}

@article{Khoury:2003aq,
    author = "Khoury, Justin and Weltman, Amanda",
    title = "{Chameleon fields: Awaiting surprises for tests of gravity in space}",
    eprint = "astro-ph/0309300",
    archivePrefix = "arXiv",
    doi = "10.1103/PhysRevLett.93.171104",
    journal = "Phys. Rev. Lett.",
    volume = "93",
    pages = "171104",
    year = "2004"
}

@article{Khoury:2003rn,
    author = "Khoury, Justin and Weltman, Amanda",
    title = "{Chameleon cosmology}",
    eprint = "astro-ph/0309411",
    archivePrefix = "arXiv",
    doi = "10.1103/PhysRevD.69.044026",
    journal = "Phys. Rev. D",
    volume = "69",
    pages = "044026",
    year = "2004"
}

@article{Hinterbichler:2010es,
    author = "Hinterbichler, Kurt and Khoury, Justin",
    title = "{Symmetron Fields: Screening Long-Range Forces Through Local Symmetry Restoration}",
    eprint = "1001.4525",
    archivePrefix = "arXiv",
    primaryClass = "hep-th",
    doi = "10.1103/PhysRevLett.104.231301",
    journal = "Phys. Rev. Lett.",
    volume = "104",
    pages = "231301",
    year = "2010"
}

@article{Hinterbichler:2011ca,
    author = "Hinterbichler, Kurt and Khoury, Justin and Levy, Aaron and Matas, Andrew",
    title = "{Symmetron Cosmology}",
    eprint = "1107.2112",
    archivePrefix = "arXiv",
    primaryClass = "astro-ph.CO",
    doi = "10.1103/PhysRevD.84.103521",
    journal = "Phys. Rev. D",
    volume = "84",
    pages = "103521",
    year = "2011"
}

@article{Fischer:2024eic,
    author = {Fischer, Hauke and K\"ading, Christian and Pitschmann, Mario},
    title = "{Screened Scalar Fields in the Laboratory and the Solar System}",
    eprint = "2405.14638",
    archivePrefix = "arXiv",
    primaryClass = "gr-qc",
    doi = "10.3390/universe10070297",
    journal = "Universe",
    volume = "10",
    number = "7",
    pages = "297",
    year = "2024"
}

@article{Brax:2010gi,
    author = "Brax, Philippe and van de Bruck, Carsten and Davis, Anne-Christine and Shaw, Douglas J.",
    title = "{The Dilaton and Modified Gravity}",
    eprint = "1005.3735",
    archivePrefix = "arXiv",
    primaryClass = "astro-ph.CO",
    doi = "10.1103/PhysRevD.82.063519",
    journal = "Phys. Rev. D",
    volume = "82",
    pages = "063519",
    year = "2010"
}

@article{Brax:2011ja,
    author = "Brax, Philippe and van de Bruck, Carsten and Davis, Anne-Christine and Li, Baojiu and Shaw, Douglas J.",
    title = "{Nonlinear Structure Formation with the Environmentally Dependent Dilaton}",
    eprint = "1102.3692",
    archivePrefix = "arXiv",
    primaryClass = "astro-ph.CO",
    doi = "10.1103/PhysRevD.83.104026",
    journal = "Phys. Rev. D",
    volume = "83",
    pages = "104026",
    year = "2011"
}

@article{Damour:1994zq,
    author = "Damour, T. and Polyakov, Alexander M.",
    title = "{The String dilaton and a least coupling principle}",
    eprint = "hep-th/9401069",
    archivePrefix = "arXiv",
    doi = "10.1016/0550-3213(94)90143-0",
    journal = "Nucl. Phys. B",
    volume = "423",
    pages = "532--558",
    year = "1994"
}

@phdthesis{Sakstein:2014jrq,
    author = "Sakstein, Jeremy",
    title = "{Astrophysical Tests of Modified Gravity}",
    eprint = "1502.04503",
    archivePrefix = "arXiv",
    primaryClass = "astro-ph.CO",
    doi = "10.17863/CAM.16133",
    school = "Cambridge U., DAMTP",
    year = "2014"
}

@article{Kading:2023mdk,
    author = {K\"ading, Christian and Pitschmann, Mario and Voith, Caroline},
    title = "{Dilaton-induced open quantum dynamics}",
    eprint = "2306.10896",
    archivePrefix = "arXiv",
    primaryClass = "hep-ph",
    doi = "10.1140/epjc/s10052-023-11939-4",
    journal = "Eur. Phys. J. C",
    volume = "83",
    number = "8",
    pages = "767",
    year = "2023"
}

@article{SupernovaSearchTeam:1998fmf,
    author = "Riess, Adam G. and others",
    collaboration = "Supernova Search Team",
    title = "{Observational evidence from supernovae for an accelerating universe and a cosmological constant}",
    eprint = "astro-ph/9805201",
    archivePrefix = "arXiv",
    doi = "10.1086/300499",
    journal = "Astron. J.",
    volume = "116",
    pages = "1009--1038",
    year = "1998"
}

@article{SupernovaCosmologyProject:1998vns,
    author = "Perlmutter, S. and others",
    collaboration = "Supernova Cosmology Project",
    title = "{Measurements of $\Omega$ and $\Lambda$ from 42 High Redshift Supernovae}",
    eprint = "astro-ph/9812133",
    archivePrefix = "arXiv",
    reportNumber = "LBNL-41801, LBL-41801",
    doi = "10.1086/307221",
    journal = "Astrophys. J.",
    volume = "517",
    pages = "565--586",
    year = "1999"
}

@article{Nadathur:2020kvq,
    author = "Nadathur, Seshadri and Percival, Will J. and Beutler, Florian and Winther, Hans",
    title = "{Testing Low-Redshift Cosmic Acceleration with Large-Scale Structure}",
    eprint = "2001.11044",
    archivePrefix = "arXiv",
    primaryClass = "astro-ph.CO",
    doi = "10.1103/PhysRevLett.124.221301",
    journal = "Phys. Rev. Lett.",
    volume = "124",
    number = "22",
    pages = "221301",
    year = "2020"
}

@article{DiValentino:2020evt,
    author = "Di Valentino, Eleonora and Gariazzo, Stefano and Mena, Olga and Vagnozzi, Sunny",
    title = "{Soundness of Dark Energy properties}",
    eprint = "2005.02062",
    archivePrefix = "arXiv",
    primaryClass = "astro-ph.CO",
    doi = "10.1088/1475-7516/2020/07/045",
    journal = "JCAP",
    volume = "07",
    number = "07",
    pages = "045",
    year = "2020"
}

@article{Carroll:2000fy,
    author = "Carroll, Sean M.",
    title = "{The Cosmological constant}",
    eprint = "astro-ph/0004075",
    archivePrefix = "arXiv",
    reportNumber = "EFI-2000-13",
    doi = "10.12942/lrr-2001-1",
    journal = "Living Rev. Rel.",
    volume = "4",
    pages = "1",
    year = "2001"
}

@article{Weinberg:1988cp,
    author = "Weinberg, Steven",
    editor = "Hsu, Jong-Ping and Fine, D.",
    title = "{The Cosmological Constant Problem}",
    reportNumber = "UTTG-12-88",
    doi = "10.1103/RevModPhys.61.1",
    journal = "Rev. Mod. Phys.",
    volume = "61",
    pages = "1--23",
    year = "1989"
}

@article{Moreno-Pulido:2022phq,
    author = "Moreno-Pulido, Cristian and Sola Peracaula, Joan",
    title = "{Renormalizing the vacuum energy in cosmological spacetime: implications for the cosmological constant problem}",
    eprint = "2201.05827",
    archivePrefix = "arXiv",
    primaryClass = "gr-qc",
    doi = "10.1140/epjc/s10052-022-10484-w",
    journal = "Eur. Phys. J. C",
    volume = "82",
    number = "6",
    pages = "551",
    year = "2022"
}

@article{SolaPeracaula:2022hpd,
    author = "Sola Peracaula, Joan",
    title = "{The cosmological constant problem and running vacuum in the expanding universe}",
    eprint = "2203.13757",
    archivePrefix = "arXiv",
    primaryClass = "gr-qc",
    doi = "10.1098/rsta.2021.0182",
    journal = "Phil. Trans. Roy. Soc. Lond. A",
    volume = "380",
    pages = "20210182",
    year = "2022"
}

@article{Carroll:1998zi,
    author = "Carroll, Sean M.",
    title = "{Quintessence and the rest of the world}",
    eprint = "astro-ph/9806099",
    archivePrefix = "arXiv",
    reportNumber = "NSF-ITP-98-063",
    doi = "10.1103/PhysRevLett.81.3067",
    journal = "Phys. Rev. Lett.",
    volume = "81",
    pages = "3067--3070",
    year = "1998"
}

@article{Amendola:1999er,
    author = "Amendola, Luca",
    title = "{Coupled quintessence}",
    eprint = "astro-ph/9908023",
    archivePrefix = "arXiv",
    doi = "10.1103/PhysRevD.62.043511",
    journal = "Phys. Rev. D",
    volume = "62",
    pages = "043511",
    year = "2000"
}

@article{Brax:2021wcv,
    author = "Brax, Philippe and Casas, Santiago and Desmond, Harry and Elder, Benjamin",
    title = "{Testing Screened Modified Gravity}",
    eprint = "2201.10817",
    archivePrefix = "arXiv",
    primaryClass = "gr-qc",
    doi = "10.3390/universe8010011",
    journal = "Universe",
    volume = "8",
    number = "1",
    pages = "11",
    year = "2021"
}

@article{Vainshtein:1972sx,
    author = "Vainshtein, A. I.",
    title = "{To the problem of nonvanishing gravitation mass}",
    doi = "10.1016/0370-2693(72)90147-5",
    journal = "Phys. Lett. B",
    volume = "39",
    pages = "393--394",
    year = "1972"
}

@article{Yin:2022geb,
    author = "Yin, Peiran and Li, Rui and Yin, Chengjiang and Xu, Xiangyu and Bian, Xiang and Xie, Han and Duan, Chang-Kui and Huang, Pu and He, Jian-hua and Du, Jiangfeng",
    title = "{Experiments with levitated force sensor challenge theories of dark energy}",
    eprint = "2405.09791",
    archivePrefix = "arXiv",
    primaryClass = "gr-qc",
    doi = "10.1038/s41567-022-01706-9",
    journal = "Nature Phys.",
    volume = "18",
    number = "10",
    pages = "1181--1185",
    year = "2022"
}

@article{Brax:2019rwf,
    author = "Brax, Philippe and van de Bruck, Carsten and Davis, Anne-Christine",
    title = "{Swampland and screened modified gravity}",
    eprint = "1911.09169",
    archivePrefix = "arXiv",
    primaryClass = "hep-th",
    doi = "10.1103/PhysRevD.101.083514",
    journal = "Phys. Rev. D",
    volume = "101",
    number = "8",
    pages = "083514",
    year = "2020"
}

@article{Bertotti:2003rm,
    author = "Bertotti, B. and Iess, L. and Tortora, P.",
    title = "{A test of general relativity using radio links with the Cassini spacecraft}",
    doi = "10.1038/nature01997",
    journal = "Nature",
    volume = "425",
    pages = "374--376",
    year = "2003"
}

@article{Fischer:2023koa,
    author = {Fischer, Hauke and K{\"a}ding, Christian and Sedmik, Ren{\'e} I. P. and Abele, Hartmut and Brax, Philippe and Pitschmann, Mario},
    title = "{Search for environment-dependent dilatons}",
    eprint = "2307.00243",
    archivePrefix = "arXiv",
    primaryClass = "gr-qc",
    doi = "10.1016/j.dark.2024.101419",
    journal = "Phys. Dark Univ.",
    volume = "43",
    pages = "101419",
    year = "2024"
}

@article{Brax:2022uyh,
    author = {Brax, Philippe and Fischer, Hauke and K{\"a}ding, Christian and Pitschmann, Mario},
    title = "{The environment dependent dilaton in the laboratory and the solar system}",
    eprint = "2203.12512",
    archivePrefix = "arXiv",
    primaryClass = "gr-qc",
    doi = "10.1140/epjc/s10052-022-10905-w",
    journal = "Eur. Phys. J. C",
    volume = "82",
    number = "10",
    pages = "934",
    year = "2022"
}

@article{Storz:2023jjx,
    author = "Storz, Simon and others",
    title = "{Loophole-free Bell inequality violation with superconducting circuits}",
    doi = "10.1038/s41586-023-05885-0",
    journal = "Nature",
    volume = "617",
    number = "7960",
    pages = "265--270",
    year = "2023"
}

@article{Chakraborty:2023kel,
    author = "Chakraborty, Sumanta and Mazumdar, Anupam and Pradhan, Ritapriya",
    title = "{Distinguishing Jordan and Einstein frames in gravity through entanglement}",
    eprint = "2310.06899",
    archivePrefix = "arXiv",
    primaryClass = "gr-qc",
    doi = "10.1103/PhysRevD.108.L121505",
    journal = "Phys. Rev. D",
    volume = "108",
    number = "12",
    pages = "L121505",
    year = "2023"
}

@article{Yin:2017ghw,
    author = "Yin, Juan and others",
    title = "{Satellite-Based Entanglement Distribution Over 1200 kilometers}",
    eprint = "1707.01339",
    archivePrefix = "arXiv",
    primaryClass = "quant-ph",
    doi = "10.1126/science.aan3211",
    journal = "Science",
    volume = "356",
    pages = "1140--1144",
    year = "2017"
}

@article{Bell:1964kc,
    author = "Bell, J. S.",
    title = "{On the Einstein-Podolsky-Rosen paradox}",
    reportNumber = "RX-1376",
    doi = "10.1103/PhysicsPhysiqueFizika.1.195",
    journal = "Physics Physique Fizika",
    volume = "1",
    pages = "195--200",
    year = "1964"
}

@article{Brunner:2013est,
    author = "Brunner, Nicolas and Cavalcanti, Daniel and Pironio, Stefano and Scarani, Valerio and Wehner, Stephanie",
    title = "{Bell nonlocality}",
    eprint = "1303.2849",
    archivePrefix = "arXiv",
    primaryClass = "quant-ph",
    doi = "10.1103/RevModPhys.86.419",
    journal = "Rev. Mod. Phys.",
    volume = "86",
    pages = "419",
    year = "2014"
}

@article{Czachor:1996cj,
    author = "Czachor, Marek",
    title = "{Einstein-Podolsky-Rosen-Bohm experiment with relativistic massive particles}",
    eprint = "quant-ph/9609022",
    archivePrefix = "arXiv",
    doi = "10.1103/PhysRevA.55.72",
    journal = "Phys. Rev. A",
    volume = "55",
    pages = "72",
    year = "1997"
}

@article{Gingrich:2002ota,
    author = "Gingrich, Robert M. and Adami, Christoph",
    title = "{Quantum Entanglement of Moving Bodies}",
    eprint = "quant-ph/0205179",
    archivePrefix = "arXiv",
    doi = "10.1103/PhysRevLett.89.270402",
    journal = "Phys. Rev. Lett.",
    volume = "89",
    pages = "270402",
    year = "2002"
}

@article{Terashima:2003ds,
    author = "Terashima, Hiroaki and Ueda, Masahito",
    title = "{Spin decoherence caused by spacetime curvature}",
    eprint = "quant-ph/0312064",
    archivePrefix = "arXiv",
    doi = "10.1088/0305-4470/38/9/013",
    journal = "J. Phys. A",
    volume = "38",
    pages = "2029--2037",
    year = "2005"
}

@article{Friis:2009va,
    author = "Friis, Nicolai and Bertlmann, Reinhold A. and Huber, Marcus and Hiesmayr, Beatrix C.",
    title = "{Relativistic entanglement of two massive particles}",
    eprint = "0912.4863",
    archivePrefix = "arXiv",
    primaryClass = "quant-ph",
    doi = "10.1103/PhysRevA.81.042114",
    journal = "Phys. Rev. A",
    volume = "81",
    pages = "042114",
    year = "2010"
}

@article{LeviSaid:2009yra,
    author = "Levi Said, Jackson and Adami, Kristian Zarb",
    title = "{The EPR correlation in Kerr-Newman spacetime}",
    eprint = "1001.0788",
    archivePrefix = "arXiv",
    primaryClass = "quant-ph",
    doi = "10.1103/PhysRevD.81.124012",
    journal = "Phys. Rev. D",
    volume = "81",
    pages = "124012",
    year = "2010"
}

@article{Einstein:1935rr,
    author = "Einstein, Albert and Podolsky, Boris and Rosen, Nathan",
    title = "{Can quantum mechanical description of physical reality be considered complete?}",
    doi = "10.1103/PhysRev.47.777",
    journal = "Phys. Rev.",
    volume = "47",
    pages = "777--780",
    year = "1935"
}

@article{Rovelli:1990pi,
    author = "Rovelli, Carlo",
    title = "{QUANTUM REFERENCE SYSTEMS}",
    reportNumber = "PITT-90-11",
    doi = "10.1088/0264-9381/8/2/012",
    journal = "Class. Quant. Grav.",
    volume = "8",
    pages = "317--332",
    year = "1991"
}

@article{Capozziello:2013wea,
    author = "Capozziello, Salvatore and Luongo, Orlando and Mancini, Stefano",
    title = "{Cosmological dark energy effects from entanglement}",
    eprint = "1302.5884",
    archivePrefix = "arXiv",
    primaryClass = "gr-qc",
    doi = "10.1016/j.physleta.2013.02.038",
    journal = "Phys. Lett. A",
    volume = "377",
    pages = "1061--1064",
    year = "2013"
}

@article{Giacomini:2017zju,
    author = "Giacomini, Flaminia and Castro-Ruiz, Esteban and Brukner, {\v{C}}aslav",
    title = "{Quantum mechanics and the covariance of physical laws in quantum reference frames}",
    eprint = "1712.07207",
    archivePrefix = "arXiv",
    primaryClass = "quant-ph",
    doi = "10.1038/s41467-018-08155-0",
    journal = "Nature Commun.",
    volume = "10",
    number = "1",
    pages = "494",
    year = "2019"
}

@article{Luongo:2019ztg,
    author = "Luongo, Orlando and Mancini, Stefano",
    title = "{Entanglement in model independent cosmological scenario}",
    eprint = "1903.05860",
    archivePrefix = "arXiv",
    primaryClass = "gr-qc",
    doi = "10.1142/S0219887819501147",
    journal = "Int. J. Geom. Meth. Mod. Phys.",
    volume = "16",
    number = "08",
    pages = "1950114",
    year = "2019"
}

@article{Belenchia:2019gcc,
    author = "Belenchia, Alessio and Wald, Robert M. and Giacomini, Flaminia and Castro-Ruiz, Esteban and Brukner, {\v{C}}aslav and Aspelmeyer, Markus",
    title = "{Information Content of the Gravitational Field of a Quantum Superposition}",
    eprint = "1905.04496",
    archivePrefix = "arXiv",
    primaryClass = "quant-ph",
    doi = "10.1142/S0218271819430016",
    journal = "Int. J. Mod. Phys. D",
    volume = "28",
    number = "14",
    pages = "1943001",
    year = "2019"
}

@article{Castro-Ruiz:2019nnl,
    author = "Castro-Ruiz, Esteban and Giacomini, Flaminia and Belenchia, Alessio and Brukner, {\v{C}}aslav",
    title = "{Quantum clocks and the temporal localisability of events in the presence of gravitating quantum systems}",
    eprint = "1908.10165",
    archivePrefix = "arXiv",
    primaryClass = "quant-ph",
    doi = "10.1038/s41467-020-16013-1",
    journal = "Nature Commun.",
    volume = "11",
    number = "1",
    pages = "2672",
    year = "2020"
}

@article{Galley:2020qsf,
    author = "Galley, Thomas D. and Giacomini, Flaminia and Selby, John H.",
    title = "{A no-go theorem on the nature of the gravitational field beyond quantum theory}",
    eprint = "2012.01441",
    archivePrefix = "arXiv",
    primaryClass = "quant-ph",
    doi = "10.22331/q-2022-08-17-779",
    journal = "Quantum",
    volume = "6",
    pages = "779",
    year = "2022"
}

@article{Giacomini:2021aof,
    author = "Giacomini, Flaminia and Brukner, {\v{C}}aslav",
    title = "{Quantum superposition of spacetimes obeys Einstein's equivalence principle}",
    eprint = "2109.01405",
    archivePrefix = "arXiv",
    primaryClass = "quant-ph",
    doi = "10.1116/5.0070018",
    journal = "AVS Quantum Sci.",
    volume = "4",
    number = "1",
    pages = "015601",
    year = "2022"
}

@article{Belfiglio:2022cnd,
    author = "Belfiglio, Alessio and Luongo, Orlando and Mancini, Stefano",
    title = "{Geometric corrections to cosmological entanglement}",
    eprint = "2201.12299",
    archivePrefix = "arXiv",
    primaryClass = "gr-qc",
    doi = "10.1103/PhysRevD.105.123523",
    journal = "Phys. Rev. D",
    volume = "105",
    number = "12",
    pages = "123523",
    year = "2022"
}

@article{Christodoulou:2022knr,
    author = "Christodoulou, Marios and Di Biagio, Andrea and Howl, Richard and Rovelli, Carlo",
    title = "{Gravity entanglement, quantum reference systems, degrees of freedom}",
    eprint = "2207.03138",
    archivePrefix = "arXiv",
    primaryClass = "quant-ph",
    doi = "10.1088/1361-6382/acb0aa",
    journal = "Class. Quant. Grav.",
    volume = "40",
    number = "4",
    pages = "047001",
    year = "2023"
}

@article{Overstreet:2022zgq,
    author = "Overstreet, Chris and Curti, Joseph and Kim, Minjeong and Asenbaum, Peter and Kasevich, Mark A. and Giacomini, Flaminia",
    title = "{Inference of gravitational field superposition from quantum measurements}",
    eprint = "2209.02214",
    archivePrefix = "arXiv",
    primaryClass = "quant-ph",
    reportNumber = "FERMILAB-PUB-22-434-SQMS-V",
    doi = "10.1103/PhysRevD.108.084038",
    journal = "Phys. Rev. D",
    volume = "108",
    number = "8",
    pages = "084038",
    year = "2023"
}

@article{Belfiglio:2022yvs,
    author = "Belfiglio, Alessio and Luongo, Orlando and Mancini, Stefano",
    title = "{Inflationary entanglement}",
    eprint = "2212.06448",
    archivePrefix = "arXiv",
    primaryClass = "gr-qc",
    doi = "10.1103/PhysRevD.107.103512",
    journal = "Phys. Rev. D",
    volume = "107",
    number = "10",
    pages = "103512",
    year = "2023"
}

@article{Galley:2023byb,
    author = "Galley, Thomas D. and Giacomini, Flaminia and Selby, John H.",
    title = "{Any consistent coupling between classical gravity and quantum matter is fundamentally irreversible}",
    eprint = "2301.10261",
    archivePrefix = "arXiv",
    primaryClass = "quant-ph",
    doi = "10.22331/q-2023-10-16-1142",
    journal = "Quantum",
    volume = "7",
    pages = "1142",
    year = "2023"
}

@article{Luongo:2023jyz,
    author = "Luongo, Orlando and Mancini, Stefano and Pierosara, Paolo",
    title = "{Entanglement entropy for spherically symmetric regular black holes}",
    eprint = "2304.06593",
    archivePrefix = "arXiv",
    primaryClass = "gr-qc",
    doi = "10.1103/PhysRevD.108.104059",
    journal = "Phys. Rev. D",
    volume = "108",
    number = "10",
    pages = "104059",
    year = "2023"
}

@article{Belfiglio:2023moe,
    author = "Belfiglio, Alessio and Luongo, Orlando and Mancini, Stefano",
    title = "{Superhorizon entanglement from inflationary particle production}",
    eprint = "2312.11419",
    archivePrefix = "arXiv",
    primaryClass = "gr-qc",
    doi = "10.1103/PhysRevD.109.123520",
    journal = "Phys. Rev. D",
    volume = "109",
    number = "12",
    pages = "123520",
    year = "2024"
}

@article{Chen:2024xvm,
    author = "Chen, Lin-Qing and Giacomini, Flaminia",
    title = "{Quantum effects in gravity beyond the Newton potential from a delocalised quantum source}",
    eprint = "2402.10288",
    archivePrefix = "arXiv",
    primaryClass = "quant-ph",
    month = "2",
    year = "2024"
}

@article{Belfiglio:2024qsa,
    author = "Belfiglio, Alessio and Luongo, Orlando and Mancini, Stefano and Tomasi, Sebastiano",
    title = "{Entanglement entropy in quantum black holes}",
    eprint = "2404.00715",
    archivePrefix = "arXiv",
    primaryClass = "gr-qc",
    doi = "10.1088/1361-6382/ad9e66",
    journal = "Class. Quant. Grav.",
    volume = "42",
    number = "3",
    pages = "035006",
    year = "2025"
}

@article{Capozziello:2024mxh,
    author = "Capozziello, Salvatore and Lapponi, Alessio and Luongo, Orlando and Mancini, Stefano",
    title = "{Preserving quantum information in f(Q) non-metric gravity cosmology}",
    eprint = "2406.19274",
    archivePrefix = "arXiv",
    primaryClass = "gr-qc",
    doi = "10.1140/epjc/s10052-024-13449-3",
    journal = "Eur. Phys. J. C",
    volume = "84",
    number = "10",
    pages = "1081",
    year = "2024"
}

@article{Belfiglio:2024wel,
    author = "Belfiglio, Alessio and Chandran, S. Mahesh and Luongo, Orlando and Mancini, Stefano",
    title = "{Horizon entanglement area law from regular black hole thermodynamics}",
    eprint = "2407.03775",
    archivePrefix = "arXiv",
    primaryClass = "gr-qc",
    doi = "10.1103/PhysRevD.111.024013",
    journal = "Phys. Rev. D",
    volume = "111",
    number = "2",
    pages = "024013",
    year = "2025"
}

@article{Cortes:2024jvb,
    author = "Cort{\^e}s, Marina and Liddle, Andrew R.",
    title = "{Hawking evaporation and the Landauer Principle}",
    eprint = "2407.08777",
    archivePrefix = "arXiv",
    primaryClass = "gr-qc",
    month = "7",
    year = "2024"
}

@article{Trivedi:2024qys,
    author = "Trivedi, Oem",
    title = "{Landauer's principle and information at the cosmological horizon}",
    eprint = "2407.15231",
    archivePrefix = "arXiv",
    primaryClass = "gr-qc",
    month = "7",
    year = "2024"
}

@article{Bagchi:2024try,
    author = "Bagchi, Bijan and Ghosh, Aritra and Sen, Sauvik",
    title = "{Landauer{\textquoteright}s principle and black hole area quantization}",
    eprint = "2408.02077",
    archivePrefix = "arXiv",
    primaryClass = "gr-qc",
    doi = "10.1007/s10714-024-03296-8",
    journal = "Gen. Rel. Grav.",
    volume = "56",
    number = "9",
    pages = "108",
    year = "2024"
}

@article{Odintsov:2024ipb,
    author = "Odintsov, Sergei D. and Paul, Tanmoy and SenGupta, Soumitra",
    title = "{Natural validation of the second law of thermodynamics in cosmology}",
    eprint = "2409.05009",
    archivePrefix = "arXiv",
    primaryClass = "gr-qc",
    doi = "10.1103/PhysRevD.111.043544",
    journal = "Phys. Rev. D",
    volume = "111",
    number = "4",
    pages = "043544",
    year = "2025"
}

@article{Abreu:2025mxa,
    author = "Abreu, Everton M. C. and Neto, Jorge Ananias",
    title = "{From modified Tsallis{\textendash}R{\'e}nyi entropy to a MOND-like force law, Bekenstein bound, and Landauer principle for black holes}",
    eprint = "2505.03061",
    archivePrefix = "arXiv",
    primaryClass = "gr-qc",
    doi = "10.1016/j.physletb.2025.139579",
    journal = "Phys. Lett. B",
    volume = "866",
    pages = "139579",
    year = "2025"
}

@article{Belfiglio:2025cst,
    author = "Belfiglio, Alessio and Luongo, Orlando and Mancini, Stefano",
    title = "{Quantum entanglement in cosmology}",
    eprint = "2506.03841",
    archivePrefix = "arXiv",
    primaryClass = "gr-qc",
    month = "6",
    year = "2025"
}

@article{Saridakis:2016mjd,
    author = "Saridakis, Emmanuel N. and Tsoukalas, Minas",
    title = "{Bi-scalar modified gravity and cosmology with conformal invariance}",
    eprint = "1602.06890",
    archivePrefix = "arXiv",
    primaryClass = "gr-qc",
    doi = "10.1088/1475-7516/2016/04/017",
    journal = "JCAP",
    volume = "04",
    pages = "017",
    year = "2016"
}

@article{Petronikolou:2023cwu,
    author = "Petronikolou, Maria and Saridakis, Emmanuel N.",
    title = "{Alleviating the H$_{0}$ Tension in Scalar{\textendash}Tensor and Bi-Scalar{\textendash}Tensor Theories}",
    eprint = "2308.16044",
    archivePrefix = "arXiv",
    primaryClass = "gr-qc",
    doi = "10.3390/universe9090397",
    journal = "Universe",
    volume = "9",
    number = "9",
    pages = "397",
    year = "2023"
}

@article{Ohashi:2015fma,
    author = "Ohashi, Seiju and Tanahashi, Norihiro and Kobayashi, Tsutomu and Yamaguchi, Masahide",
    title = "{The most general second-order field equations of bi-scalar-tensor theory in four dimensions}",
    eprint = "1505.06029",
    archivePrefix = "arXiv",
    primaryClass = "gr-qc",
    reportNumber = "KEK-TH-1826, RUP-15-11",
    doi = "10.1007/JHEP07(2015)008",
    journal = "JHEP",
    volume = "07",
    pages = "008",
    year = "2015"
}

@article{Cecchini:2024xoq,
    author = "Cecchini, Chiara and De Angelis, Mariaveronica and Giar{\`e}, William and Rinaldi, Massimiliano and Vagnozzi, Sunny",
    title = "{Testing scale-invariant inflation against cosmological data}",
    eprint = "2403.04316",
    archivePrefix = "arXiv",
    primaryClass = "astro-ph.CO",
    doi = "10.1088/1475-7516/2024/07/058",
    journal = "JCAP",
    volume = "07",
    pages = "058",
    year = "2024"
}

@article{Rinaldi:2014gha,
    author = "Rinaldi, Massimiliano and Cognola, Guido and Vanzo, Luciano and Zerbini, Sergio",
    title = "{Inflation in scale-invariant theories of gravity}",
    eprint = "1410.0631",
    archivePrefix = "arXiv",
    primaryClass = "gr-qc",
    doi = "10.1103/PhysRevD.91.123527",
    journal = "Phys. Rev. D",
    volume = "91",
    number = "12",
    pages = "123527",
    year = "2015"
}

@article{Ratra:1987rm,
    author = "Ratra, Bharat and Peebles, P. J. E.",
    title = "{Cosmological Consequences of a Rolling Homogeneous Scalar Field}",
    reportNumber = "PUPT-1072",
    doi = "10.1103/PhysRevD.37.3406",
    journal = "Phys. Rev. D",
    volume = "37",
    pages = "3406",
    year = "1988"
}

@article{Wetterich:1987fm,
    author = "Wetterich, C.",
    title = "{Cosmology and the Fate of Dilatation Symmetry}",
    eprint = "1711.03844",
    archivePrefix = "arXiv",
    primaryClass = "hep-th",
    reportNumber = "PRINT-87-0756, DESY-87-123",
    doi = "10.1016/0550-3213(88)90193-9",
    journal = "Nucl. Phys. B",
    volume = "302",
    pages = "668--696",
    year = "1988"
}

@article{Caldwell:1997ii,
    author = "Caldwell, R. R. and Dave, Rahul and Steinhardt, Paul J.",
    title = "{Cosmological imprint of an energy component with general equation of state}",
    eprint = "astro-ph/9708069",
    archivePrefix = "arXiv",
    doi = "10.1103/PhysRevLett.80.1582",
    journal = "Phys. Rev. Lett.",
    volume = "80",
    pages = "1582--1585",
    year = "1998"
}

@article{Zlatev:1998tr,
    author = "Zlatev, Ivaylo and Wang, Li-Min and Steinhardt, Paul J.",
    title = "{Quintessence, cosmic coincidence, and the cosmological constant}",
    eprint = "astro-ph/9807002",
    archivePrefix = "arXiv",
    doi = "10.1103/PhysRevLett.82.896",
    journal = "Phys. Rev. Lett.",
    volume = "82",
    pages = "896--899",
    year = "1999"
}

@article{Capozziello:2002rd,
    author = "Capozziello, Salvatore",
    title = "{Curvature quintessence}",
    eprint = "gr-qc/0201033",
    archivePrefix = "arXiv",
    doi = "10.1142/S0218271802002025",
    journal = "Int. J. Mod. Phys. D",
    volume = "11",
    pages = "483--492",
    year = "2002"
}

@article{Li:2004rb,
    author = "Li, Miao",
    title = "{A Model of holographic dark energy}",
    eprint = "hep-th/0403127",
    archivePrefix = "arXiv",
    doi = "10.1016/j.physletb.2004.10.014",
    journal = "Phys. Lett. B",
    volume = "603",
    pages = "1",
    year = "2004"
}

@article{Nojiri:2006ri,
    author = "Nojiri, Shin'ichi and Odintsov, Sergei D.",
    editor = "Borowiec, Andrzej",
    title = "{Introduction to modified gravity and gravitational alternative for dark energy}",
    eprint = "hep-th/0601213",
    archivePrefix = "arXiv",
    reportNumber = "KARP-2006-06",
    doi = "10.1142/S0219887807001928",
    journal = "eConf",
    volume = "C0602061",
    pages = "06",
    year = "2006"
}

@article{Bamba:2012cp,
    author = "Bamba, Kazuharu and Capozziello, Salvatore and Nojiri, Shin'ichi and Odintsov, Sergei D.",
    title = "{Dark energy cosmology: the equivalent description via different theoretical models and cosmography tests}",
    eprint = "1205.3421",
    archivePrefix = "arXiv",
    primaryClass = "gr-qc",
    doi = "10.1007/s10509-012-1181-8",
    journal = "Astrophys. Space Sci.",
    volume = "342",
    pages = "155--228",
    year = "2012"
}

@article{Chamseddine:2013kea,
    author = "Chamseddine, Ali H. and Mukhanov, Viatcheslav",
    title = "{Mimetic Dark Matter}",
    eprint = "1308.5410",
    archivePrefix = "arXiv",
    primaryClass = "astro-ph.CO",
    doi = "10.1007/JHEP11(2013)135",
    journal = "JHEP",
    volume = "11",
    pages = "135",
    year = "2013"
}

@article{Rinaldi:2014yta,
    author = "Rinaldi, Massimiliano",
    title = "{Higgs Dark Energy}",
    eprint = "1404.0532",
    archivePrefix = "arXiv",
    primaryClass = "astro-ph.CO",
    doi = "10.1088/0264-9381/32/4/045002",
    journal = "Class. Quant. Grav.",
    volume = "32",
    pages = "045002",
    year = "2015"
}

@article{Sebastiani:2016ras,
    author = "Sebastiani, L. and Vagnozzi, S. and Myrzakulov, R.",
    title = "{Mimetic gravity: a review of recent developments and applications to cosmology and astrophysics}",
    eprint = "1612.08661",
    archivePrefix = "arXiv",
    primaryClass = "gr-qc",
    doi = "10.1155/2017/3156915",
    journal = "Adv. High Energy Phys.",
    volume = "2017",
    pages = "3156915",
    year = "2017"
}

@article{Nojiri:2017ncd,
    author = "Nojiri, S. and Odintsov, S. D. and Oikonomou, V. K.",
    title = "{Modified Gravity Theories on a Nutshell: Inflation, Bounce and Late-time Evolution}",
    eprint = "1705.11098",
    archivePrefix = "arXiv",
    primaryClass = "gr-qc",
    reportNumber = "PHYS.REPT.-692-(2017)-1-104, Phys.Rept. 692 (2017) 1-104",
    doi = "10.1016/j.physrep.2017.06.001",
    journal = "Phys. Rept.",
    volume = "692",
    pages = "1--104",
    year = "2017"
}

@article{Dutta:2017fjw,
    author = "Dutta, Jibitesh and Khyllep, Wompherdeiki and Saridakis, Emmanuel N. and Tamanini, Nicola and Vagnozzi, Sunny",
    title = "{Cosmological dynamics of mimetic gravity}",
    eprint = "1711.07290",
    archivePrefix = "arXiv",
    primaryClass = "gr-qc",
    doi = "10.1088/1475-7516/2018/02/041",
    journal = "JCAP",
    volume = "02",
    pages = "041",
    year = "2018"
}

@article{Casalino:2018tcd,
    author = "Casalino, Alessandro and Rinaldi, Massimiliano and Sebastiani, Lorenzo and Vagnozzi, Sunny",
    title = "{Mimicking dark matter and dark energy in a mimetic model compatible with GW170817}",
    eprint = "1803.02620",
    archivePrefix = "arXiv",
    primaryClass = "gr-qc",
    doi = "10.1016/j.dark.2018.10.001",
    journal = "Phys. Dark Univ.",
    volume = "22",
    pages = "108",
    year = "2018"
}

@article{Saridakis:2018unr,
    author = "Saridakis, Emmanuel N. and Bamba, Kazuharu and Myrzakulov, R. and Anagnostopoulos, Fotios K.",
    title = "{Holographic dark energy through Tsallis entropy}",
    eprint = "1806.01301",
    archivePrefix = "arXiv",
    primaryClass = "gr-qc",
    reportNumber = "FU-PCG-36",
    doi = "10.1088/1475-7516/2018/12/012",
    journal = "JCAP",
    volume = "12",
    pages = "012",
    year = "2018"
}

@article{Visinelli:2018utg,
    author = "Visinelli, Luca and Vagnozzi, Sunny",
    title = "{Cosmological window onto the string axiverse and the supersymmetry breaking scale}",
    eprint = "1809.06382",
    archivePrefix = "arXiv",
    primaryClass = "hep-ph",
    reportNumber = "NORDITA-2018-051",
    doi = "10.1103/PhysRevD.99.063517",
    journal = "Phys. Rev. D",
    volume = "99",
    number = "6",
    pages = "063517",
    year = "2019"
}

@article{Odintsov:2019evb,
    author = "Odintsov, S. D. and Oikonomou, V. K.",
    title = "{Unification of Inflation with Dark Energy in $f(R)$ Gravity and Axion Dark Matter}",
    eprint = "1905.03496",
    archivePrefix = "arXiv",
    primaryClass = "gr-qc",
    doi = "10.1103/PhysRevD.99.104070",
    journal = "Phys. Rev. D",
    volume = "99",
    number = "10",
    pages = "104070",
    year = "2019"
}

@article{Saridakis:2020zol,
    author = "Saridakis, Emmanuel N.",
    title = "{Barrow holographic dark energy}",
    eprint = "2005.04115",
    archivePrefix = "arXiv",
    primaryClass = "gr-qc",
    doi = "10.1103/PhysRevD.102.123525",
    journal = "Phys. Rev. D",
    volume = "102",
    number = "12",
    pages = "123525",
    year = "2020"
}

@article{Odintsov:2020zct,
    author = "Odintsov, S. D. and Oikonomou, V. K. and Paul, Tanmoy",
    title = "{From a Bounce to the Dark Energy Era with $F(R)$ Gravity}",
    eprint = "2009.09947",
    archivePrefix = "arXiv",
    primaryClass = "gr-qc",
    doi = "10.1088/1361-6382/abbc47",
    journal = "Class. Quant. Grav.",
    volume = "37",
    number = "23",
    pages = "235005",
    year = "2020"
}

@article{Oikonomou:2020qah,
    author = "Oikonomou, V. K.",
    title = "{Unifying inflation with early and late dark energy epochs in axion $F(R)$ gravity}",
    eprint = "2012.00586",
    archivePrefix = "arXiv",
    primaryClass = "astro-ph.CO",
    doi = "10.1103/PhysRevD.103.044036",
    journal = "Phys. Rev. D",
    volume = "103",
    number = "4",
    pages = "044036",
    year = "2021"
}

@article{Motta:2021hvl,
    author = "Motta, Ver{\'o}nica and Garc{\'\i}a-Aspeitia, Miguel A. and Hern{\'a}ndez-Almada, Alberto and Maga{\~n}a, Juan and Verdugo, Tom{\'a}s",
    title = "{Taxonomy of Dark Energy Models}",
    eprint = "2104.04642",
    archivePrefix = "arXiv",
    primaryClass = "astro-ph.CO",
    doi = "10.3390/universe7060163",
    journal = "Universe",
    volume = "7",
    number = "6",
    pages = "163",
    year = "2021"
}

@article{Drepanou:2021jiv,
    author = "Drepanou, Niki and Lymperis, Andreas and Saridakis, Emmanuel N. and Yesmakhanova, Kuralay",
    title = "{Kaniadakis holographic dark energy and cosmology}",
    eprint = "2109.09181",
    archivePrefix = "arXiv",
    primaryClass = "gr-qc",
    doi = "10.1140/epjc/s10052-022-10415-9",
    journal = "Eur. Phys. J. C",
    volume = "82",
    number = "5",
    pages = "449",
    year = "2022"
}

@article{Oikonomou:2022wuk,
    author = "Oikonomou, V. K. and Giannakoudi, Ifigeneia",
    title = "{A panorama of viable F(R) gravity dark energy models}",
    eprint = "2205.08599",
    archivePrefix = "arXiv",
    primaryClass = "gr-qc",
    doi = "10.1142/S0218271822500754",
    journal = "Int. J. Mod. Phys. D",
    volume = "31",
    number = "09",
    pages = "2250075",
    year = "2022"
}

@article{Trivedi:2024inb,
    author = "Trivedi, Oem and Bidlan, Ayush and Moniz, Paulo",
    title = "{Fractional holographic dark energy}",
    eprint = "2407.16685",
    archivePrefix = "arXiv",
    primaryClass = "gr-qc",
    reportNumber = "PLB{\_}139074",
    doi = "10.1016/j.physletb.2024.139074",
    journal = "Phys. Lett. B",
    volume = "858",
    pages = "139074",
    year = "2024"
}

@article{Lin:2025gne,
    author = "Lin, Weikang and Visinelli, Luca and Yanagida, Tsutomu T.",
    title = "{Testing Quintessence Axion Dark Energy with Recent Cosmological Results}",
    eprint = "2504.17638",
    archivePrefix = "arXiv",
    primaryClass = "astro-ph.CO",
    month = "4",
    year = "2025"
}

@article{Aspect:1982fx,
    author = "Aspect, Alain and Dalibard, Jean and Roger, Gerard",
    title = "{Experimental test of Bell's inequalities using time varying analyzers}",
    doi = "10.1103/PhysRevLett.49.1804",
    journal = "Phys. Rev. Lett.",
    volume = "49",
    pages = "1804--1807",
    year = "1982"
}

@article{Abel:1992kz,
    author = "Abel, S. A. and Dittmar, M. and Dreiner, Herbert K.",
    title = "{Testing locality at colliders via Bell's inequality?}",
    reportNumber = "OUTP-91-40-P",
    doi = "10.1016/0370-2693(92)90071-B",
    journal = "Phys. Lett. B",
    volume = "280",
    pages = "304--312",
    year = "1992"
}

@article{Ancochea:1998nx,
    author = "Ancochea, B. and Bramon, A. and Nowakowski, M.",
    title = "{Bell inequalities for K0 anti-K0 pairs from phi resonance decays}",
    eprint = "hep-ph/9811404",
    archivePrefix = "arXiv",
    reportNumber = "UAB-FT-459-98, IFT-P-087-98",
    doi = "10.1103/PhysRevD.60.094008",
    journal = "Phys. Rev. D",
    volume = "60",
    pages = "094008",
    year = "1999"
}

@article{Bramon:1998nz,
    author = "Bramon, Albert and Nowakowski, Marek",
    title = "{Bell inequalities for entangled pairs of neutral kaons}",
    eprint = "hep-ph/9811406",
    archivePrefix = "arXiv",
    reportNumber = "UAB-FT-458-98, IFT-P-086-98",
    doi = "10.1103/PhysRevLett.83.1",
    journal = "Phys. Rev. Lett.",
    volume = "83",
    pages = "1--5",
    year = "1999"
}

@article{Hiesmayr:2000rm,
    author = "Hiesmayr, Beatrix C.",
    title = "{A Generalized Bell Inequality and Decoherence for the K$^{0}$K$^{O}$}",
    eprint = "hep-ph/0010108",
    archivePrefix = "arXiv",
    reportNumber = "UWTHPH-2000-41",
    doi = "10.1023/A:1013457210230",
    journal = "Found. Phys. Lett.",
    volume = "14",
    number = "3",
    pages = "231--245",
    year = "2001"
}

@article{Bertlmann:2001sk,
    author = "Bertlmann, Reinhold A. and Hiesmayr, Beatrix C.",
    title = "{Bell inequalities for entangled kaons and their unitary time evolution}",
    eprint = "hep-ph/0101356",
    archivePrefix = "arXiv",
    reportNumber = "UWTHPH-2001-3",
    doi = "10.1103/PhysRevA.63.062112",
    journal = "Phys. Rev. A",
    volume = "63",
    pages = "062112",
    year = "2001"
}

@article{Genovese:2001pk,
    author = "Genovese, M. and Novero, C. and Predazzi, E.",
    title = "{Can experimental tests of Bell inequalities performed with pseudoscalar mesons be definitive?}",
    eprint = "hep-ph/0103298",
    archivePrefix = "arXiv",
    doi = "10.1016/S0370-2693(01)00580-9",
    journal = "Phys. Lett. B",
    volume = "513",
    pages = "401--405",
    year = "2001"
}

@article{Gallicchio:2013iva,
    author = "Gallicchio, Jason and Friedman, Andrew S. and Kaiser, David I.",
    title = "{Testing Bell{\textquoteright}s Inequality with Cosmic Photons: Closing the Setting-Independence Loophole}",
    eprint = "1310.3288",
    archivePrefix = "arXiv",
    primaryClass = "quant-ph",
    doi = "10.1103/PhysRevLett.112.110405",
    journal = "Phys. Rev. Lett.",
    volume = "112",
    number = "11",
    pages = "110405",
    year = "2014"
}

@article{Maldacena:2015bha,
    author = "Maldacena, Juan",
    title = "{A model with cosmological Bell inequalities}",
    eprint = "1508.01082",
    archivePrefix = "arXiv",
    primaryClass = "hep-th",
    doi = "10.1002/prop.201500097",
    journal = "Fortsch. Phys.",
    volume = "64",
    pages = "10--23",
    year = "2016"
}

@article{Choudhury:2016cso,
    author = "Choudhury, Sayantan and Panda, Sudhakar and Singh, Rajeev",
    title = "{Bell violation in the Sky}",
    eprint = "1607.00237",
    archivePrefix = "arXiv",
    primaryClass = "hep-th",
    reportNumber = "TIFR-TH-16-19",
    doi = "10.1140/epjc/s10052-016-4553-3",
    journal = "Eur. Phys. J. C",
    volume = "77",
    number = "2",
    pages = "60",
    year = "2017"
}

@article{Chen:2016xqz,
    author = "Chen, Jiunn-Wei and Sun, Sichun and Zhang, Yun-Long",
    title = "{Bell inequality in the holographic EPR pair}",
    eprint = "1612.09513",
    archivePrefix = "arXiv",
    primaryClass = "hep-th",
    doi = "10.1016/j.physletb.2019.02.012",
    journal = "Phys. Lett. B",
    volume = "791",
    pages = "73--79",
    year = "2019"
}

@article{Choudhury:2016pfr,
    author = "Choudhury, Sayantan and Panda, Sudhakar and Singh, Rajeev",
    editor = {D{\k{a}}browski, Mariusz P. and Kr{\"a}mer, Manuel and Salzano, Vincenzo},
    title = "{Bell violation in primordial cosmology}",
    eprint = "1612.09445",
    archivePrefix = "arXiv",
    primaryClass = "hep-th",
    reportNumber = "TIFR-TH-16-49",
    doi = "10.3390/universe3010013",
    journal = "Universe",
    volume = "3",
    number = "1",
    pages = "13",
    year = "2017"
}

@article{Kanno:2017dci,
    author = "Kanno, Sugumi and Soda, Jiro",
    title = "{Infinite violation of Bell inequalities in inflation}",
    eprint = "1705.06199",
    archivePrefix = "arXiv",
    primaryClass = "hep-th",
    reportNumber = "KOBE-COSMO-17-08",
    doi = "10.1103/PhysRevD.96.083501",
    journal = "Phys. Rev. D",
    volume = "96",
    number = "8",
    pages = "083501",
    year = "2017"
}

@article{Martin:2017zxs,
    author = "Martin, Jerome and Vennin, Vincent",
    title = "{Obstructions to Bell CMB Experiments}",
    eprint = "1706.05001",
    archivePrefix = "arXiv",
    primaryClass = "astro-ph.CO",
    doi = "10.1103/PhysRevD.96.063501",
    journal = "Phys. Rev. D",
    volume = "96",
    number = "6",
    pages = "063501",
    year = "2017"
}

@article{Choudhury:2017bou,
    author = "Choudhury, Sayantan and Panda, Sudhakar",
    title = "{Entangled de Sitter from stringy axionic Bell pair I: an analysis using Bunch{\textendash}Davies vacuum}",
    eprint = "1708.02265",
    archivePrefix = "arXiv",
    primaryClass = "hep-th",
    reportNumber = "TIFR-TH-17-01",
    doi = "10.1140/epjc/s10052-017-5503-4",
    journal = "Eur. Phys. J. C",
    volume = "78",
    number = "1",
    pages = "52",
    year = "2018"
}

@article{Feng:2018ebt,
    author = "Feng, Jun and Huang, Xiaoyang and Zhang, Yao-Zhong and Fan, Heng",
    title = "{Bell inequalities violation within non-Bunch{\textendash}Davies states}",
    eprint = "1806.08923",
    archivePrefix = "arXiv",
    primaryClass = "hep-th",
    doi = "10.1016/j.physletb.2018.10.020",
    journal = "Phys. Lett. B",
    volume = "786",
    pages = "403--409",
    year = "2018"
}

@article{Ali:2021jch,
    author = "Ali, Md Sabir and Bhattacharya, Sourav and Chakrabortty, Shankhadeep and Kaushal, Shagun",
    title = "{Fermionic Bell violation in the presence of background electromagnetic fields in the cosmological de Sitter spacetime}",
    eprint = "2102.11745",
    archivePrefix = "arXiv",
    primaryClass = "hep-th",
    doi = "10.1103/PhysRevD.104.125012",
    journal = "Phys. Rev. D",
    volume = "104",
    number = "12",
    pages = "125012",
    year = "2021"
}

@article{Fabbrichesi:2021npl,
    author = "Fabbrichesi, M. and Floreanini, R. and Panizzo, G.",
    title = "{Testing Bell Inequalities at the LHC with Top-Quark Pairs}",
    eprint = "2102.11883",
    archivePrefix = "arXiv",
    primaryClass = "hep-ph",
    doi = "10.1103/PhysRevLett.127.161801",
    journal = "Phys. Rev. Lett.",
    volume = "127",
    number = "16",
    pages = "161801",
    year = "2021"
}

@article{Choudhury:2021mht,
    author = "Choudhury, Sayantan",
    title = "{Cosmological Geometric Phase From Pure Quantum States: A Study without/with having Bell's Inequality Violation}",
    eprint = "2105.06254",
    archivePrefix = "arXiv",
    primaryClass = "gr-qc",
    doi = "10.1002/prop.202100144",
    journal = "Fortsch. Phys.",
    volume = "70",
    number = "9-10",
    pages = "2100144",
    year = "2022"
}

@article{Barr:2021zcp,
    author = "Barr, Alan J.",
    title = "{Testing Bell inequalities in Higgs boson decays}",
    eprint = "2106.01377",
    archivePrefix = "arXiv",
    primaryClass = "hep-ph",
    doi = "10.1016/j.physletb.2021.136866",
    journal = "Phys. Lett. B",
    volume = "825",
    pages = "136866",
    year = "2022"
}

@article{Takubo:2021sdk,
    author = "Takubo, Yosuke and Ichikawa, Tsubasa and Higashino, Satoshi and Mori, Yuichiro and Nagano, Kunihiro and Tsutsui, Izumi",
    title = "{Feasibility of Bell inequality violation at the ATLAS experiment with flavor entanglement of B0B{\textasciimacron}0 pairs from pp collisions}",
    eprint = "2106.07399",
    archivePrefix = "arXiv",
    primaryClass = "hep-ph",
    reportNumber = "KEK-TH-2330",
    doi = "10.1103/PhysRevD.104.056004",
    journal = "Phys. Rev. D",
    volume = "104",
    number = "5",
    pages = "056004",
    year = "2021"
}

@article{Gong:2021bcp,
    author = "Gong, Wenjie and Parida, Ganesh and Tu, Zhoudunming and Venugopalan, Raju",
    title = "{Measurement of Bell-type inequalities and quantum entanglement from {\ensuremath{\Lambda}}-hyperon spin correlations at high energy colliders}",
    eprint = "2107.13007",
    archivePrefix = "arXiv",
    primaryClass = "hep-ph",
    doi = "10.1103/PhysRevD.106.L031501",
    journal = "Phys. Rev. D",
    volume = "106",
    number = "3",
    pages = "L031501",
    year = "2022"
}

@article{Severi:2021cnj,
    author = "Severi, Claudio and Boschi, Cristian Degli Esposti and Maltoni, Fabio and Sioli, Maximiliano",
    title = "{Quantum tops at the LHC: from entanglement to Bell inequalities}",
    eprint = "2110.10112",
    archivePrefix = "arXiv",
    primaryClass = "hep-ph",
    doi = "10.1140/epjc/s10052-022-10245-9",
    journal = "Eur. Phys. J. C",
    volume = "82",
    number = "4",
    pages = "285",
    year = "2022"
}

@article{Espinosa-Portales:2022yok,
    author = "Espinosa-Portal{\'e}s, Lloren{\c{c}} and Vennin, Vincent",
    title = "{Real-space Bell inequalities in de~Sitter}",
    eprint = "2203.03505",
    archivePrefix = "arXiv",
    primaryClass = "quant-ph",
    reportNumber = "IFT-UAM/CSIC-22-17",
    doi = "10.1088/1475-7516/2022/07/037",
    journal = "JCAP",
    volume = "07",
    number = "07",
    pages = "037",
    year = "2022"
}

@article{Aguilar-Saavedra:2022uye,
    author = "Aguilar-Saavedra, J. A. and Casas, J. A.",
    title = "{Improved tests of entanglement and Bell inequalities with LHC tops}",
    eprint = "2205.00542",
    archivePrefix = "arXiv",
    primaryClass = "hep-ph",
    reportNumber = "IFT-UAM/CSIC-22-45",
    doi = "10.1140/epjc/s10052-022-10630-4",
    journal = "Eur. Phys. J. C",
    volume = "82",
    number = "8",
    pages = "666",
    year = "2022"
}

@article{Aguilar-Saavedra:2022wam,
    author = "Aguilar-Saavedra, J. A. and Bernal, A. and Casas, J. A. and Moreno, J. M.",
    title = "{Testing entanglement and Bell inequalities in H{\textrightarrow}ZZ}",
    eprint = "2209.13441",
    archivePrefix = "arXiv",
    primaryClass = "hep-ph",
    doi = "10.1103/PhysRevD.107.016012",
    journal = "Phys. Rev. D",
    volume = "107",
    number = "1",
    pages = "016012",
    year = "2023"
}

@article{Sinha:2022crx,
    author = "Sinha, Aninda and Zahed, Ahmadullah",
    title = "{Bell inequalities in 2-2 scattering}",
    eprint = "2212.10213",
    archivePrefix = "arXiv",
    primaryClass = "hep-th",
    doi = "10.1103/PhysRevD.108.025015",
    journal = "Phys. Rev. D",
    volume = "108",
    number = "2",
    pages = "025015",
    year = "2023"
}

@article{Peruzzo:2022tog,
    author = "Peruzzo, Giovani and Sorella, Silvio Paolo",
    title = "{Feynman path integral formulation of the Bell-Clauser-Horne-Shimony-Holt inequality in quantum field theory}",
    eprint = "2212.14407",
    archivePrefix = "arXiv",
    primaryClass = "hep-th",
    doi = "10.1103/PhysRevD.107.105001",
    journal = "Phys. Rev. D",
    volume = "107",
    number = "10",
    pages = "105001",
    year = "2023"
}

@article{Choudhury:2022mch,
    author = "Choudhury, Sayantan",
    title = "{Entanglement Negativity in de Sitter Biverse from Stringy Axionic Bell Pair: An Analysis Using Bunch-Davies Vacuum}",
    eprint = "2301.05203",
    archivePrefix = "arXiv",
    primaryClass = "hep-th",
    doi = "10.1002/prop.202300063",
    journal = "Fortsch. Phys.",
    volume = "72",
    number = "1",
    pages = "2300063",
    year = "2024"
}

@article{Fabbrichesi:2023cev,
    author = "Fabbrichesi, Marco and Floreanini, Roberto and Gabrielli, Emidio and Marzola, Luca",
    title = "{Bell inequalities and quantum entanglement in weak gauge boson production at the LHC and future colliders}",
    eprint = "2302.00683",
    archivePrefix = "arXiv",
    primaryClass = "hep-ph",
    doi = "10.1140/epjc/s10052-023-11935-8",
    journal = "Eur. Phys. J. C",
    volume = "83",
    number = "9",
    pages = "823",
    year = "2023"
}

@article{Dale:2023fnp,
    author = "Dale, Roberto and Lapiedra, Ramon and Morales-Lladosa, Juan Antonio",
    title = "{Cosmic primordial density fluctuations and Bell inequalities}",
    eprint = "2302.05125",
    archivePrefix = "arXiv",
    primaryClass = "gr-qc",
    doi = "10.1103/PhysRevD.107.023506",
    journal = "Phys. Rev. D",
    volume = "107",
    number = "2",
    pages = "023506",
    year = "2023"
}

@article{Ghosh:2023rpj,
    author = "Ghosh, Diptimoy and Sharma, Rajat",
    title = "{Bell violation in 2 {\textrightarrow} 2 scattering in photon, gluon and graviton EFTs}",
    eprint = "2303.03375",
    archivePrefix = "arXiv",
    primaryClass = "hep-th",
    doi = "10.1007/JHEP08(2023)146",
    journal = "JHEP",
    volume = "08",
    pages = "146",
    year = "2023"
}

@article{Dudal:2023pbc,
    author = "Dudal, David and De Fabritiis, Philipe and Guimaraes, Marcelo Santos and Peruzzo, Giovani and Sorella, Silvio Paolo",
    title = "{BRST invariant formulation of the Bell-CHSH inequality in gauge field theories}",
    eprint = "2304.01028",
    archivePrefix = "arXiv",
    primaryClass = "hep-th",
    doi = "10.21468/SciPostPhys.15.5.201",
    journal = "SciPost Phys.",
    volume = "15",
    number = "5",
    pages = "201",
    year = "2023"
}

@article{Petruzziello:2023xhb,
    author = "Petruzziello, Luciano and Illuminati, Fabrizio",
    title = "{Gravitationally modulated quantum correlations: Discriminating classical and quantum models of ultra-compact objects with Bell nonlocality}",
    eprint = "2304.10868",
    archivePrefix = "arXiv",
    primaryClass = "gr-qc",
    doi = "10.1016/j.physletb.2024.138652",
    journal = "Phys. Lett. B",
    volume = "853",
    pages = "138652",
    year = "2024"
}

@article{Fabbrichesi:2023idl,
    author = "Fabbrichesi, M. and Floreanini, R. and Gabrielli, E. and Marzola, L.",
    title = "{Bell inequality is violated in B0{\textrightarrow}J/{\ensuremath{\psi}}K*(892)0 decays}",
    eprint = "2305.04982",
    archivePrefix = "arXiv",
    primaryClass = "hep-ph",
    doi = "10.1103/PhysRevD.109.L031104",
    journal = "Phys. Rev. D",
    volume = "109",
    number = "3",
    pages = "L031104",
    year = "2024"
}

@article{Dong:2023xiw,
    author = "Dong, Zhongtian and Gon{\c{c}}alves, Dorival and Kong, Kyoungchul and Navarro, Alberto",
    title = "{Entanglement and Bell inequalities with boosted tt{\textasciimacron}}",
    eprint = "2305.07075",
    archivePrefix = "arXiv",
    primaryClass = "hep-ph",
    doi = "10.1103/PhysRevD.109.115023",
    journal = "Phys. Rev. D",
    volume = "109",
    number = "11",
    pages = "115023",
    year = "2024"
}

@article{Morales:2023gow,
    author = "Morales, R. A.",
    title = "{Exploring Bell inequalities and quantum entanglement in vector boson scattering}",
    eprint = "2306.17247",
    archivePrefix = "arXiv",
    primaryClass = "hep-ph",
    doi = "10.1140/epjp/s13360-023-04784-7",
    journal = "Eur. Phys. J. Plus",
    volume = "138",
    number = "12",
    pages = "1157",
    year = "2023"
}

@article{Dudal:2023mij,
    author = "Dudal, David and De Fabritiis, Philipe and Guimaraes, Marcelo S. and Roditi, Itzhak and Sorella, Silvio P.",
    title = "{Maximal violation of the Bell-Clauser-Horne-Shimony-Holt inequality via bumpified Haar wavelets}",
    eprint = "2307.04611",
    archivePrefix = "arXiv",
    primaryClass = "hep-th",
    doi = "10.1103/PhysRevD.108.L081701",
    journal = "Phys. Rev. D",
    volume = "108",
    pages = "L081701",
    year = "2023"
}

@article{Moradpour:2023gsn,
    author = "Moradpour, Hooman and Jalalzadeh, Shahram and Tebyanian, Hamid",
    title = "{The shadows of quantum gravity on Bell{\textquoteright}s inequality}",
    eprint = "2307.13006",
    archivePrefix = "arXiv",
    primaryClass = "quant-ph",
    reportNumber = "2550004",
    doi = "10.1142/S021773232550004X",
    journal = "Mod. Phys. Lett. A",
    volume = "40",
    number = "07n08",
    pages = "2550004",
    year = "2025"
}

@article{Fabbri:2023ncz,
    author = "Fabbri, Federica and Howarth, James and Maurin, Theo",
    title = "{Isolating semi-leptonic $H\rightarrow WW^{*}$decays for Bell inequality tests}",
    eprint = "2307.13783",
    archivePrefix = "arXiv",
    primaryClass = "hep-ph",
    doi = "10.1140/epjc/s10052-023-12371-4",
    journal = "Eur. Phys. J. C",
    volume = "84",
    number = "1",
    pages = "20",
    year = "2024"
}

@article{Bernal:2023ruk,
    author = "Bernal, Alexander and Caban, Pawe{\l} and Rembieli{\'n}ski, Jakub",
    title = "{Entanglement and Bell inequalities violation in $H\rightarrow ZZ$ with anomalous coupling}",
    eprint = "2307.13496",
    archivePrefix = "arXiv",
    primaryClass = "hep-ph",
    doi = "10.1140/epjc/s10052-023-12216-0",
    journal = "Eur. Phys. J. C",
    volume = "83",
    number = "11",
    pages = "1050",
    year = "2023"
}

@article{Bi:2023uop,
    author = "Bi, Qi and Cao, Qing-Hong and Cheng, Kun and Zhang, Hao",
    title = "{New observables for testing Bell inequalities in W boson pair production}",
    eprint = "2307.14895",
    archivePrefix = "arXiv",
    primaryClass = "hep-ph",
    doi = "10.1103/PhysRevD.109.036022",
    journal = "Phys. Rev. D",
    volume = "109",
    number = "3",
    pages = "036022",
    year = "2024"
}

@article{DeFabritiis:2023tkh,
    author = "De Fabritiis, P. and Guedes, F. M. and Guimaraes, M. S. and Peruzzo, G. and Roditi, I. and Sorella, S. P.",
    title = "{Weyl operators, Tomita-Takesaki theory, and Bell-Clauser-Horne-Shimony-Holt inequality violations}",
    eprint = "2309.02941",
    archivePrefix = "arXiv",
    primaryClass = "hep-th",
    doi = "10.1103/PhysRevD.108.085026",
    journal = "Phys. Rev. D",
    volume = "108",
    number = "8",
    pages = "085026",
    year = "2023"
}

@article{Ma:2023yvd,
    author = "Ma, Kai and Li, Tong",
    title = "{Testing Bell inequality through ${\boldsymbol h{\bf\rightarrow}\boldsymbol\tau\boldsymbol\tau }$ at CEPC*}",
    eprint = "2309.08103",
    archivePrefix = "arXiv",
    primaryClass = "hep-ph",
    doi = "10.1088/1674-1137/ad62d8",
    journal = "Chin. Phys. C",
    volume = "48",
    number = "10",
    pages = "103105",
    year = "2024"
}

@article{Tselentis:2023kwh,
    author = {Tselentis, Eleftherios-Ermis and Baumeler, {\"A}min},
    title = {{M{\"o}bius games and other Bell tests for relativity}},
    eprint = "2309.15752",
    archivePrefix = "arXiv",
    primaryClass = "gr-qc",
    doi = "10.1103/PhysRevA.111.052211",
    journal = "Phys. Rev. A",
    volume = "111",
    number = "5",
    pages = "052211",
    year = "2025"
}

@article{Han:2023fci,
    author = "Han, Tao and Low, Matthew and Wu, Tong Arthur",
    title = "{Quantum entanglement and Bell inequality violation in semi-leptonic top decays}",
    eprint = "2310.17696",
    archivePrefix = "arXiv",
    primaryClass = "hep-ph",
    reportNumber = "PITT-PACC-2316",
    doi = "10.1007/JHEP07(2024)192",
    journal = "JHEP",
    volume = "07",
    pages = "192",
    year = "2024"
}

@article{Cheng:2023qmz,
    author = "Cheng, Kun and Han, Tao and Low, Matthew",
    title = "{Optimizing fictitious states for Bell inequality violation in bipartite qubit systems with applications to the tt{\textasciimacron} system}",
    eprint = "2311.09166",
    archivePrefix = "arXiv",
    primaryClass = "hep-ph",
    reportNumber = "PITT-PACC-2321",
    doi = "10.1103/PhysRevD.109.116005",
    journal = "Phys. Rev. D",
    volume = "109",
    number = "11",
    pages = "116005",
    year = "2024"
}

@article{Ehataht:2023zzt,
    author = {Ehat{\"a}ht, Karl and Fabbrichesi, Marco and Marzola, Luca and Veelken, Christian},
    title = "{Probing entanglement and testing Bell inequality violation with e+e-{\textrightarrow}{\ensuremath{\tau}}+{\ensuremath{\tau}}- at Belle II}",
    eprint = "2311.17555",
    archivePrefix = "arXiv",
    primaryClass = "hep-ph",
    doi = "10.1103/PhysRevD.109.032005",
    journal = "Phys. Rev. D",
    volume = "109",
    number = "3",
    pages = "032005",
    year = "2024"
}

@article{Li:2024luk,
    author = "Li, Song and Shen, Wei and Yang, Jin Min",
    title = "{Can Bell inequalities be tested via scattering cross-section at colliders ?}",
    eprint = "2401.01162",
    archivePrefix = "arXiv",
    primaryClass = "hep-th",
    doi = "10.1140/epjc/s10052-024-13584-x",
    journal = "Eur. Phys. J. C",
    volume = "84",
    number = "11",
    pages = "1195",
    year = "2024"
}

@article{Guedes:2024tcq,
    author = "Guedes, Fillipe M. and Guimaraes, Marcelo S. and Roditi, Itzhak and Sorella, Silvio P.",
    title = "{Unruh-De Witt detectors, Bell-CHSH inequality and Tomita-Takesaki theory}",
    eprint = "2401.03313",
    archivePrefix = "arXiv",
    primaryClass = "hep-th",
    doi = "10.1007/JHEP06(2024)031",
    journal = "JHEP",
    volume = "06",
    pages = "031",
    year = "2024"
}

@article{Barr:2024djo,
    author = "Barr, Alan J. and Fabbrichesi, Marco and Floreanini, Roberto and Gabrielli, Emidio and Marzola, Luca",
    title = "{Quantum entanglement and Bell inequality violation at colliders}",
    eprint = "2402.07972",
    archivePrefix = "arXiv",
    primaryClass = "hep-ph",
    doi = "10.1016/j.ppnp.2024.104134",
    journal = "Prog. Part. Nucl. Phys.",
    volume = "139",
    pages = "104134",
    year = "2024"
}

@article{Morales:2024jhj,
    author = "Morales, R. A.",
    title = "{Tripartite entanglement and Bell non-locality in loop-induced Higgs boson decays}",
    eprint = "2403.18023",
    archivePrefix = "arXiv",
    primaryClass = "hep-ph",
    doi = "10.1140/epjc/s10052-024-12921-4",
    journal = "Eur. Phys. J. C",
    volume = "84",
    number = "6",
    pages = "581",
    year = "2024"
}

@article{Sou:2024tjv,
    author = "Sou, Chon Man and Wang, Junqi and Wang, Yi",
    title = "{Cosmological Bell tests with decoherence effects}",
    eprint = "2405.07141",
    archivePrefix = "arXiv",
    primaryClass = "hep-th",
    doi = "10.1088/1475-7516/2024/10/084",
    journal = "JCAP",
    volume = "10",
    pages = "084",
    year = "2024"
}

@article{Wu:2024qhd,
    author = "Wu, Shu-Min and Wang, Rui-Di and Huang, Xiao-Li and Wang, Zejun",
    title = "{Does gravitational wave assist vacuum steering and Bell nonlocality?}",
    eprint = "2405.07235",
    archivePrefix = "arXiv",
    primaryClass = "gr-qc",
    doi = "10.1007/JHEP07(2024)155",
    journal = "JHEP",
    volume = "07",
    pages = "155",
    year = "2024"
}

@article{Fabbrichesi:2024wcd,
    author = "Fabbrichesi, M. and Marzola, L.",
    title = "{Quantum tomography with {\ensuremath{\tau}} leptons at the FCC-ee: Entanglement, Bell inequality violation, sin{\ensuremath{\theta}}W, and anomalous couplings}",
    eprint = "2405.09201",
    archivePrefix = "arXiv",
    primaryClass = "hep-ph",
    doi = "10.1103/PhysRevD.110.076004",
    journal = "Phys. Rev. D",
    volume = "110",
    number = "7",
    pages = "076004",
    year = "2024"
}

@article{Bernal:2024xhm,
    author = "Bernal, Alexander and Caban, Pawe{\l} and Rembieli{\'n}ski, Jakub",
    title = "{Entanglement and Bell inequality violation in vector diboson systems produced in decays of spin-0 particles}",
    eprint = "2405.16525",
    archivePrefix = "arXiv",
    primaryClass = "hep-ph",
    doi = "10.1038/s41598-025-07747-3",
    journal = "Sci. Rep.",
    volume = "15",
    number = "1",
    pages = "23410",
    year = "2025"
}

@article{DeFabritiis:2024jfy,
    author = "De Fabritiis, Philipe and Guimaraes, Marcelo S. and Roditi, Itzhak and Sorella, Silvio P.",
    title = "{Numerical approach to the Bell-Clauser-Horne-Shimony-Holt inequality in quantum field theory}",
    eprint = "2406.20033",
    archivePrefix = "arXiv",
    primaryClass = "hep-th",
    doi = "10.1103/PhysRevD.110.065006",
    journal = "Phys. Rev. D",
    volume = "110",
    number = "6",
    pages = "065006",
    year = "2024"
}

@article{Guo:2024jch,
    author = "Guo, Yuxun and Liu, Xiaohui and Yuan, Feng and Zhu, Hua Xing",
    title = "{Long-Range Azimuthal Correlation, Entanglement, and Bell Inequality Violation by Spinning Gluons at the Large Hadron Collider}",
    eprint = "2406.05880",
    archivePrefix = "arXiv",
    primaryClass = "hep-ph",
    doi = "10.34133/research.0552",
    journal = "Research",
    volume = "2025",
    pages = "0552",
    year = "2025"
}

@article{Wu:2024asu,
    author = "Wu, Sihao and Qian, Chen and Wang, Qun and Zhou, Xiao-Rong",
    title = "{Bell nonlocality and entanglement in e+e-{\textrightarrow}YY{\textasciimacron} at BESIII}",
    eprint = "2406.16298",
    archivePrefix = "arXiv",
    primaryClass = "hep-ph",
    doi = "10.1103/PhysRevD.110.054012",
    journal = "Phys. Rev. D",
    volume = "110",
    number = "5",
    pages = "054012",
    year = "2024"
}

@article{Fabbrichesi:2024rec,
    author = "Fabbrichesi, M. and Floreanini, R. and Gabrielli, E. and Marzola, L.",
    title = "{Bell inequality is violated in charmonium decays}",
    eprint = "2406.17772",
    archivePrefix = "arXiv",
    primaryClass = "hep-ph",
    doi = "10.1103/PhysRevD.110.053008",
    journal = "Phys. Rev. D",
    volume = "110",
    number = "5",
    pages = "053008",
    year = "2024"
}

@article{Cheng:2024btk,
    author = "Cheng, Kun and Han, Tao and Low, Matthew",
    title = "{Optimizing entanglement and Bell inequality violation in top antitop events}",
    eprint = "2407.01672",
    archivePrefix = "arXiv",
    primaryClass = "hep-ph",
    reportNumber = "PITT-PACC-2401",
    doi = "10.1103/PhysRevD.111.033004",
    journal = "Phys. Rev. D",
    volume = "111",
    number = "3",
    pages = "033004",
    year = "2025"
}

@article{Chen:2024drt,
    author = "Chen, Kaiwen and Xing, Zhi-Peng and Zhu, Ruilin",
    title = "{Test of bell locality violation in flavor entangled neutral meson pair}",
    eprint = "2407.19242",
    archivePrefix = "arXiv",
    primaryClass = "hep-ph",
    doi = "10.1016/j.nuclphysa.2025.123199",
    journal = "Nucl. Phys. A",
    volume = "1063",
    pages = "123199",
    year = "2025"
}

@article{Gabrielli:2024kbz,
    author = "Gabrielli, Emidio and Marzola, Luca",
    title = "{Entanglement and Bell Inequality Violation in B {\textrightarrow} {\ensuremath{\phi}}{\ensuremath{\phi}} Decays}",
    eprint = "2408.05010",
    archivePrefix = "arXiv",
    primaryClass = "hep-ph",
    doi = "10.3390/sym16081036",
    journal = "Symmetry",
    volume = "16",
    number = "8",
    pages = "1036",
    year = "2024"
}

@article{Ruzi:2024cbt,
    author = "Ruzi, Alim and Wu, Youpeng and Ding, Ran and Qian, Sitian and Levin, Andrew Micheal and Li, Qiang",
    title = "{Testing Bell inequalities and probing quantum entanglement at a muon collider}",
    eprint = "2408.05429",
    archivePrefix = "arXiv",
    primaryClass = "hep-ph",
    doi = "10.1007/JHEP10(2024)211",
    journal = "JHEP",
    volume = "10",
    pages = "211",
    year = "2024"
}

@article{Du:2024sly,
    author = "Du, Yong and He, Xiao-Gang and Liu, Chia-Wei and Ma, Jian-Ping",
    title = "{Impact of parity violation on quantum entanglement and Bell nonlocality}",
    eprint = "2409.15418",
    archivePrefix = "arXiv",
    primaryClass = "hep-ph",
    month = "9",
    year = "2024"
}

@article{Wu:2024ovc,
    author = "Wu, Youpeng and Jiang, Ruobing and Ruzi, Alim and Ban, Yong and Yan, Xueqing and Li, Qiang",
    title = "{Testing Bell inequalities and probing quantum entanglement at CEPC}",
    eprint = "2410.17025",
    archivePrefix = "arXiv",
    primaryClass = "hep-ph",
    doi = "10.1103/PhysRevD.111.036008",
    journal = "Phys. Rev. D",
    volume = "111",
    number = "3",
    pages = "036008",
    year = "2025"
}

@article{Grabarczyk:2024wnk,
    author = "Grabarczyk, Rados{\l}aw",
    title = "{An improved Bell-CHSH observable for gauge boson pairs}",
    eprint = "2410.18022",
    archivePrefix = "arXiv",
    primaryClass = "hep-ph",
    month = "10",
    year = "2024"
}

@article{Cheng:2025cuv,
    author = "Cheng, Kun and Yan, Bin",
    title = "{Bell Inequality Violation of Light Quarks in Dihadron Pair Production at Lepton Colliders}",
    eprint = "2501.03321",
    archivePrefix = "arXiv",
    primaryClass = "hep-ph",
    reportNumber = "PITT-PACC-2414",
    doi = "10.1103/gmqz-v4cl",
    journal = "Phys. Rev. Lett.",
    volume = "135",
    number = "1",
    pages = "011902",
    year = "2025"
}

@article{Han:2025ewp,
    author = "Han, Tao and Low, Matthew and Su, Youle",
    title = "{Entanglement and Bell Nonlocality in $\tau^+ \tau^-$ at the BEPC}",
    eprint = "2501.04801",
    archivePrefix = "arXiv",
    primaryClass = "hep-ph",
    reportNumber = "PITT-PACC-2412",
    month = "1",
    year = "2025"
}

@article{Varma:2025gkp,
    author = "Varma, Mira and Baker, Oliver K.",
    title = "{Bell Inequalities and Quantum Correlations in $H \rightarrow ZZ \rightarrow 2e\,2\mu$}",
    eprint = "2502.03219",
    archivePrefix = "arXiv",
    primaryClass = "hep-ph",
    month = "2",
    year = "2025"
}

@article{Dale:2025nhc,
    author = "Dale, Roberto and Gand{\'\i}a, Jes{\'u}s M. and Morales-Lladosa, Juan Antonio and Lapiedra, Ramon",
    title = "{Violation of Bell inequalities from Cosmic Microwave Background data}",
    eprint = "2502.13846",
    archivePrefix = "arXiv",
    primaryClass = "gr-qc",
    doi = "10.1088/1475-7516/2025/07/044",
    journal = "JCAP",
    volume = "07",
    pages = "044",
    year = "2025"
}

@article{Zhang:2025mmm,
    author = "Zhang, Yulei and Zhou, Bai-Hong and Liu, Qi-Bin and Li, Shu and Hsu, Shih-Chieh and Han, Tao and Low, Matthew and Wu, Tong Arthur",
    title = "{Entanglement and Bell Nonlocality in $\tau^+ \tau^-$ at the LHC using Machine Learning for Neutrino Reconstruction}",
    eprint = "2504.01496",
    archivePrefix = "arXiv",
    primaryClass = "hep-ph",
    month = "4",
    year = "2025"
}

@article{Afik:2025grr,
    author = "Afik, Yoav and Kats, Yevgeny and de Nova, Juan Ram{\'o}n Mu{\~n}oz and Soffer, Abner and Uzan, David",
    title = "{Entanglement and Bell nonlocality with bottom-quark pairs at hadron colliders}",
    eprint = "2406.04402",
    archivePrefix = "arXiv",
    primaryClass = "hep-ph",
    doi = "10.1103/fhkc-kfhr",
    journal = "Phys. Rev. D",
    volume = "111",
    number = "11",
    pages = "L111902",
    year = "2025"
}

@article{Qi:2025onf,
    author = "Qi, Wei and Guo, Zijing and Xiao, Bo-Wen",
    title = "{Studying Maximal Entanglement and Bell Nonlocality at an Electron-Ion Collider}",
    eprint = "2506.12889",
    archivePrefix = "arXiv",
    primaryClass = "hep-ph",
    month = "6",
    year = "2025"
}

@article{Lin:2025eci,
    author = "Lin, Shi-Jia and Liu, Ming-Jun and Shao, Ding Yu and Wei, Shu-Yi",
    title = "{Spin correlations and Bell nonlocality in $Λ\barΛ$ pair production from $e^+e^-$ collisions with a thrust cut}",
    eprint = "2507.15387",
    archivePrefix = "arXiv",
    primaryClass = "hep-ph",
    month = "7",
    year = "2025"
}

@article{Abel:2025skj,
    author = "Abel, Steven A. and Dreiner, Herbi K. and Sengupta, Rhitaja and Ubaldi, Lorenzo",
    title = "{Colliders are Testing neither Locality via Bell's Inequality nor Entanglement versus Non-Entanglement}",
    eprint = "2507.15949",
    archivePrefix = "arXiv",
    primaryClass = "hep-ph",
    reportNumber = "BONN-TH-2025-22, IPPP/25/48",
    month = "7",
    year = "2025"
}

@book{Fujii:2003pa,
    author = "Fujii, Y. and Maeda, K.",
    title = "{The scalar-tensor theory of gravitation}",
    doi = "10.1017/CBO9780511535093",
    isbn = "978-0-521-03752-5, 978-0-521-81159-0, 978-0-511-02988-2",
    publisher = "Cambridge University Press",
    series = "Cambridge Monographs on Mathematical Physics",
    month = "7",
    year = "2007"
}

@article{Magnano:1993bd,
    author = "Magnano, Guido and Sokolowski, Leszek M.",
    title = "{On physical equivalence between nonlinear gravity theories and a general relativistic selfgravitating scalar field}",
    eprint = "gr-qc/9312008",
    archivePrefix = "arXiv",
    reportNumber = "TO-JLL-P-3-93",
    doi = "10.1103/PhysRevD.50.5039",
    journal = "Phys. Rev. D",
    volume = "50",
    pages = "5039--5059",
    year = "1994"
}

@article{Capozziello:1996xg,
    author = "Capozziello, S. and de Ritis, R. and Marino, Alma Angela",
    title = "{Some aspects of the cosmological conformal equivalence between 'Jordan frame' and 'Einstein frame'}",
    eprint = "gr-qc/9612053",
    archivePrefix = "arXiv",
    doi = "10.1088/0264-9381/14/12/010",
    journal = "Class. Quant. Grav.",
    volume = "14",
    pages = "3243--3258",
    year = "1997"
}

@article{Faraoni:1999hp,
    author = "Faraoni, V. and Gunzig, E.",
    title = "{Einstein frame or Jordan frame?}",
    eprint = "astro-ph/9910176",
    archivePrefix = "arXiv",
    doi = "10.1023/A:1026645510351",
    journal = "Int. J. Theor. Phys.",
    volume = "38",
    pages = "217--225",
    year = "1999"
}

@article{Faraoni:2006fx,
    author = "Faraoni, Valerio and Nadeau, Shahn",
    title = "{The (pseudo)issue of the conformal frame revisited}",
    eprint = "gr-qc/0612075",
    archivePrefix = "arXiv",
    doi = "10.1103/PhysRevD.75.023501",
    journal = "Phys. Rev. D",
    volume = "75",
    pages = "023501",
    year = "2007"
}

@article{Capozziello:2010sc,
    author = "Capozziello, S. and Martin-Moruno, P. and Rubano, C.",
    title = "{Physical non-equivalence of the Jordan and Einstein frames}",
    eprint = "1003.5394",
    archivePrefix = "arXiv",
    primaryClass = "gr-qc",
    doi = "10.1016/j.physletb.2010.04.058",
    journal = "Phys. Lett. B",
    volume = "689",
    pages = "117--121",
    year = "2010"
}

@article{Steinwachs:2011zs,
    author = "Steinwachs, Christian F. and Kamenshchik, Alexander Yu.",
    title = "{One-loop divergences for gravity non-minimally coupled to a multiplet of scalar fields: calculation in the Jordan frame. I. The main results}",
    eprint = "1101.5047",
    archivePrefix = "arXiv",
    primaryClass = "gr-qc",
    doi = "10.1103/PhysRevD.84.024026",
    journal = "Phys. Rev. D",
    volume = "84",
    pages = "024026",
    year = "2011"
}

@article{Ren:2014sya,
    author = "Ren, Jing and Xianyu, Zhong-Zhi and He, Hong-Jian",
    title = "{Higgs Gravitational Interaction, Weak Boson Scattering, and Higgs Inflation in Jordan and Einstein Frames}",
    eprint = "1404.4627",
    archivePrefix = "arXiv",
    primaryClass = "gr-qc",
    reportNumber = "KCL-PH-TH-2014-25",
    doi = "10.1088/1475-7516/2014/06/032",
    journal = "JCAP",
    volume = "06",
    pages = "032",
    year = "2014"
}

@article{Postma:2014vaa,
    author = "Postma, Marieke and Volponi, Marco",
    title = "{Equivalence of the Einstein and Jordan frames}",
    eprint = "1407.6874",
    archivePrefix = "arXiv",
    primaryClass = "astro-ph.CO",
    doi = "10.1103/PhysRevD.90.103516",
    journal = "Phys. Rev. D",
    volume = "90",
    number = "10",
    pages = "103516",
    year = "2014"
}

@article{Kamenshchik:2014waa,
    author = "Kamenshchik, Alexander Yu. and Steinwachs, Christian F.",
    title = "{Question of quantum equivalence between Jordan frame and Einstein frame}",
    eprint = "1408.5769",
    archivePrefix = "arXiv",
    primaryClass = "gr-qc",
    reportNumber = "FR-PHENO-2014-011",
    doi = "10.1103/PhysRevD.91.084033",
    journal = "Phys. Rev. D",
    volume = "91",
    number = "8",
    pages = "084033",
    year = "2015"
}

@article{Domenech:2015qoa,
    author = "Dom{\`e}nech, Guillem and Sasaki, Misao",
    title = "{Conformal Frame Dependence of Inflation}",
    eprint = "1501.07699",
    archivePrefix = "arXiv",
    primaryClass = "gr-qc",
    reportNumber = "YITP-15-6",
    doi = "10.1088/1475-7516/2015/04/022",
    journal = "JCAP",
    volume = "04",
    pages = "022",
    year = "2015"
}

@article{Myrzakulov:2015qaa,
    author = "Myrzakulov, R. and Sebastiani, L. and Vagnozzi, S.",
    title = "{Inflation in $f(R,\phi )$ -theories and mimetic gravity scenario}",
    eprint = "1504.07984",
    archivePrefix = "arXiv",
    primaryClass = "gr-qc",
    doi = "10.1140/epjc/s10052-015-3672-6",
    journal = "Eur. Phys. J. C",
    volume = "75",
    pages = "444",
    year = "2015"
}

@article{Quiros:2015bfa,
    author = "Quiros, Israel and Garc{\'\i}a-Salcedo, Ricardo and Gonzalez, Tame and Horta-Rangel, F. Antonio",
    title = "{Chameleon effect in the Jordan frame of the Brans-Dicke theory}",
    eprint = "1506.05420",
    archivePrefix = "arXiv",
    primaryClass = "gr-qc",
    doi = "10.1103/PhysRevD.92.044055",
    journal = "Phys. Rev. D",
    volume = "92",
    number = "4",
    pages = "044055",
    year = "2015"
}

@article{Sakstein:2015jca,
    author = "Sakstein, Jeremy and Verner, Sarunas",
    title = "{Disformal Gravity Theories: A Jordan Frame Analysis}",
    eprint = "1509.05679",
    archivePrefix = "arXiv",
    primaryClass = "gr-qc",
    doi = "10.1103/PhysRevD.92.123005",
    journal = "Phys. Rev. D",
    volume = "92",
    number = "12",
    pages = "123005",
    year = "2015"
}

@article{vandeBruck:2015gjd,
    author = "van de Bruck, Carsten and Longden, Chris",
    title = "{Higgs Inflation with a Gauss-Bonnet term in the Jordan Frame}",
    eprint = "1512.04768",
    archivePrefix = "arXiv",
    primaryClass = "hep-ph",
    doi = "10.1103/PhysRevD.93.063519",
    journal = "Phys. Rev. D",
    volume = "93",
    number = "6",
    pages = "063519",
    year = "2016"
}

@article{Banerjee:2016lco,
    author = "Banerjee, Narayan and Majumder, Barun",
    title = "{A question mark on the equivalence of Einstein and Jordan frames}",
    eprint = "1601.06152",
    archivePrefix = "arXiv",
    primaryClass = "gr-qc",
    doi = "10.1016/j.physletb.2016.01.022",
    journal = "Phys. Lett. B",
    volume = "754",
    pages = "129--134",
    year = "2016"
}

@article{Kamenshchik:2016gcy,
    author = "Kamenshchik, A. Yu. and Pozdeeva, E. O. and Vernov, Sergey Yu. and Tronconi, A. and Venturi, G.",
    title = "{Transformations between Jordan and Einstein frames: Bounces, antigravity, and crossing singularities}",
    eprint = "1602.07192",
    archivePrefix = "arXiv",
    primaryClass = "gr-qc",
    doi = "10.1103/PhysRevD.94.063510",
    journal = "Phys. Rev. D",
    volume = "94",
    number = "6",
    pages = "063510",
    year = "2016"
}

@article{Bahamonde:2016wmz,
    author = "Bahamonde, Sebastian and Odintsov, S. D. and Oikonomou, V. K. and Wright, Matthew",
    title = "{Correspondence of F(R) gravity singularities in Jordan and Einstein frames}",
    eprint = "1603.05113",
    archivePrefix = "arXiv",
    primaryClass = "gr-qc",
    doi = "10.1016/j.aop.2016.06.020",
    journal = "Annals Phys.",
    volume = "373",
    pages = "96--114",
    year = "2016"
}

@article{Pandey:2016unk,
    author = "Pandey, Sachin and Banerjee, Narayan",
    title = "{Equivalence of Jordan and Einstein frames at the quantum level}",
    eprint = "1610.00584",
    archivePrefix = "arXiv",
    primaryClass = "gr-qc",
    doi = "10.1140/epjp/i2017-11385-0",
    journal = "Eur. Phys. J. Plus",
    volume = "132",
    number = "3",
    pages = "107",
    year = "2017"
}

@article{Jarv:2016sow,
    author = {J{\"a}rv, Laur and Kannike, Kristjan and Marzola, Luca and Racioppi, Antonio and Raidal, Martti and R{\"u}nkla, Mihkel and Saal, Margus and Veerm{\"a}e, Hardi},
    title = "{Frame-Independent Classification of Single-Field Inflationary Models}",
    eprint = "1612.06863",
    archivePrefix = "arXiv",
    primaryClass = "hep-ph",
    doi = "10.1103/PhysRevLett.118.151302",
    journal = "Phys. Rev. Lett.",
    volume = "118",
    number = "15",
    pages = "151302",
    year = "2017"
}

@article{Mathew:2017lvh,
    author = "Mathew, Jose and Johnson, Joseph P. and Shankaranarayanan, S.",
    title = "{Inflation with $f(R,\phi)$ in Jordan frame}",
    eprint = "1705.07945",
    archivePrefix = "arXiv",
    primaryClass = "gr-qc",
    doi = "10.1007/s10714-018-2410-4",
    journal = "Gen. Rel. Grav.",
    volume = "50",
    number = "7",
    pages = "90",
    year = "2018"
}

@article{Karam:2017zno,
    author = "Karam, Alexandros and Pappas, Thomas and Tamvakis, Kyriakos",
    title = "{Frame-dependence of higher-order inflationary observables in scalar-tensor theories}",
    eprint = "1707.00984",
    archivePrefix = "arXiv",
    primaryClass = "gr-qc",
    doi = "10.1103/PhysRevD.96.064036",
    journal = "Phys. Rev. D",
    volume = "96",
    number = "6",
    pages = "064036",
    year = "2017"
}

@article{Azri:2018gsz,
    author = "Azri, Hemza",
    title = "{Are there really conformal frames? Uniqueness of affine inflation}",
    eprint = "1802.01247",
    archivePrefix = "arXiv",
    primaryClass = "gr-qc",
    doi = "10.1142/S0218271818300069",
    journal = "Int. J. Mod. Phys. D",
    volume = "27",
    number = "09",
    pages = "1830006",
    year = "2018"
}

@article{Rinaldi:2018qpu,
    author = "Rinaldi, Massimiliano",
    title = "{On the equivalence of Jordan and Einstein frames in scale-invariant gravity}",
    eprint = "1808.08154",
    archivePrefix = "arXiv",
    primaryClass = "gr-qc",
    doi = "10.1140/epjp/i2018-12213-9",
    journal = "Eur. Phys. J. Plus",
    volume = "133",
    number = "10",
    pages = "408",
    year = "2018"
}

@article{Falls:2018olk,
    author = "Falls, Kevin and Herrero-Valea, Mario",
    title = "{Frame (In)equivalence in Quantum Field Theory and Cosmology}",
    eprint = "1812.08187",
    archivePrefix = "arXiv",
    primaryClass = "hep-th",
    doi = "10.1140/epjc/s10052-019-7070-3",
    journal = "Eur. Phys. J. C",
    volume = "79",
    number = "7",
    pages = "595",
    year = "2019"
}

@article{Nashed:2019yto,
    author = "Nashed, G. G. L. and El Hanafy, W. and Odintsov, S. D. and Oikonomou, V. K.",
    title = "{Thermodynamical correspondence of $f(R)$ gravity in the Jordan and Einstein frames}",
    eprint = "1912.03897",
    archivePrefix = "arXiv",
    primaryClass = "gr-qc",
    doi = "10.1142/S021827182050090X",
    journal = "Int. J. Mod. Phys. D",
    volume = "29",
    number = "13",
    pages = "2050090",
    year = "2020"
}

@article{Giacomini:2020grc,
    author = "Giacomini, Alex and Leon, Genly and Paliathanasis, Andronikos and Pan, Supriya",
    title = "{Cosmological Evolution of Two-Scalar fields Cosmology in the Jordan frame}",
    eprint = "2001.02414",
    archivePrefix = "arXiv",
    primaryClass = "gr-qc",
    doi = "10.1140/epjc/s10052-020-7730-3",
    journal = "Eur. Phys. J. C",
    volume = "80",
    number = "3",
    pages = "184",
    year = "2020"
}

@article{Elizalde:2020icc,
    author = "Elizalde, E. and Nashed, G. G. L. and Nojiri, S. and Odintsov, S. D.",
    title = "{Spherically symmetric black holes with electric and magnetic charge in extended gravity: physical properties, causal structure, and stability analysis in Einstein{\textquoteright}s and Jordan{\textquoteright}s frames}",
    eprint = "2001.11357",
    archivePrefix = "arXiv",
    primaryClass = "gr-qc",
    doi = "10.1140/epjc/s10052-020-7686-3",
    journal = "Eur. Phys. J. C",
    volume = "80",
    number = "2",
    pages = "109",
    year = "2020"
}

@article{Oikonomou:2020oex,
    author = "Oikonomou, V. K.",
    title = "{Rescaled Einstein-Hilbert Gravity from $f(R)$ Gravity: Inflation, Dark Energy and the Swampland Criteria}",
    eprint = "2012.01312",
    archivePrefix = "arXiv",
    primaryClass = "gr-qc",
    doi = "10.1103/PhysRevD.103.124028",
    journal = "Phys. Rev. D",
    volume = "103",
    number = "12",
    pages = "124028",
    year = "2021"
}

@article{Bamonti:2021jmg,
    author = "Bamonti, Nicola and Costantini, Andrea and Montani, Giovanni",
    title = "{Features of the primordial Universe in f(R)-gravity as viewed in the Jordan frame}",
    eprint = "2103.17063",
    archivePrefix = "arXiv",
    primaryClass = "gr-qc",
    doi = "10.1088/1361-6382/ac7694",
    journal = "Class. Quant. Grav.",
    volume = "39",
    number = "17",
    pages = "175011",
    year = "2022"
}

@article{Copeland:2021qby,
    author = "Copeland, Edmund J. and Millington, Peter and Mu{\~n}oz, Sergio Sevillano",
    title = "{Fifth forces and broken scale symmetries in the Jordan frame}",
    eprint = "2111.06357",
    archivePrefix = "arXiv",
    primaryClass = "hep-th",
    doi = "10.1088/1475-7516/2022/02/016",
    journal = "JCAP",
    volume = "02",
    number = "02",
    pages = "016",
    year = "2022"
}

@article{Racioppi:2021jai,
    author = "Racioppi, Antonio and Vasar, Martin",
    title = "{On the number of e-folds in the Jordan and Einstein frames}",
    eprint = "2111.09677",
    archivePrefix = "arXiv",
    primaryClass = "gr-qc",
    doi = "10.1140/epjp/s13360-022-02853-x",
    journal = "Eur. Phys. J. Plus",
    volume = "137",
    number = "5",
    pages = "637",
    year = "2022"
}

@article{Galaverni:2021jcy,
    author = "Galaverni, Matteo and Gionti S. J., Gabriele and J., S.",
    title = "{Jordan and Einstein Frames Hamiltonian Analysis for FLRW Brans-Dicke Theory}",
    eprint = "2112.02098",
    archivePrefix = "arXiv",
    primaryClass = "gr-qc",
    doi = "10.3390/universe8010014",
    journal = "Universe",
    volume = "8",
    number = "1",
    pages = "14",
    year = "2021"
}

@article{Shtanov:2022wpr,
    author = "Shtanov, Yuri",
    title = "{On the Conformal Frames in f(R) Gravity}",
    eprint = "2202.00818",
    archivePrefix = "arXiv",
    primaryClass = "gr-qc",
    doi = "10.3390/universe8020069",
    journal = "Universe",
    volume = "8",
    number = "2",
    pages = "69",
    year = "2022"
}

@article{Paliathanasis:2022tmt,
    author = "Paliathanasis, Andronikos",
    title = "{Hyperbolic Inflation in the Jordan Frame}",
    eprint = "2203.14610",
    archivePrefix = "arXiv",
    primaryClass = "gr-qc",
    doi = "10.3390/universe8040199",
    journal = "Universe",
    volume = "8",
    number = "4",
    pages = "199",
    year = "2022"
}

@article{Nojiri:2022ski,
    author = "Nojiri, S. and Odintsov, S. D. and Oikonomou, V. K.",
    title = "{Integral F(R) gravity and saddle point condition as a remedy for the H0-tension}",
    eprint = "2205.11681",
    archivePrefix = "arXiv",
    primaryClass = "gr-qc",
    doi = "10.1016/j.nuclphysb.2022.115850",
    journal = "Nucl. Phys. B",
    volume = "980",
    pages = "115850",
    year = "2022"
}

@article{Paliathanasis:2022akr,
    author = "Paliathanasis, Andronikos",
    title = "{Analytic Solution and Noether Symmetries for the Hyperbolic Inflationary Model in the Jordan Frame}",
    eprint = "2206.03302",
    archivePrefix = "arXiv",
    primaryClass = "gr-qc",
    doi = "10.3390/universe8060325",
    journal = "Universe",
    volume = "8",
    number = "6",
    pages = "325",
    year = "2022"
}

@article{Karciauskas:2022jzd,
    author = "Kar{\v{c}}iauskas, Mindaugas and D{\'\i}az, Jos{\'e} Jaime Terente",
    title = "{Slow-roll inflation in the Jordan frame}",
    eprint = "2206.08677",
    archivePrefix = "arXiv",
    primaryClass = "gr-qc",
    doi = "10.1103/PhysRevD.106.083526",
    journal = "Phys. Rev. D",
    volume = "106",
    number = "8",
    pages = "083526",
    year = "2022"
}

@article{DeAngelis:2022qhm,
    author = "De Angelis, Mariaveronica and Montani, Giovanni",
    title = "{On the emergence of a classical isotropic universe from a quantum f(R) Bianchi cosmology in the Jordan frame}",
    eprint = "2207.14683",
    archivePrefix = "arXiv",
    primaryClass = "gr-qc",
    doi = "10.1140/epjc/s10052-023-11454-6",
    journal = "Eur. Phys. J. C",
    volume = "83",
    number = "4",
    pages = "285",
    year = "2023"
}

@article{Bamber:2022eoy,
    author = "Bamber, Jamie",
    title = "{Fifth forces and frame invariance}",
    eprint = "2210.06396",
    archivePrefix = "arXiv",
    primaryClass = "gr-qc",
    doi = "10.1103/PhysRevD.107.024013",
    journal = "Phys. Rev. D",
    volume = "107",
    number = "2",
    pages = "024013",
    year = "2023"
}

@article{Odintsov:2022bpg,
    author = "Odintsov, S. D. and Oikonomou, V. K.",
    title = "{Generalized Rp-attractor cosmology in the Jordan and Einstein frames: New type of attractors and revisiting standard Jordan frame Rp inflation}",
    eprint = "2210.11351",
    archivePrefix = "arXiv",
    primaryClass = "gr-qc",
    doi = "10.1142/S0218271822501358",
    journal = "Int. J. Mod. Phys. D",
    volume = "32",
    number = "01",
    pages = "2250135",
    year = "2023"
}

@article{Schiavone:2022wvq,
    author = "Schiavone, Tiziano and Montani, Giovanni and Bombacigno, Flavio",
    title = "{f(R) gravity in the Jordan frame as a paradigm for the Hubble tension}",
    eprint = "2211.16737",
    archivePrefix = "arXiv",
    primaryClass = "gr-qc",
    doi = "10.1093/mnrasl/slad041",
    journal = "Mon. Not. Roy. Astron. Soc.",
    volume = "522",
    number = "1",
    pages = "L72--L77",
    year = "2023"
}

@article{Oikonomou:2023dgu,
    author = "Oikonomou, Vasilis K.",
    title = "{Rp attractors static neutron star phenomenology}",
    eprint = "2301.12136",
    archivePrefix = "arXiv",
    primaryClass = "gr-qc",
    doi = "10.1093/mnras/stad326",
    journal = "Mon. Not. Roy. Astron. Soc.",
    volume = "520",
    number = "2",
    pages = "2934--2941",
    year = "2023"
}

@article{Velasquez:2023jld,
    author = "Vel{\'a}squez, Joel and Hortua, H{\'e}ctor J. and Casta{\~n}eda, Leonardo",
    title = "{About Jordan and Einstein Frames: A Study in Inflationary Magnetogenesis}",
    eprint = "2303.01301",
    archivePrefix = "arXiv",
    primaryClass = "gr-qc",
    doi = "10.3390/universe10090350",
    journal = "Universe",
    volume = "10",
    number = "9",
    pages = "350",
    year = "2024"
}

@article{Diaz:2023tma,
    author = "D{\'\i}az, Jos{\'e} Jaime Terente and Kar{\v{c}}iauskas, Mindaugas",
    title = "{Comoving curvature perturbation in Jordan and Einstein frames}",
    eprint = "2305.15326",
    archivePrefix = "arXiv",
    primaryClass = "gr-qc",
    doi = "10.1103/PhysRevD.108.083535",
    journal = "Phys. Rev. D",
    volume = "108",
    number = "8",
    pages = "083535",
    year = "2023"
}

@article{Luongo:2024opv,
    author = "Luongo, Orlando and Mengoni, Tommaso",
    title = "{Generalized K-essence inflation in Jordan and Einstein frames}",
    eprint = "2309.03065",
    archivePrefix = "arXiv",
    primaryClass = "gr-qc",
    doi = "10.1088/1361-6382/ad3ac9",
    journal = "Class. Quant. Grav.",
    volume = "41",
    number = "10",
    pages = "105006",
    year = "2024"
}

@article{GiontiSJ:2023tgx,
    author = "Gionti S. J., Gabriele and Galaverni, Matteo",
    title = "{On the canonical equivalence between Jordan and Einstein frames}",
    eprint = "2310.09539",
    archivePrefix = "arXiv",
    primaryClass = "gr-qc",
    doi = "10.1140/epjc/s10052-024-12586-z",
    journal = "Eur. Phys. J. C",
    volume = "84",
    number = "3",
    pages = "265",
    year = "2024"
}

@article{Seleim:2023enf,
    author = "Seleim, Karim H. and Arya, Richa and Jor{\'a}s, Sergio E.",
    title = {{Parametric resonance in the Einstein frame: The Jordan-frame Doppelg{\"a}nger}},
    eprint = "2312.13689",
    archivePrefix = "arXiv",
    primaryClass = "gr-qc",
    doi = "10.1016/j.dark.2024.101751",
    journal = "Phys. Dark Univ.",
    volume = "47",
    pages = "101751",
    year = "2025"
}

@article{Belfiglio:2024swy,
    author = "Belfiglio, Alessio and Luongo, Orlando and Mengoni, Tommaso",
    title = "{Comparing geometric and gravitational particle production in Jordan and Einstein frames}",
    eprint = "2411.11130",
    archivePrefix = "arXiv",
    primaryClass = "gr-qc",
    doi = "10.1103/yf63-knqx",
    journal = "Phys. Rev. D",
    volume = "111",
    number = "12",
    pages = "123512",
    year = "2025"
}

@article{Galaverni:2025acu,
    author = "Galaverni, Matteo and Gionti, Gabriele",
    title = "{Spherically symmetric geometrodynamics in Jordan and Einstein frames}",
    eprint = "2501.08364",
    archivePrefix = "arXiv",
    primaryClass = "gr-qc",
    doi = "10.1140/epjc/s10052-025-14447-9",
    journal = "Eur. Phys. J. C",
    volume = "85",
    number = "7",
    pages = "740",
    year = "2025"
}

@article{Wang:2025ger,
    author = "Wang, Hang and Liu, Shuang and Li, Yu and Wang, Yao-chuan",
    title = "{The equivalence between Einstein and Jordan frames: a study based on the inflationary magnetogenesis model}",
    eprint = "2504.18005",
    archivePrefix = "arXiv",
    primaryClass = "astro-ph.CO",
    month = "4",
    year = "2025"
}

@article{Brax:2004ym,
    author = "Brax, Ph. and van de Bruck, Carsten and Davis, A. C.",
    title = "{Is the radion a chameleon?}",
    eprint = "astro-ph/0408464",
    archivePrefix = "arXiv",
    doi = "10.1088/1475-7516/2004/11/004",
    journal = "JCAP",
    volume = "11",
    pages = "004",
    year = "2004"
}

@article{Brax:2004qh,
    author = "Brax, Philippe and van de Bruck, Carsten and Davis, Anne-Christine and Khoury, Justin and Weltman, Amanda",
    title = "{Detecting dark energy in orbit: The cosmological chameleon}",
    eprint = "astro-ph/0408415",
    archivePrefix = "arXiv",
    doi = "10.1103/PhysRevD.70.123518",
    journal = "Phys. Rev. D",
    volume = "70",
    pages = "123518",
    year = "2004"
}

@article{Capozziello:2007eu,
    author = "Capozziello, Salvatore and Tsujikawa, Shinji",
    title = "{Solar system and equivalence principle constraints on f(R) gravity by chameleon approach}",
    eprint = "0712.2268",
    archivePrefix = "arXiv",
    primaryClass = "gr-qc",
    doi = "10.1103/PhysRevD.77.107501",
    journal = "Phys. Rev. D",
    volume = "77",
    pages = "107501",
    year = "2008"
}

@article{Brax:2008hh,
    author = "Brax, Philippe and van de Bruck, Carsten and Davis, Anne-Christine and Shaw, Douglas J.",
    title = "{f(R) Gravity and Chameleon Theories}",
    eprint = "0806.3415",
    archivePrefix = "arXiv",
    primaryClass = "astro-ph",
    doi = "10.1103/PhysRevD.78.104021",
    journal = "Phys. Rev. D",
    volume = "78",
    pages = "104021",
    year = "2008"
}

@article{Brax:2010xq,
    author = "Brax, Philippe and Zioutas, Konstantin",
    title = "{Solar Chameleons}",
    eprint = "1004.1846",
    archivePrefix = "arXiv",
    primaryClass = "astro-ph.SR",
    doi = "10.1103/PhysRevD.82.043007",
    journal = "Phys. Rev. D",
    volume = "82",
    pages = "043007",
    year = "2010"
}

@article{Brax:2010kv,
    author = "Brax, Philippe and van de Bruck, Carsten and Mota, David F. and Nunes, Nelson J. and Winther, Hans A.",
    title = "{Chameleons with Field Dependent Couplings}",
    eprint = "1006.2796",
    archivePrefix = "arXiv",
    primaryClass = "astro-ph.CO",
    doi = "10.1103/PhysRevD.82.083503",
    journal = "Phys. Rev. D",
    volume = "82",
    pages = "083503",
    year = "2010"
}

@article{Gannouji:2010fc,
    author = "Gannouji, Radouane and Moraes, Bruno and Mota, David F. and Polarski, David and Tsujikawa, Shinji and Winther, Hans A.",
    title = "{Chameleon dark energy models with characteristic signatures}",
    eprint = "1010.3769",
    archivePrefix = "arXiv",
    primaryClass = "astro-ph.CO",
    doi = "10.1103/PhysRevD.82.124006",
    journal = "Phys. Rev. D",
    volume = "82",
    pages = "124006",
    year = "2010"
}

@article{Brax:2011wp,
    author = "Brax, Philippe and Lindner, Axel and Zioutas, Konstantin",
    title = "{Detection prospects for solar and terrestrial chameleons}",
    eprint = "1110.2583",
    archivePrefix = "arXiv",
    primaryClass = "hep-ph",
    doi = "10.1103/PhysRevD.85.043014",
    journal = "Phys. Rev. D",
    volume = "85",
    pages = "043014",
    year = "2012"
}

@article{Wang:2012kj,
    author = "Wang, Junpu and Hui, Lam and Khoury, Justin",
    title = "{No-Go Theorems for Generalized Chameleon Field Theories}",
    eprint = "1208.4612",
    archivePrefix = "arXiv",
    primaryClass = "astro-ph.CO",
    doi = "10.1103/PhysRevLett.109.241301",
    journal = "Phys. Rev. Lett.",
    volume = "109",
    pages = "241301",
    year = "2012"
}

@article{Khoury:2013yya,
    author = "Khoury, Justin",
    title = "{Chameleon Field Theories}",
    eprint = "1306.4326",
    archivePrefix = "arXiv",
    primaryClass = "astro-ph.CO",
    doi = "10.1088/0264-9381/30/21/214004",
    journal = "Class. Quant. Grav.",
    volume = "30",
    pages = "214004",
    year = "2013"
}

@article{Erickcek:2013dea,
    author = "Erickcek, Adrienne L. and Barnaby, Neil and Burrage, Clare and Huang, Zhiqi",
    title = "{Chameleons in the Early Universe: Kicks, Rebounds, and Particle Production}",
    eprint = "1310.5149",
    archivePrefix = "arXiv",
    primaryClass = "astro-ph.CO",
    doi = "10.1103/PhysRevD.89.084074",
    journal = "Phys. Rev. D",
    volume = "89",
    number = "8",
    pages = "084074",
    year = "2014"
}

@article{Elder:2016yxm,
    author = {Elder, Benjamin and Khoury, Justin and Haslinger, Philipp and Jaffe, Matt and M{\"u}ller, Holger and Hamilton, Paul},
    title = "{Chameleon Dark Energy and Atom Interferometry}",
    eprint = "1603.06587",
    archivePrefix = "arXiv",
    primaryClass = "astro-ph.CO",
    doi = "10.1103/PhysRevD.94.044051",
    journal = "Phys. Rev. D",
    volume = "94",
    number = "4",
    pages = "044051",
    year = "2016"
}

@article{Brax:2016did,
    author = "Brax, Philippe and Burrage, Clare and Englert, Christoph and Spannowsky, Michael",
    title = "{LHC Signatures Of Scalar Dark Energy}",
    eprint = "1604.04299",
    archivePrefix = "arXiv",
    primaryClass = "hep-ph",
    reportNumber = "IPPP-16-31, DCPT-16-62",
    doi = "10.1103/PhysRevD.94.084054",
    journal = "Phys. Rev. D",
    volume = "94",
    number = "8",
    pages = "084054",
    year = "2016"
}

@article{Burrage:2016bwy,
    author = "Burrage, Clare and Sakstein, Jeremy",
    title = "{A Compendium of Chameleon Constraints}",
    eprint = "1609.01192",
    archivePrefix = "arXiv",
    primaryClass = "astro-ph.CO",
    doi = "10.1088/1475-7516/2016/11/045",
    journal = "JCAP",
    volume = "11",
    pages = "045",
    year = "2016"
}

@article{Burrage:2017shh,
    author = "Burrage, Clare and Copeland, Edmund J. and Moss, Adam and Stevenson, James A.",
    title = "{The shape dependence of chameleon screening}",
    eprint = "1711.02065",
    archivePrefix = "arXiv",
    primaryClass = "astro-ph.CO",
    doi = "10.1088/1475-7516/2018/01/056",
    journal = "JCAP",
    volume = "01",
    pages = "056",
    year = "2018"
}

@article{Burrage:2018pyg,
    author = {Burrage, Clare and K{\"a}ding, Christian and Millington, Peter and Min{\'a}{\v{r}}, Ji{\v{r}}{\'\i}},
    title = "{Open quantum dynamics induced by light scalar fields}",
    eprint = "1812.08760",
    archivePrefix = "arXiv",
    primaryClass = "hep-th",
    doi = "10.1103/PhysRevD.100.076003",
    journal = "Phys. Rev. D",
    volume = "100",
    number = "7",
    pages = "076003",
    year = "2019"
}

@article{Katsuragawa:2019uto,
    author = "Katsuragawa, Taishi and Nakamura, Tomohiro and Ikeda, Taishi and Capozziello, Salvatore",
    title = "{Gravitational Waves in $F(R)$ Gravity: Scalar Waves and the Chameleon Mechanism}",
    eprint = "1902.02494",
    archivePrefix = "arXiv",
    primaryClass = "gr-qc",
    doi = "10.1103/PhysRevD.99.124050",
    journal = "Phys. Rev. D",
    volume = "99",
    number = "12",
    pages = "124050",
    year = "2019"
}

@article{Sakstein:2019qgn,
    author = "Sakstein, Jeremy and Desmond, Harry and Jain, Bhuvnesh",
    title = "{Screened Fifth Forces Mediated by Dark Matter--Baryon Interactions: Theory and Astrophysical Probes}",
    eprint = "1907.03775",
    archivePrefix = "arXiv",
    primaryClass = "astro-ph.CO",
    doi = "10.1103/PhysRevD.100.104035",
    journal = "Phys. Rev. D",
    volume = "100",
    number = "10",
    pages = "104035",
    year = "2019"
}

@article{Desmond:2019ygn,
    author = "Desmond, Harry and Jain, Bhuvnesh and Sakstein, Jeremy",
    title = "{Local resolution of the Hubble tension: The impact of screened fifth forces on the cosmic distance ladder}",
    eprint = "1907.03778",
    archivePrefix = "arXiv",
    primaryClass = "astro-ph.CO",
    doi = "10.1103/PhysRevD.100.043537",
    journal = "Phys. Rev. D",
    volume = "100",
    number = "4",
    pages = "043537",
    year = "2019",
    note = "[Erratum: Phys.Rev.D 101, 069904 (2020), Erratum: Phys.Rev.D 101, 129901 (2020)]"
}

@article{Hartley:2019wzu,
    author = {Hartley, Daniel and K{\"a}ding, Christian and Howl, Richard and Fuentes, Ivette},
    title = "{Quantum-enhanced screened dark energy detection}",
    eprint = "1909.02272",
    archivePrefix = "arXiv",
    primaryClass = "gr-qc",
    doi = "10.1140/epjc/s10052-023-12360-7",
    journal = "Eur. Phys. J. C",
    volume = "84",
    number = "1",
    pages = "49",
    year = "2024"
}

@article{Vagnozzi:2019kvw,
    author = "Vagnozzi, Sunny and Visinelli, Luca and Mena, Olga and Mota, David F.",
    title = "{Do we have any hope of detecting scattering between dark energy and baryons through cosmology?}",
    eprint = "1911.12374",
    archivePrefix = "arXiv",
    primaryClass = "gr-qc",
    doi = "10.1093/mnras/staa311",
    journal = "Mon. Not. Roy. Astron. Soc.",
    volume = "493",
    number = "1",
    pages = "1139--1152",
    year = "2020"
}

@article{Cai:2021wgv,
    author = "Cai, Rong-Gen and Guo, Zong-Kuan and Li, Li and Wang, Shao-Jiang and Yu, Wang-Wei",
    title = "{Chameleon dark energy can resolve the Hubble tension}",
    eprint = "2102.02020",
    archivePrefix = "arXiv",
    primaryClass = "astro-ph.CO",
    doi = "10.1103/PhysRevD.103.L121302",
    journal = "Phys. Rev. D",
    volume = "103",
    number = "12",
    pages = "121302",
    year = "2021"
}

@article{Vagnozzi:2021quy,
    author = "Vagnozzi, Sunny and Visinelli, Luca and Brax, Philippe and Davis, Anne-Christine and Sakstein, Jeremy",
    title = "{Direct detection of dark energy: The XENON1T excess and future prospects}",
    eprint = "2103.15834",
    archivePrefix = "arXiv",
    primaryClass = "hep-ph",
    doi = "10.1103/PhysRevD.104.063023",
    journal = "Phys. Rev. D",
    volume = "104",
    number = "6",
    pages = "063023",
    year = "2021"
}

@article{Karwal:2021vpk,
    author = "Karwal, Tanvi and Raveri, Marco and Jain, Bhuvnesh and Khoury, Justin and Trodden, Mark",
    title = "{Chameleon early dark energy and the Hubble tension}",
    eprint = "2106.13290",
    archivePrefix = "arXiv",
    primaryClass = "astro-ph.CO",
    doi = "10.1103/PhysRevD.105.063535",
    journal = "Phys. Rev. D",
    volume = "105",
    number = "6",
    pages = "063535",
    year = "2022"
}

@article{Katsuragawa:2021wmw,
    author = "Katsuragawa, Taishi and Matsuzaki, Shinya and Homma, Kensuke",
    title = "{Hunting dark energy with pressure-dependent photon-photon scattering}",
    eprint = "2107.00478",
    archivePrefix = "arXiv",
    primaryClass = "gr-qc",
    doi = "10.1103/PhysRevD.106.044011",
    journal = "Phys. Rev. D",
    volume = "106",
    number = "4",
    pages = "044011",
    year = "2022"
}

@article{Dima:2021pwx,
    author = "Dima, Alexandru and Bezares, Miguel and Barausse, Enrico",
    title = "{Dynamical chameleon neutron stars: Stability, radial oscillations, and scalar radiation in spherical symmetry}",
    eprint = "2107.04359",
    archivePrefix = "arXiv",
    primaryClass = "gr-qc",
    doi = "10.1103/PhysRevD.104.084017",
    journal = "Phys. Rev. D",
    volume = "104",
    number = "8",
    pages = "084017",
    year = "2021"
}

@article{Benisty:2021cmq,
    author = "Benisty, David and Davis, Anne-Christine",
    title = "{Dark energy interactions near the Galactic Center}",
    eprint = "2108.06286",
    archivePrefix = "arXiv",
    primaryClass = "astro-ph.CO",
    doi = "10.1103/PhysRevD.105.024052",
    journal = "Phys. Rev. D",
    volume = "105",
    number = "2",
    pages = "024052",
    year = "2022"
}

@article{Tamosiunas:2021kth,
    author = "Tamosiunas, Andrius and Briddon, Chad and Burrage, Clare and Cui, Weiguang and Moss, Adam",
    title = "{Chameleon screening depends on the shape and structure of NFW halos}",
    eprint = "2108.10364",
    archivePrefix = "arXiv",
    primaryClass = "gr-qc",
    doi = "10.1088/1475-7516/2022/04/047",
    journal = "JCAP",
    volume = "04",
    number = "04",
    pages = "047",
    year = "2022"
}

@article{Briddon:2021etm,
    author = "Briddon, Chad and Burrage, Clare and Moss, Adam and Tamosiunas, Andrius",
    title = "{SELCIE: a tool for investigating the chameleon field of arbitrary sources}",
    eprint = "2110.11917",
    archivePrefix = "arXiv",
    primaryClass = "gr-qc",
    doi = "10.1088/1475-7516/2021/12/043",
    journal = "JCAP",
    volume = "12",
    number = "12",
    pages = "043",
    year = "2021"
}

@article{Brax:2021owd,
    author = "Brax, Philippe and Davis, Anne-Christine and Elder, Benjamin",
    title = "{(g-2){\ensuremath{\mu}} and screened modified gravity}",
    eprint = "2111.01188",
    archivePrefix = "arXiv",
    primaryClass = "hep-ph",
    doi = "10.1103/PhysRevD.106.044040",
    journal = "Phys. Rev. D",
    volume = "106",
    number = "4",
    pages = "044040",
    year = "2022"
}

@article{Ferlito:2022mok,
    author = "Ferlito, Fulvio and Vagnozzi, Sunny and Mota, David F. and Baldi, Marco",
    title = "{Cosmological direct detection of dark energy: Non-linear structure formation signatures of dark energy scattering with visible matter}",
    eprint = "2201.04528",
    archivePrefix = "arXiv",
    primaryClass = "astro-ph.CO",
    doi = "10.1093/mnras/stac649",
    journal = "Mon. Not. Roy. Astron. Soc.",
    volume = "512",
    number = "2",
    pages = "1885--1905",
    year = "2022"
}

@article{Yuan:2022cpw,
    author = "Yuan, Guan-Wen and Zu, Lei and Feng, Lei and Cai, Yi-Fu and Fan, Yi-Zhong",
    title = "{Is the W-boson mass enhanced by the axion-like particle, dark photon, or chameleon dark energy?}",
    eprint = "2204.04183",
    archivePrefix = "arXiv",
    primaryClass = "hep-ph",
    doi = "10.1007/s11433-022-2011-8",
    journal = "Sci. China Phys. Mech. Astron.",
    volume = "65",
    number = "12",
    pages = "129512",
    year = "2022"
}

@article{Chakrabarti:2022zvv,
    author = "Chakrabarti, Soumya and Dutta, Koushik and Said Levi, Jackson",
    title = "{Screening mechanism and late-time cosmology: Role of a Chameleon{\textendash}Brans{\textendash}Dicke scalar field}",
    eprint = "2205.03789",
    archivePrefix = "arXiv",
    primaryClass = "gr-qc",
    doi = "10.1093/mnras/stac1321",
    journal = "Mon. Not. Roy. Astron. Soc.",
    volume = "514",
    number = "1",
    pages = "427--439",
    year = "2022"
}

@article{Tamosiunas:2022tic,
    author = "Tamosiunas, Andrius and Briddon, Chad and Burrage, Clare and Cutforth, Alan and Moss, Adam and Vincent, Thomas",
    title = "{Chameleon screening in cosmic voids}",
    eprint = "2206.06480",
    archivePrefix = "arXiv",
    primaryClass = "gr-qc",
    doi = "10.1088/1475-7516/2022/11/056",
    journal = "JCAP",
    volume = "11",
    pages = "056",
    year = "2022"
}

@article{Brax:2022olf,
    author = "Brax, Philippe and Davis, Anne-Christine and Elder, Benjamin",
    title = "{Screened scalar fields in hydrogen and muonium}",
    eprint = "2207.11633",
    archivePrefix = "arXiv",
    primaryClass = "hep-ph",
    doi = "10.1103/PhysRevD.107.044008",
    journal = "Phys. Rev. D",
    volume = "107",
    number = "4",
    pages = "044008",
    year = "2023"
}

@article{Benisty:2022lox,
    author = "Benisty, David and Brax, Philippe and Davis, Anne-Christine",
    title = "{Stringent pulsar timing bounds on light scalar couplings to matter}",
    eprint = "2212.03098",
    archivePrefix = "arXiv",
    primaryClass = "gr-qc",
    reportNumber = "CERN-TH-2022-207",
    doi = "10.1103/PhysRevD.107.064049",
    journal = "Phys. Rev. D",
    volume = "107",
    number = "6",
    pages = "064049",
    year = "2023"
}

@article{Boumechta:2023qhd,
    author = "Boumechta, Yacer and Haridasu, Balakrishna S. and Pizzuti, Lorenzo and Butt, Minahil Adil and Baccigalupi, Carlo and Lapi, Andrea",
    title = "{Constraining chameleon screening using galaxy cluster dynamics}",
    eprint = "2303.02074",
    archivePrefix = "arXiv",
    primaryClass = "astro-ph.CO",
    doi = "10.1103/PhysRevD.108.044007",
    journal = "Phys. Rev. D",
    volume = "108",
    number = "4",
    pages = "044007",
    year = "2023"
}

@article{Elder:2023oar,
    author = "Elder, Benjamin and Sakstein, Jeremy",
    title = "{Constraining the chameleon-photon coupling with atomic spectroscopy}",
    eprint = "2305.15638",
    archivePrefix = "arXiv",
    primaryClass = "hep-ph",
    doi = "10.1103/PhysRevD.109.124007",
    journal = "Phys. Rev. D",
    volume = "109",
    number = "12",
    pages = "124007",
    year = "2024"
}

@article{Benisty:2023dkn,
    author = "Benisty, David and Brax, Philippe and Davis, Anne-Christine",
    title = "{Multiscale constraints on scalar-field couplings to matter: The geodetic and frame-dragging effects}",
    eprint = "2305.19977",
    archivePrefix = "arXiv",
    primaryClass = "gr-qc",
    doi = "10.1103/PhysRevD.108.063031",
    journal = "Phys. Rev. D",
    volume = "108",
    number = "6",
    pages = "063031",
    year = "2023"
}

@article{Paliathanasis:2023ttu,
    author = "Paliathanasis, Andronikos",
    title = "{Dynamical Analysis in Chameleon Dark Energy}",
    eprint = "2306.03880",
    archivePrefix = "arXiv",
    primaryClass = "gr-qc",
    doi = "10.1002/prop.202300088",
    journal = "Fortsch. Phys.",
    volume = "71",
    number = "8-9",
    pages = "2300088",
    year = "2023"
}

@article{Morales-Navarrete:2025lzp,
    author = "Morales-Navarrete, Gerardo and Cervantes-Cota, Jorge L.",
    title = "{Fullshape power spectrum for the Symmetron modified gravity model}",
    eprint = "2506.09304",
    archivePrefix = "arXiv",
    primaryClass = "astro-ph.CO",
    month = "6",
    year = "2025"
}

\end{document}